\documentclass{ws-rv9x6}
\usepackage{subfigure}   
\usepackage{ws-rv-thm}   
\usepackage{ws-rv-van}   
\usepackage{url} 
\makeindex

\begin{document}

\chapter[High-Power Single-Frequency Fiber Amplifiers]{High-Power Single-Frequency Fiber Amplifiers} \label{ra_ch1}

\author[C.-W.~Chen]{Chun-Wei~Chen}

\address{Department of Physics, University of Bath, Bath BA2 7AY, UK \\
Department of Applied Physics, Yale University, New Haven, CT, USA \\ 
Edward L.\ Ginzton Laboratory, Stanford University, Stanford, CA, USA 
\\ 
For correspondence: \texttt{cwc218@bath.ac.uk} }

\begin{abstract}
High-power single-frequency fiber amplifiers are increasingly dominant in technologies that require high optical power, stability, and coherence simultaneously. 
In this chapter, we first provide an overview of their key characteristics, applications, architectures, and components.
We then review the nonlinear acousto- and thermo-optical instabilities that limit power scaling and summarize the primary mitigation techniques.
Finally, we discuss multimode excitation with wavefront shaping and anti-Stokes fluorescence cooling as two emerging strategies to circumvent the limitations of current approaches. 
\end{abstract}


\body


\section{Introduction}\label{ra_sec1}
Fiber lasers bring together compactness, high efficiency, low cost, and superior thermal management, making them a compelling and increasingly dominant light source across diverse technologies, from optical communications\cite{kikuchi2015fundamentals} and materials processing\cite{kawahito2018ultra, bakhtari2024review} to remote sensing\cite{cariou2006laser, canat2016high, wang20201645}, directed energy\cite{grandidier2021laser, karr2024new, holten2024beam}, and space photonics\cite{rochat2001fiber, anthony2019fiber, karr2024synthetic}. 
Many of these applications demand laser powers ranging from watts (W) to kilowatts (kW) or higher. To produce high-power output while maintaining diffraction-limited performance (with a beam quality factor of $M^2 \approx 1$) and the desired optical properties, a master-oscillator power-amplifier (MOPA) comprising double-clad active fibers has so far proven to be one of the most effective architectures.\cite{digonnet2001rare, zenteno2002high, richardson2010high, jauregui2013high, zervas2014high, dong2025past} (Note: $M^2$ is defined as the product of a beam's width and divergence angle relative to that of an ideal Gaussian beam.)

Among the various laser types achievable with MOPA, this chapter focuses on high-power single-frequency fiber amplifiers, which are in high demand for technologies that require high coherence, stability, and/or spectral purity at elevated power levels.\cite{fu2017review, li2023high} Key applications include directed energy, coherent lidar, and gravitational-wave detection. The ideal concept of a ``single-frequency'' source outputs \textit{monochromatic} light, meaning zero spectral linewidth, absence of phase and frequency noise,\cite{di2010simple} and infinitely high coherence. However, as no real-world laser is truly monochromatic, the term practically refers to lasers with an extremely narrow linewidth, typically below one megahertz (MHz). A linewidth of 1~MHz in frequency is equivalent to several femtometers (fm) in wavelength in the near-infrared regime (with a center wavelength around 1--2~\textmu m) and corresponds to a longitudinal (temporal) coherence length $L_{\rm c}$ on the order of 100~meters.

\begin{figure}
\centerline{\includegraphics[width=\textwidth]{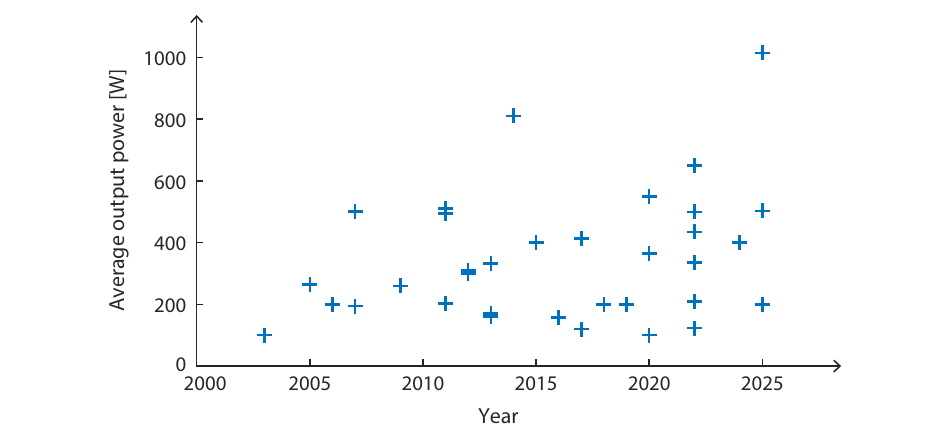}}
\caption{
\textbf{Power scaling of single-frequency fiber amplifiers.} Overview of average output powers reported for single-frequency Yb-doped fiber amplifiers over the past 15~years. All data points exceed 100~W and are taken from publications cited herein.  
}
\label{fig:yearPower} 
\end{figure}

Compared to bulk gain media (e.g., laser crystals), fiber amplifiers benefit from light guiding, which provides excellent beam quality, a long gain length for high pump efficiency, and a large surface-area-to-volume ratio for effective heat dissipation. However, concentrating so much power in a small core ($\sim$10~\textmu m in diameter) yields significantly high intensity, readily triggering nonlinear optical effects that limit power scaling of fiber lasers.  In single-frequency fiber amplifiers, the nonlinear effects with the lowest thresholds are stimulated Brillouin scattering (SBS) and transverse mode instability (TMI).\cite{lee2015transverse, panbhiharwala2018investigation, young2022tradeoff} SBS is the nonlinear backscattering of light by sound waves induced by optical forces,\cite{kobyakov2009stimulated} while TMI is the thermo-optically induced dynamic coupling between the fundamental and higher-order transverse modes.\cite{jauregui2020transverse} Consequently, recent advancements in single-frequency fiber amplifiers have focused on mitigating these nonlinear effects to push the maximum output power, which currently stands at 1.0~kW [Fig.~\ref{fig:yearPower}].\cite{li2025functional}

\subsection{Linewidth and power needed}
Here, we outline the power and linewidth requirements for three example applications of high-power single-frequency fiber amplifiers.

\textbf{Directed energy.} Delivering a laser beam with sufficiently high power (0.1--1~MW) to a small distant target often requires the coherent combination of multiple laser beams through optical interference, ensuring their powers add up constructively within the target spot.\cite{fathi2021towards, linslal2022challenges, holten2024beam, karr2024new} The laser linewidth $\Delta \nu_{\rm s}$ determines the longitudinal coherence length $L_{\rm c}$, which defines the distance over which multiple laser beams can interfere with each other. $L_{\rm c} = F c/(n\Delta \nu_{\rm s})$, where $F$ is a prefactor that depends on the lineshape,\cite{akcay2002estimation} $c$ is the speed of light in vacuum, and $n$ is the refractive index. When the coherence length is short, the optical path lengths of the laser beams must be matched with high precision. Balancing power and linewidth requirements, the linewidth of most fiber lasers used in directed energy is on the order of 1--10~GHz ($L_{\rm c} \sim$ 1--10~cm) at kilowatts.\cite{creeden2018advanced, nicholson20235, edgecumbe2024fiber, holten2024beam} 

\textbf{Coherent lidar.} An object's distance and motion can be measured by interfering the backscattered signal from the object with a reference signal from the local oscillator (optical heterodyne detection).\cite{cariou2006laser, canat2016high, wang20201645} The coherence length, which is inversely proportional to the laser linewidth, places a strong constraint on the maximum detectable range.\cite{bayer2021single, atalar20243d} Since the intensity of the backscattered signal decreases rapidly with propagation distance, this range is also limited by the laser power determining how much light is available to maintain a sufficient signal-to-noise ratio (SNR). For continuous-wave lidar, kilometer-scale detection typically requires laser powers on the order of 0.1--10~W and linewidths of about 10--100~kHz.\cite{cariou2006laser, bayer2021single, atalar20243d} Furthermore, a moving object’s velocity is obtained from the Doppler shift of the backscattered signal relative to the local oscillator frequency, so the laser linewidth determines the minimum resolvable frequency shift and therefore the lowest measurable velocity. For example, a 100-kHz laser can detect objects moving as slowly as $\sim$10~cm/s.\cite{bayer2021single} As detection ranges continue to increase, lasers with narrower linewidths and higher output powers are required. A particularly demanding application is ground-based, high-resolution imaging of space objects at distances of $\sim$100~km to $\sim$100,000~km (toward the Moon and possibly beyond).\cite{karr2024synthetic} In this context, the terms \textit{ladar} and \textit{laser radar} are commonly used to emphasize the functional parallels with radio-frequency or microwave radars.\cite{richmond2010ladar} Such applications may require laser linewidths down to the hertz level and output powers reaching 100~kW or higher.\cite{karr2024synthetic} 

\textbf{Gravitational-wave detection.} These detectors are Michelson interferometers with each arms being a 3--4-km-long Fabry--Perot cavity,\cite{lyons2000shot} which effectively increase the optical path length and power. They require lasers with exceptionally high coherence to resolve mirror displacements far smaller than the width of a proton, and even small amounts of laser phase noise can obscure the true gravitational-wave signal.\cite{willke2008stabilized, willke2010stabilized, kwee2012stabilized} Therefore, extremely high phase stability is needed, demanding a laser with an intrinsic kHz linewidth, which is then actively stabilized to achieve a linewidth of 0.1--1~Hz.\cite{hall2016laser, steinke2017single, de2017single, buikema2019narrow, wellmann2019high, kapasi2020tunable, hochheim2021single, wellmann2021low, cahillane2021laser, meylahn2022stabilized} To reach the required signal-to-noise ratio, the injected laser power is $\sim$10--100~W with ultralow power noise, which is then built up to nearly 1~MW circulating in the arm cavities.\cite{willke2008stabilized, willke2010stabilized, kwee2012stabilized, jia2021point} For such long cavities, high beam quality and pointing stability are also critical to enable efficient power recycling and precise mode matching.\cite{kwee2012stabilized, hall2016laser}

\subsection{Master-oscillator power-amplifier (MOPA)}
\begin{figure}
\centerline{\includegraphics[width=\textwidth]{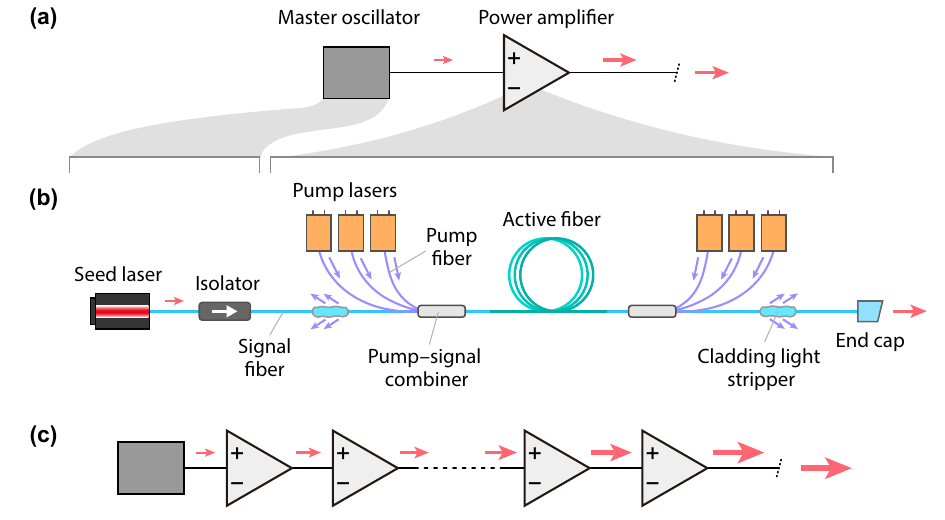}}
\caption{
\textbf{Master-oscillator power-amplifier (MOPA).} 
(a) Conceptual illustration, where a laser signal generated by the master oscillator is amplified by a power amplifier.  
(b) Fiber-based implementation. 
(c) Cascaded amplifier stages for power scaling.
}
\label{fig:MOPA} 
\end{figure}

MOPA\cite{koester1964amplification} begins with a seed laser (the master oscillator) that generates an optical signal with desired spectral and temporal properties but at relatively low power (up to $\sim$0.1~W) [Fig.~\ref{fig:MOPA}(a,b)]. This signal first passes through an optical isolator, which protects the source from linear and nonlinear backscattering that could otherwise destabilize or damage the seed laser. The signal then enters a pump--signal combiner, where it is combined with pump light and launched into a rare-earth-doped fiber (the power amplifier). The pump light excites the rare-earth ions, and the signal photons then trigger these excited ions to release identical photons in a chain of stimulated-emission events, thereby amplifying the signal power. The active fiber's output end (where the amplified signal enters free space) is usually angle-cleaved by $\gtrsim 4^\circ$ or end-capped to minimize the back-reflection of both the signal and amplified spontaneous emission (ASE) into the doped core.\cite{sinha2007investigation, aydin2019endcapping, nicholson2015axicons, han2019fiber, chen2023functional} This prevents strong (re-amplified) backward signal and parasitic lasing (self-lasing), both of which could damage upstream components. While angled cleaving is suitable for powers up to a few tens of watts, end-capping is generally required to handle higher powers.   

Minimizing unwanted light is crucial throughout the fiber amplifier, not just at the output. Cladding-light strippers are often integrated within the amplifier on the side opposite the pump combiners, as well as between the pump lasers and combiners.\cite{boyd2016co2, yan2017kilowatt, liu20202, chen2023functional} They are implemented by first stripping a section of the fiber, then either coating the bare cladding with a higher-index epoxy\cite{mermelstein2007all, yan2017kilowatt} or roughening the cladding surface\cite{boyd2016co2, liu20202}. The strippers remove residual pump light to avoid thermal damage to surrounding components and free-space optics, and prevent backscattered signal and ASE from re-entering the core and becoming re-amplified. 

To effectively suppress ASE in a high-power fiber amplifier, a key strategy is to saturate the optical gain with sufficiently strong seed signal. When the initial seed power is low, it is common to cascade power amplifiers to ensure gain saturation throughout the system [Fig.~\ref{fig:MOPA}(c)]. Gain saturation also suppresses the relative intensity noise (RIN) transferred from the seed laser to the output signal\cite{hildebrandt2008brillouin, zeng2023simultaneous} and transverse mode instability (see Sec.~\ref{sec:tmiGS}). To enhance ASE suppression, tilted fiber Bragg gratings can be spliced to the active fibers to couple them out of the core.\cite{chen2023functional, morasse2024efficient, leleux2025high} 

Effective thermal management of all MOPA components is essential for stable operation and to prevent damage.\cite{davis1998thermal, brown2001thermal, wang2004thermal, galvanauskas2004high, li20053, lapointe2009thermal, stolov2009thermal, hansen2011thermo, fan2011thermal, dong2016thermal, daniel2016metal, charles2016diode, zervas2019transverse, ballato2024prospects} For output powers above a few tens of watts, active cooling using liquid-cooled metal baseplates is often required.\cite{galvanauskas2004high, lapointe2009thermal, fan2011thermal} 

\subsection{Double-clad fiber} \label{sec:dcf}
A \textit{single-clad} single-mode fiber amplifier produces diffraction-limited signal output (below the nonlinear instability thresholds), making it attractive for many applications. However, efficient coupling of the pump light into the core requires a single-mode pump laser, whose output power is typically limited to $\sim$100~mW. To use pump lasers with much higher power, the fiber core can be enlarged, but this introduces higher-order modes. As a result, the signal output often becomes speckled, leading to poor beam quality.

\begin{figure}
\centerline{\includegraphics[width=\textwidth]{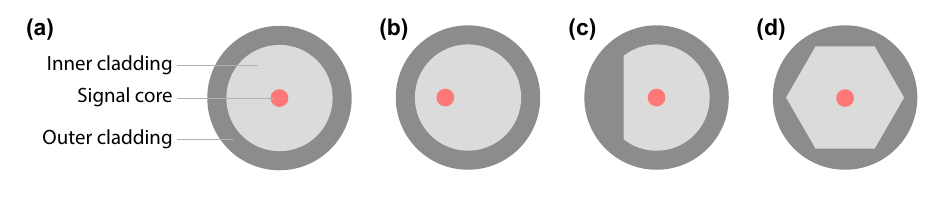}}
\caption{
\textbf{Common cross-sectional geometries of double-clad fibers.} (a) Centered core, (b) off-centered core, (c) D-shaped inner cladding, and (d) hexagonal inner cladding. 
}
\label{fig:DCF} 
\end{figure}

\textit{Double-clad} fibers were pioneered by Snitzer \textit{et al.} to decouple the core size for pump light from that for signal light.\cite{snitzer1988double, po1989double} These fibers consist of a rare-earth-doped single-mode core surrounded by two undoped claddings, with the highest refractive index in the core and the lowest in the outer cladding [Fig.~\ref{fig:DCF}(a)]. In a double-clad fiber amplifier, the signal is coupled into the core and the pump light into the inner cladding. This ensures that the signal stays confined within the doped core, while the pump light propagates through \textit{a larger core}, formed by the inner cladding and doped core. The central portion of the pump light overlaps with the core and is thus absorbed. Since the numerical aperture (NA) and size of the inner cladding can be much greater than that of the core, it becomes possible to efficiently couple substantial pump power from low-brightness sources, such as multimode (single-emitter) or multi-emitter diode lasers. In a multimode laser, all spatial modes that lase do so independently, causing their intensity profiles to superimpose incoherently in space. As a result, the intensity distribution of the pump beam becomes nearly uniform across the inner cladding and doped core. The same applies to multi-emitter lasers, where all emitters lase independently. The absorption coefficient of a cladding-pumped fiber is lower than in the core-pumped case and scales roughly with the core-to-cladding area ratio.  

The shape of the inner cladding is critical for maximizing pump absorption efficiency [Fig.~\ref{fig:DCF}]. The simplest design, featuring circular symmetry for all three layers (core, inner cladding, and outer cladding) centered on the fiber axis, is straightforward to fabricate and facilitates pump light coupling. However, this structure supports many higher-order pump modes that have almost no power overlap with the core and thus unabsorbed.\cite{leproux2001modeling, kouznetsov2001efficiency1, mortensen2007air} Significant pump power is wasted. 
One approach to enhance pump absorption is to induce mode scrambling through bending, twisting, or squeezing the fiber.\cite{kovska2016enhanced} For instance, kidney-shaped coiling has proven effective.\cite{li2004high, tunnermann2010fiber, kovska2016enhanced} Nevertheless, the fibers for high-power amplifiers typically have large outer diameters, making them more rigid and thus challenging to manipulate mechanically. 
A more effective approach is to break the circular symmetry. One way to do this is by shifting the doped core off-center, which allows more modes to overlap with it.\cite{snitzer1988double, po1989double, kouznetsov2002efficiency2} Non-circular inner claddings also enhance pump absorption by forcing the pump modes to extend across almost the entire cross-section, particularly the chaotic ones like D (circular with a straight cut) or stadium shapes.\cite{doya2001optimized, leproux2001modeling, kouznetsov2001efficiency1, kouznetsov2002efficiency2, kouznetsov2002efficiency3, li2004high, mortensen2007air, zervas2014high} However, irregular shapes pose challenges in launching pump light with low insertion loss, so relatively symmetric shapes, like hexagon or octagon, are more widely used.\cite{li2004high, mortensen2007air, tunnermann2010fiber, jackson2012towards, kovska2016enhanced} 

\subsection{Core composition}
Fiber amplifiers are primarily based on glass fibers doped with rare-earth ions in the core, such as Yb$^{3+}$, Er$^{3+}$, Tm$^{3+}$, Ho$^{3+}$, Nd$^{3+}$, Pr$^{3+}$, Dy$^{3+}$, Sm$^{3+}$, and Tb$^{3+}$.\cite{digonnet2001rare, tanabe2002rare, zhu2010high, jackson2012towards, chen2023functional, bernier2025get} Under optical pumping, these ions absorb pump light and are excited from their ground state to a higher energy state. They then relax to lower energy states, emitting light with photon energy lower than the pump through either spontaneous emission (fluorescence) or stimulated emission (signal amplification). Ytterbium (Yb$^{3+}$) is commonly used for amplification in the 1.0~\textmu m region,\cite{paschotta2002ytterbium} erbium (Er$^{3+}$) for the 1.5~\textmu m region,\cite{becker1999erbium} and thulium (Tm$^{3+}$) for the 2.0~\textmu m region.\cite{sincore2017high, gaida2018ultrafast} 
If a primary rare-earth dopant is chosen for its emission spectrum but suffers from weak pump absorption, it is often co-doped with another rare-earth element that has strong pump absorption and facilitates efficient energy transfer to the primary dopant.\cite{fermann1988efficient, pollnau2008advances} For example, Er$^{3+}$ is often co-doped with Yb$^{3+}$ to enhance pump absorption per unit length, enabling more efficient amplification in the telecom band.\cite{nilsson2003high} 

The host glass composition is similarly critical. It determines the fiber's core refractive index\cite{gray2006high, li2007ge} and spectral transparency window\cite{lucas1999infrared, pollnau2008advances}, modifies the spectroscopic properties of the rare-earth ions (e.g., emission/absorption cross-sections, spectral shapes, and transition lifetimes),\cite{dragic2018materials} limits the maximum doping concentration (due to quenching),\cite{dragic2018materials} and affects other factors relevant to laser performance, such as resistivity to photodarkening\cite{engholm2009improved}, optical nonlinearities\cite{ballato2018unified, dragic2018unifieda, dragic2018unifiedb, cavillon2018unified, hawkins2021kilowatt}, and high-energy radiation\cite{xiang2025fully}. 
Silicates are the most widely used host glass in fibers due to their exceptional transparency in the visible and near-infrared regions, mechanical strength, chemical stability, and mature fabrication techniques.\cite{ballato2013rethinking} Pure silica (SiO$_2$) exhibits poor solubility for rare-earth ions, requiring the use of co-dopants, such as aluminum (Al) and phosphorus (P), to reduce ion clustering and resulting quenching.\cite{chen2025emerging} Co-dopants can also be used to modify other properties, such as using cerium (Ce) or sodium (Na) to mitigate photodarkening.\cite{engholm2009improved, zhao2017mitigation}  
Beyond the initial host composition, the processes of preform fabrication and subsequent fiber drawing can significantly influence the compositional profiles across the core and, consequently, the properties mentioned above.\cite{schuster2014material, meehan2025insights}

\subsection{Seed laser}
\begin{figure}
\centerline{\includegraphics[width=\textwidth]{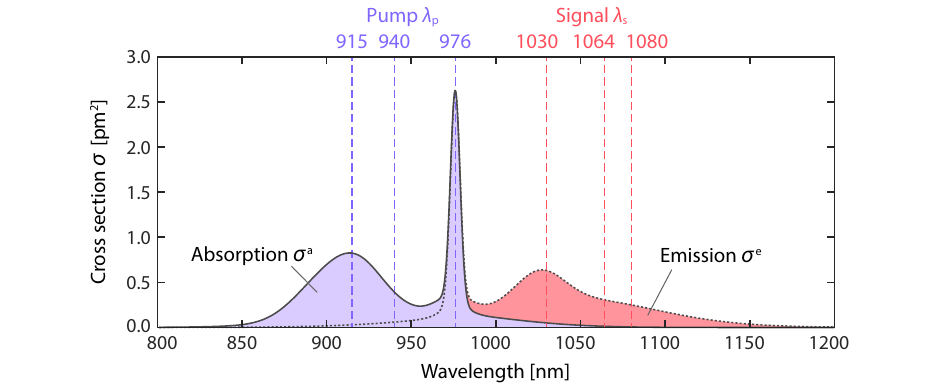}}
\caption{
\textbf{Absorption and emission spectra of Yb-doped silicate fiber.} Solid and dotted curves are example spectra of absorption cross-section ($\sigma^{\rm a}$) and emission cross-section ($\sigma^{\rm e}$), respectively. Dashed lines mark the common pump wavelengths ($\lambda_{\rm p}$) and signal wavelengths ($\lambda_{\rm s}$). Spectra were generated using open-source codes developed by Luke Rumbaugh.\cite{rumbaugh2013codes}  
}
\label{fig:spectra} 
\end{figure}

To ensure the linewidth is sufficiently narrow to be considered single-frequency (typically on the order of 1~MHz or below), the seed laser produces a low-noise continuous-wave (CW) signal or pulses longer than $\sim$100~ns, with a time--bandwidth product close to the Fourier transform limit. Typical sources include diode lasers with distributed feedback (DFB), distributed Bragg reflector (DBR), or external-cavity configurations,\cite{strzelecki2002investigation, salhi2006single, jimenez2017narrow} fiber lasers using DFB, DBR, or ring cavities,\cite{loh19951, agger2004single, bernier2014all, fu2017review, xie2020single, fu2021diode, tao2022high, zhang20222, li2023high1} and nonplanar ring oscillators (NPROs) based on shaped laser crystals.\cite{kane1987frequency, steinke2017single, willke2010stabilized} The selection of the seed laser generally depends on the required wavelength, spectral purity, amplitude and phase noise spectra, modulation capability, and output power, which can vary substantially with the specific application or experimental requirements. For Yb-doped fiber amplifiers, 1064~nm is the standard seed wavelength, with several studies targeting operation at 1030 and 1080~nm [Fig.~\ref{fig:spectra}].\cite{mermelstein2007all, dixneuf2020ultra, lai2020550, shi2022700, li2025functional}

These sources tend to be low power for architectural reasons. For instance, in a DFB fiber laser, a Bragg grating written in the active fiber incorporates a $\pi$-phase shift at its center to create a single ultranarrow resonance at the center of the photonic bandgap. The effective gain length is consequently very short, limiting the maximum output power to $\sim$100~mW.\cite{tao2022high} In DBR fiber lasers (comprising two Bragg gratings sandwiching an active fiber), the cavity must be short enough to support only one longitudinal mode, yet relatively long ($\sim$1~cm) to enable higher power ($\sim$1~W).\cite{fu2021diode} Ring-cavity fiber lasers can employ meter-scale gain lengths and incorporate sub-cavities or saturable absorbers to enforce single-longitudinal-mode operation.\cite{zhang20222} Their output can exceed that of the previous two types, but typically remains at the watt level. 
 
For oscillators with adjustable laser frequency, the linewidth can be pushed to the hertz or sub-hertz regime through self-injection locking or Pound--Drever--Hall (PDH) locking to a reference cavity with an ultrahigh quality factor.\cite{drever1983laser, salomon1988laser, lee2013spiral, hirata2014sub, liang2015ultralow, morton2018high, jin2021hertz, zhang2025narrow} Preserving the linewidth at this level through a high-power fiber amplifier has not yet been demonstrated and can be challenging, as additional noise can be introduced in the amplification stages.

\subsection{Pump laser} 
Fiber amplifiers are commonly pumped by fiber-coupled diode lasers.\cite{samson2007diode, zucker2014advancements, gapontsev2017highly} These diode lasers can be single emitters or one-dimensional emitter arrays (laser bars), selected according to the required power, brightness, and coupling configuration. When pumping the core directly, diffraction-limited single-mode diodes are needed to couple the pump light efficiently into a single-mode active fiber, but their average output powers are relatively low (typically $\sim$1--100~mW). 

To achieve higher power output, multimode pump diodes, such as broad-area chips or laser bars, are the most practical option.\cite{samson2007diode, zucker2014advancements, gapontsev2017highly} 
However, the presence of higher-order modes results in both a large beam size and increased divergence. This challenge is overcome through the use of double-clad active fibers, which allow the pump light to be efficiently coupled into a large inner cladding with high NA, as discussed earlier in Sec.~\ref{sec:dcf}. 

High-power pump modules typically deliver $\sim$0.1--1~kW of power, with wall-plug (electrical-to-optical) efficiencies around 50--80\%.\cite{berishev2005algainas, kanskar200573} The power loss turns into heat, which often requires active dissipation through liquid-cooled metal plates (cold plates) or thermoelectric coolers (TEC). Heating also shifts the pump wavelength by $\sim$0.1--1~nm/W.\cite{volodin2004wavelength, talbot2022wavelength} This can be addressed through active temperature control or passive wavelength stabilization by coupling the diode laser to a volume Bragg grating or fiber Bragg grating ($\sim$10\% reflectivity, $\sim$0.1--1-nm bandwidth). Alternatively, pumping within a broad flat absorption band can reduce sensitivity to wavelength drift. For example, in Yb-doped fiber lasers and amplifiers, it is common to pump around 915 or 940~nm instead of 976~nm (with maximum absorption but spectrally narrow) [Fig.~\ref{fig:spectra}].\cite{jiang2022650, chen2024915} 

To reduce quantum-defect heating in a fiber amplifier and improve its slope efficiency by pumping closer to the signal wavelength, some systems use fiber lasers as pumps when high-power diodes are unavailable at the desired wavelength.\cite{richardson2010high, zhou2017high, ma2018high, baer2024ultra} This approach is known as tandem pumping. Examples include pumping a 2.09-\textmu m Ho-doped fiber amplifier with a 1.95-\textmu m Tm-doped fiber laser\cite{baer2024ultra}, pumping a 1.55-\textmu m Er-doped fiber amplifier with a 1.48-\textmu m Raman fiber laser\cite{kotov2022high}, and pumping a 1.56-\textmu m Er/Yb-codoped fiber amplifier with a 1.02-\textmu m Yb-doped fiber laser\cite{varona2018all}.

\label{pump}

\subsection{Pump--signal combiner} 
A pump--signal combiner is the component that couples pump light into the active fiber, either in an all-fiber configuration or via free-space optics.\cite{braglia2014architectures, zervas2014high2} It may be placed at the fiber input (co-pumping), at the output (counter-pumping), at both ends (bidirectional), or mid-span along the active fiber to extend the effective gain length (distributed pumping) [Fig.~\ref{fig:MOPA}(b)].\cite{hildebrandt2008brillouin, wang2004thermal, zervas2014high, li2017experimental}  

\begin{figure}
\centerline{\includegraphics[width=\textwidth]{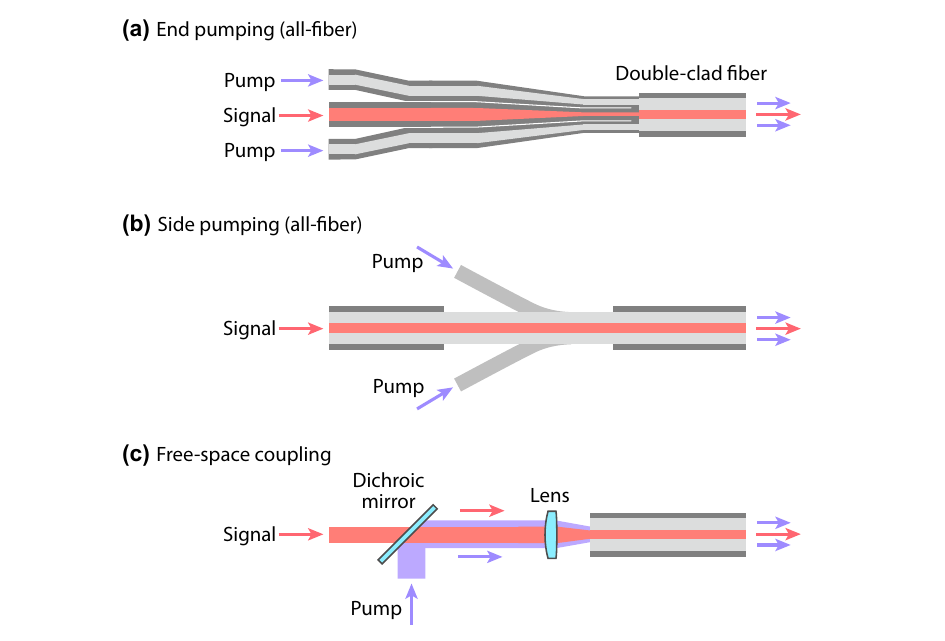}}
\caption{
\textbf{Pump--signal combiners.} 
(a) All-fiber end pumping, 
(b) all-fiber side pumping, 
and (c) free-space coupling using dichroic mirror and lens.
}
\label{fig:PumpCombiner} 
\end{figure}

In all-fiber configurations, pump light can be coupled from either the fiber end or the side [Fig.~\ref{fig:PumpCombiner}(a,b)]. End-pumping employs a multi-port combiner: several pump fibers surround a central signal fiber to form a bundle, which is then tapered and fusion-spliced to a passive double-clad fiber matched in geometry and NA to the active fiber [Fig.~\ref{fig:PumpCombiner}(a)].\cite{stachowiak2018high, liu2021high, majumder2022design} Local tapering in this arrangement alters the signal-core diameter along the combiner, potentially degrading performance in certain applications. End-pump combiners can also be built for core-pumping a single-clad active fiber, but this requires single-mode pump fibers, and the small core of the output fiber limits the number of pump inputs.\cite{kotov2022high, tao2023over} 

In side-pump combiners, a section of the matching passive fiber's outer cladding is removed, and pump fibers are spliced to a bridging structure (e.g., tapered coreless fibers) attached to the stripped matching fiber to couple light at an angle into the inner cladding [Fig.~\ref{fig:PumpCombiner}(b)].\cite{stachowiak2018high, jauregui2010side, magnan2020fuseless, brockmuller2025side} They support co-, counter-, and distributed pumping and maintain a constant signal-core diameter throughout the device. 

As all-fiber combiners may contain several splice points through which high pump power is injected, scattering of the pump light can occur at these interfaces. Subsequent absorption of this stray light causes significant heating, making efficient heat dissipation essential to prevent damage.\cite{lapointe2009thermal} With appropriate combiner design and thermal management, they can handle multi-kilowatt power.\cite{shcherbakov2013industrial, sun2022100} It’s also worth noting that, if no matching passive fiber is available, the active fiber can be integrated directly with the combiner. However, substantial pump absorption and the ensuing heat load limit the power-handling capability of active pump--signal combiners compared with passive ones.

Free-space coupling is often used when the active fiber uses a custom geometry or NA for which all-fiber combiners are unavailable [Fig.~\ref{fig:PumpCombiner}(c)].\cite{jeong2005single, hildebrandt2008brillouin, robin2014modal, beier2017single, smith2025wavefront} For co-pumping in free space, the pump and signal are combined onto a common optical path with a dichroic mirror and then focused by a lens into the input end of the active fiber. Counter-pumping uses the same optics in a mirrored arrangement at the output end. Implementing distributed pumping with free-space optics is generally impractical (alignment-sensitive), bulkier, and offering little benefit compared with all-fiber solutions. 

\label{combiner}

\section{Mitigation of Stimulated Brillouin Scattering (SBS)}

\subsection{What is SBS?}
\begin{figure}
\centerline{\includegraphics[width=\textwidth]{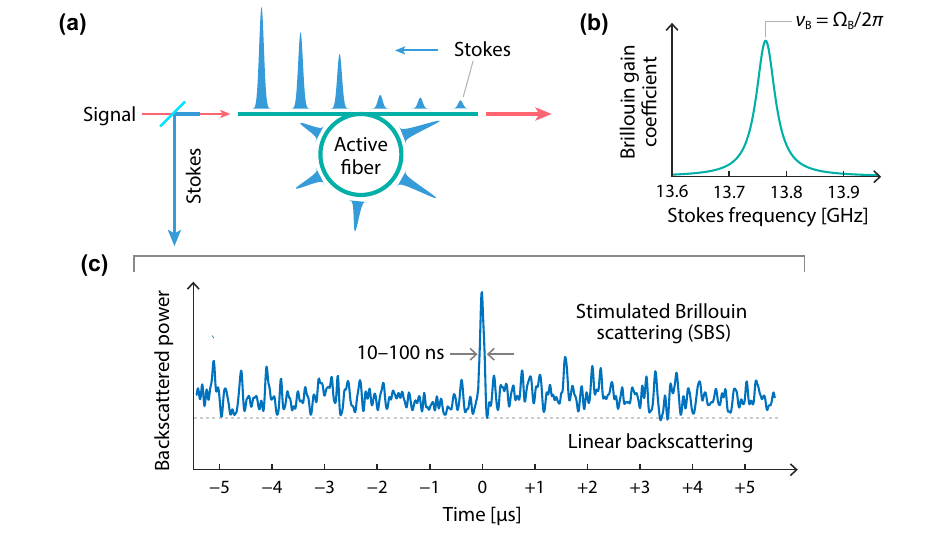}}
\caption{
\textbf{Stimulated Brillouin scattering (SBS) in fiber amplifier.}
(a) Backward-propagating Stokes field gains power from the forward signal via SBS and from stimulated emission.
(b) Typical Brillouin gain spectrum $g_{\rm B}(\nu)$ in single-mode silicate fibers, centered at $\nu_{\rm B} = \Omega_{\rm B}/2\pi \approx$ 10--20~GHz from the signal frequency. The Stokes shift is expressed in linear frequency $\nu$ to follow experimental literature conventions. Data were generated using open-source codes developed by Wisal~\textit{et al.}\cite{wisal2023codes} 
(c) Example time trace of backscattered power near the onset of SBS (synthetic data),\cite{panbhiharwala2018investigation, rothe2025wavefront} showing amplified Stokes fluctuations with intermittent strong pulses (10--100-ns duration) on top of constant linear backscattering from Fresnel reflection and Rayleigh scattering.
}
\label{fig:SBS} 
\end{figure}

Stimulated Brillouin scattering (SBS) is a nonlinear interaction between light and acoustic waves.\cite{agrawal2000nonlinear, kobyakov2009stimulated, eggleton2013inducing} In an optical fiber, thermally excited acoustic waves are always present and can scatter the signal light, generating Stokes-shifted light. This spontaneous Brillouin scattering provides the initial Stokes seed for the SBS process. The efficiency of SBS is governed by the phase-matching condition, which is most readily satisfied for counter-propagating optical waves coupled through a guided acoustic wave: $q = \beta_{\rm s} - \beta_{\rm St} \approx 2\beta_{\rm s}$, where $q$, $\beta_{\rm s}$, and $\beta_{\rm St}$ are the propagation constants for the acoustic, signal, and Stokes waves, respectively. This Stokes wave propagates in the backward direction and has a slightly lower frequency of $\omega_{\rm St}=\omega_{\rm s}-\Omega$, where $\omega_{\rm s}$ and $\Omega = 2\pi\nu$ are the signal and acoustic frequencies, respectively. The interference between the signal and Stokes waves forms a traveling intensity grating with velocity $\Omega/q$, which in turn drives the guided acoustic wave through electrostriction. The acoustic wave then scatters more signal power into the Stokes, triggering a stimulated process. As a result, the Stokes field grows exponentially in the backward direction along the fiber [Fig.~\ref{fig:SBS}(a)].

SBS is the primary nonlinear effect limiting the power-scaling of single-frequency fiber amplifiers for two main reasons. Both stem from the characteristics of the \textit{Brillouin gain spectrum} $g_{\rm B}(\Omega)$, which determines the exponential growth rate of the Stokes field [Fig.~\ref{fig:SBS}(b)]. 

First, this spectrum is very narrow. In silicate fibers, acoustic waves decay exponentially as ${\rm exp}(-t/\Gamma_{\rm B})$, where the phonon decay rate $\Gamma_{\rm B}$ is typically on the order of 10--100~MHz. This decay rate determines the Brillouin gain bandwidth $\Delta\nu_{\rm B}=\Gamma_{\rm B}/2\pi$, which falls within the same frequency range.\cite{boyd1990noise, gaeta1991stochastic, agrawal2000nonlinear} Because the linewidth of single-frequency lasers ($\lesssim 1$~MHz) is well below this gain bandwidth, virtually all frequency components of the laser light contribute coherently to the SBS process. As a result, the laser power is efficiently transferred into the backward Stokes field.

Second, the peak frequency is very low. The Brillouin gain spectrum is peaked at frequency $\Omega_{\rm B} = q V_{\rm a}$, where $V_{\rm a}$ is the longitudinal acoustic velocity in glass. Using the phase-matching condition $q \approx 2\beta_{\rm s}$, this frequency can then be written as $\Omega_{\rm B} = 2\beta_{\rm s}V_{\rm a} = 4\pi n V_{\rm a}/\lambda_{\rm s}$, where $n$ is the refractive index and $\lambda_{\rm s}$ is the signal wavelength in vacuum. For silicate fibers operating in the near-infrared, the resulting Stokes shift $\nu_{\rm B}$ (often preferred in experimental contexts, where $\nu_{\rm B} = \Omega_{\rm B}/2\pi$) is only about 10 to 20~GHz, which is less than 100~pm in wavelength.\cite{agrawal2000nonlinear, kobyakov2009stimulated, hawkins2021kilowatt}  Consequently, the Stokes field remains well within the gain bandwidth of the rare-earth dopants, allowing it to draw energy not only through SBS but also from the population inversion. This causes the Stokes power to grow more aggressively than in passive fibers, potentially leading to catastrophic damage to upstream components.

SBS in fiber amplifiers is commonly characterized by measuring the backscattered power (as a time trace or a time-averaged value) at the fiber input against the pump power [Fig.~\ref{fig:SBS}(c)].\cite{panbhiharwala2018investigation} At low pump powers, the backscattered Stokes field first appears as random noise fluctuating on the 10--100-ns scale (associated with the phonon lifetime $\tau_{\rm B}=1/\Gamma_{\rm B}$).\cite{agrawal2000nonlinear} With increasing pump power, the Stokes fluctuations are amplified, leading to the emergence of intermittent high-power pulsing with 10--100-ms intervals. The interval shortens as the pump power continues to rise, eventually evolving into quasi-periodic pulsing. At higher pump powers, occasional giant pulses with kW-level peak power can occur, depleting most of the available optical gain and followed by a 10-\textmu s-scaled gain recovery. For safe and stable amplifier operation, the pump power should be limited below the onset of quasi-periodic pulsing. Even if no physical damage occurs, the strong noise added by SBS can severely degrade signal coherence.

Because the Stokes power generally increases exponentially with pump power, there is no strict SBS threshold. In amplifiers especially, deleterious effects from intense Stokes pulses may arise even when the average backscattered power is still too low to measurably affect the linear scaling of output power with pump power. Therefore, different studies use varying practical definitions, such as the ratio of average backscattered power to output signal power (e.g., 0.1\%),\cite{gray2006high, gray2007502} the peak backscattered power relative to the background,\cite{rothe2025wavefront} or the relative intensity noise at the output.\cite{hildebrandt2008brillouin, hochheim2020single, hochheim2021single} This makes direct comparison between published results challenging. Nevertheless, regardless of the definition used, the SBS threshold $P_{\rm SBS}$ scales proportionally with the effective mode area $A_{\rm eff}$, and inversely with the effective Brillouin gain coefficient coefficient $g_{\rm B, eff}$ and the effective fiber length $L_{\rm eff} = \int_{0}^{L} P_{\rm s}(z)/P_{\rm s}(L)\,{\rm d}z$ (accounting for the variation in signal power along the amplifier): 
\begin{equation}
P_{\rm SBS} \propto \frac{A_{\rm eff}}{g_{\rm B,eff} L_{\rm eff}}.
\end{equation}

Single-frequency fiber amplifiers are typically designed with the shortest possible length to suppress SBS, so we won’t delve into the length effect or related methods in the following subsections. However, it's worth noting that some studies use high rare-earth doping to significantly reduce the required active length.\cite{kruska2025power} While SBS is greatly suppressed, this high gain can lead to intense localized heating, which carries the risk of triggering unwanted thermal effects or even damaging the fiber. 

\subsection{Intensity reduction with a larger core size}
\label{sec:largecore}
Increasing the core area $A_{\rm core}$ leads to a larger effective mode area $A_{\rm eff}$, thus suppressing SBS by reducing the intensity for a given power. However, a large core typically supports higher-order modes, whose interference produces a speckled output beam.\cite{kobyakov2009stimulated} Therefore, various fiber designs have been developed to increase the core size while maintaining single-mode operation ($V<2.405$) to preserve beam quality. Here $V=2\pi a\, {\rm NA}/\lambda_{\rm s}$, where $a$ is the core radius, NA is the numerical aperture, and $\lambda_{\rm s}$ is the signal wavelength. 

\begin{figure}
\centerline{\includegraphics[width=\textwidth]{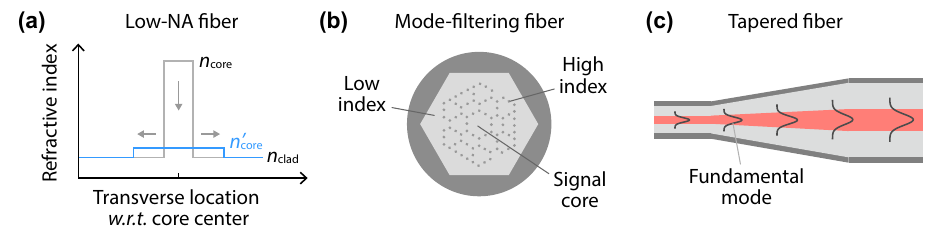}}
\caption{
\textbf{Large-core fiber designs for single-mode operation.} (a) Refractive-index profiles of low-NA versus standard step-index fiber. (b) Cross section of mode-filtering photonic crystal fiber.\cite{matniyaz2022high} (c) Tapered fiber for selective excitation of the fundamental mode in a multimode core through adiabatic core expansion. 
}
\label{fig:LargeCore} 
\end{figure}

\textbf{Low NA.} 
One common approach is to reduce the NA of the core to counterbalance the increase in core area ($A_{\rm core}=\pi a^2$), since the number of modes per polarization in a step-index fiber is roughly $V^2/4 = \pi A_{\rm core} {\rm NA}^2/\lambda_{\rm s}^2$ [Fig.~\ref{fig:LargeCore}(a)].\cite{liem2003100, jeong2005single, mermelstein2007all, wang2012310, ma2013single, dixneuf2020ultra, kruska2025power} This is achieved by carefully adjusting the glass composition\cite{gray2006high, li2007ge} so that the core refractive index is brought exceptionally close to that of the cladding. Currently, the minimum achievable NA is $\approx$0.03, which supports single-mode core diameters up to around 30~\textmu m.\cite{li2025functional} Fabrication of these fibers is challenging because achieving such a low NA requires the core–cladding index difference to be on the order of $10^{-4}$ with high transverse and longitudinal uniformity. This demands high-precision control over dopant distributions throughout preform and fiber-drawing stages. Moreover, the fundamental mode is only weakly confined, so the fiber should be kept straight or coiled with a large diameter to minimize bend loss. 
\newline

\textbf{Mode filtering.} 
Another route is to employ mode-filtering fiber designs, including photonic crystals\cite{dajani2009experimental, robin2011acoustically, limpert2012yb, robin2014modal, pulford2015400, matniyaz2022high}, chirally coupled cores\cite{zhu2011single, hochheim2020single, hochheim2021single}, leaky channels\cite{dong2007leakage}, and anti-resonances\cite{cheng2024design} [Fig.~\ref{fig:LargeCore}(b)]. In these fibers, the cladding structure is engineered so that higher order modes in the core are resonant with cladding modes and are thus preferentially stripped away. Similar to low-NA fibers, these microstructured designs are also susceptible to bending, as tight coiling can induce unwanted power transfer from the low-loss fundamental mode to high-loss higher order modes. In addition to structural mode-filtering, the gain profile can also be tailored to minimize its overlap with higher order modes and consequently suppress their growth.\cite{sousa1999multimode, cooper2022confined, li2022confined, kruska2024high} This approach is known as confined doping.  

A new record for the highest output power in single-frequency fiber amplifiers was recently established (August 2025), reaching 1,015~W, using a large-core fiber with a bat-shaped index profile.\cite{li2025functional}

\textbf{Taper.}
To overcome the core size limit imposed by the single-mode condition in step-index fibers ($V < 2.405$), \textit{tapered} fibers have been explored for high-power single-frequency amplifiers, reaching up to 650~W.\cite{trikshev2013160, pierre2018200, lai2020550, jiang2022650} These tapered fibers begin with a small-core section for signal input, adiabatically taper to a large core, and end with a uniform large-core section [Fig.~\ref{fig:LargeCore}(c)]. In the front section, only the fundamental mode is excited due to the single-mode core. As the signal propagates through the taper, the fundamental mode area gradually expands. A tapered fiber suppresses SBS through two effects: (1) The intensity remains low throughout the fiber as the signal reaches its maximum power in the large-core section. (2) The frequency of the peak Brillouin gain (known as the Brillouin frequency $\nu_{\rm B}$) shifts with the fiber diameter. This leads to an effective broadening of the Brillouin gain spectrum (increasing $\Delta\nu_{\rm B,eff}$), which in turn reduces its peak value.\cite{shiraki1995suppression} This approach allows for the use of a much larger core diameter, up to around 60~\textmu m, with the SBS threshold typically in the kilowatt range. However, because higher-order modes are supported in the large core, signal power can couple from the fundamental mode to these modes through dynamic thermo-optical scattering.\cite{kholaif2025influence} As a result, the power scaling of tapered single-frequency fiber amplifiers is often limited by TMI. 

\subsection{Tailoring acousto-optic interactions}

In addition to the signal intensity, the SBS threshold also depends on the response of the acousto-optic medium (i.e., the fiber core) to the signal light, which is characterized by the effective Brillouin gain coefficient $g_{\rm B,eff}$. This coefficient scales proportionally with the material's intrinsic Brillouin gain coefficient $g_{\rm B,0}$ and the overlap between the acoustic and optical mode profiles $O$, while scaling inversely with the effective Brillouin gain bandwidth $\Delta\nu_{\rm B,eff}$. Below are three strategies that can be used to tailor the Brillouin gain in fiber amplifiers: (1) compositional engineering to reduce $g_{\rm B,0}$,\cite{dragic2018unifiedb, cavillon2018unified, hawkins2021kilowatt, dragic2024low} (2) structural engineering to minimize $O$, and (3) applying a longitudinal temperature or strain gradient to increase $\Delta\nu_{\rm B,eff}$.  

\textbf{Compositional engineering.} 
The intrinsic Brillouin gain coefficient $g_{\rm B,0}$ at the Brillouin frequency $\nu_{\rm B}$ is related to material properties by
\begin{equation}
g_{\rm B,0}(\nu_{\rm B}) = \frac{2 \pi^2 n^7 p_{12}^2}{c \lambda_{\rm s}^2 \rho V_{\rm a} \Delta\nu_{\rm B,0}},
\end{equation}
where $n$ is the refractive index, $p_{12}$ the transverse photoelastic coefficient, $c$ the speed of light in vacuum, $\lambda_{\rm s}$ the signal wavelength in vacuum, $\rho$ the density, $V_{\rm a}$ the longitudinal acoustic velocity, and $\Delta\nu_{\rm B,0}$ the intrinsic Brillouin gain bandwidth.\cite{melloni1998direct, agrawal2000nonlinear, hawkins2021kilowatt} These quantities can be adjusted by co-doping silica (SiO$_2$) with Al, P, B, Ba, Sr, and/or other elements [Table~\ref{tab:SBS}].\cite{dragic2018unifiedb, cavillon2018unified, hawkins2021kilowatt, dragic2024low} For instance, many dopants introduce stronger acoustic damping and thus broaden the Brillouin gain bandwidth $\Delta\nu_{\rm B,0}$ (e.g., B$_2$O$_3$ has a particularly large $\Delta\nu_{\rm B,0}$ of about 428~MHz, compared with 17~MHz for pure SiO$_2$).\cite{dragic2011brillouin, dragic2013pockels, dragic2018unifiedb} The $p_{12}$ of Al$_2$O$_3$, BaO, SrO are negative,\cite{dragic2013pockels, dragic2013brillouin2, dragic2013brillouin, cavillon2016brillouin} which can offset the positive contributions from SiO$_2$, P$_2$O$_5$, and B$_2$O$_3$,\cite{bertholds2002determination, melloni1998direct, law2011acoustic, dragic2011brillouin, dragic2013pockels} and the $p_{12}$ of AlPO$_4$ is close to zero.\cite{yu2019alpo} Careful selection of co-dopants and optimization of their concentrations are required to suppress the peak Brillouin gain while preserving the desired optical properties.\cite{ballato2018unified, dragic2018unifieda, dragic2018unifiedb, cavillon2018unified} Beyond co-doping in silica, crystal-derived glasses provide an alternative route toward exceptionally low Brillouin gain.\cite{dragic2010brillouin, dragic2012sapphire}

\begin{table}[ht]
\tbl{Effects of various dopants on the physical properties of silicate glass relevant to SBS.\cite{dragic2018unifiedb, cavillon2018unified, hawkins2021kilowatt} Arrows indicate increases ($\uparrow$), decreases ($\downarrow$), or negligible changes ($\approx$) relative to SiO$_2$ upon doping.}
{\begin{tabular}{@{}lccccc@{}} \toprule
Compound & $n$ & $\rho$ & $V_{\text a}$ & $\Delta \nu_{\text B}$ & $p_{12}$* \\ \colrule
SiO$_2$ & 1.45 & 2200~kg/m$^3$ & 5970~m/s & 17~MHz & +0.226 \\
Al$_2$O$_3$ & $\uparrow$ & $\uparrow$ & $\uparrow$ & $\uparrow$ & $\downarrow$ \\
P$_2$O$_5$ & $\uparrow$ & $\uparrow$ & $\downarrow$ & $\uparrow$ &  $\uparrow$ \\
AlPO$_4$ & $\approx$ & $\approx$ & $\downarrow$ & $\uparrow$ & $\downarrow$ \\
GeO$_2$ & $\uparrow$ & $\uparrow$ & $\downarrow$ & $\uparrow$ & $\uparrow$ \\
B$_2$O$_3$ & $\downarrow$ & $\downarrow$ & $\downarrow$ & $\uparrow$ &  $\uparrow$ \\
BaO & $\uparrow$ & $\uparrow$ & $\downarrow$ & $\uparrow$ & $\downarrow$ \\ 
SrO & $\uparrow$ & $\uparrow$ & $\downarrow$ & $\uparrow$ & $\downarrow$ \\
Yb$_2$O$_3$ & $\uparrow$ & $\uparrow$ & $\downarrow$ & $\uparrow$ & $\downarrow$ \\ \botrule
\end{tabular}
}
\begin{tabnote}
*In the $p_{12}$ column, downward arrows incidentally correspond to negative values, except for AlPO$_4$ ($p_{12} \approx 0.0$). 
\end{tabnote}
\label{tab:SBS}
\end{table}

\begin{figure}
\centerline{\includegraphics[width=\textwidth]{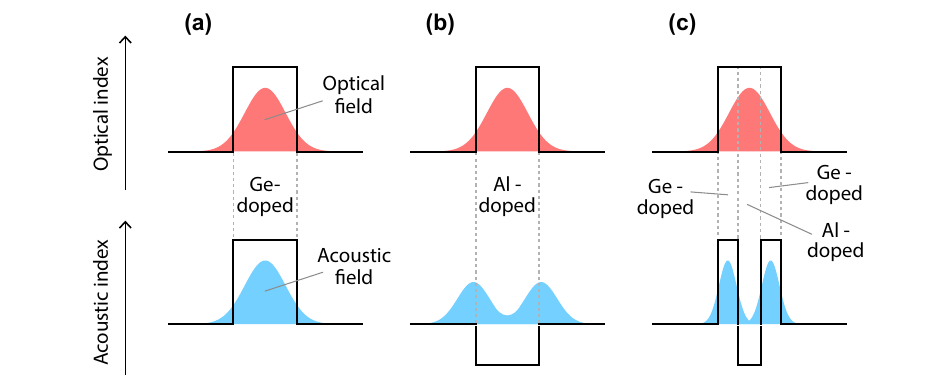}}
\caption{
\textbf{Fiber designs for reducing acousto-optic overlap.}\cite{li2007ge} 
Co-dopants germanium (Ge) and aluminum (Al) both increase the optical index, but Ge raises the acoustic index, whereas Al reduces it. (a) Standard step-index fiber with a Ge-doped core, guiding both optical (upper) and acoustic (lower) fields with significant overlap. (b) Acoustically anti-guiding design using an Al-doped core. (c) Segmented doping of Al and Ge in the core, demonstrating that the acoustic and optical profiles can be independently tailored.
}
\label{fig:AO} 
\end{figure}

\textbf{Minimizing acousto-optic overlap.}
The effective brillouin gain coefficient $g_{\rm B,eff}$ is the intrinsic Brillouin gain coefficient $g_{\rm B,0}$ weighted by the acousto-optic overlap $|O|^2$.\cite{agrawal2000nonlinear, kobyakov2009stimulated, wisal2024theorysbs} 
\begin{equation}
O = \langle (\vec{\psi}_{{\rm s}} \cdot \vec{\psi}_{{\rm St}}^{*}) \xi^* \rangle,
\end{equation}
where $\vec{\psi}_{\rm s}$, $\vec{\psi}_{\rm St}$, and $\xi$ are the mode profiles of the signal, Stokes, and acoustic fields, respectively, and $\langle \, . \, \rangle$ is the integral over the entire fiber cross-section. In a single-mode fiber, both the signal and Stokes fields propagate in the fundamental mode, with their profiles nearly identical due to a spectral separation of only $\sim$10~GHz. There could be multiple acoustic modes in the fiber, and $\xi$ corresponds to the one that has the greatest overlap with the optical modes.

In step-index fibers, the uniform material composition of the core causes the optical and acoustic waves to share the same core, which maximizes the acousto-optic interaction. To suppress this interaction, a key strategy is to design microstructured fibers that decouple the acoustic and optical modes. This is made possible by the fact that co-dopants affect the optical and acoustic properties of silica glass differently [Table~\ref{tab:SBS}].\cite{gray2006high, li2007ge, dragic2018unifiedb, hawkins2021kilowatt} For example, both Ge and Al doping increase the optical refractive index. However, while Ge doping raises the acoustic index (which is inversely proportional to $V_{\rm a}$), Al doping lowers it. By strategically incorporating two or more co-dopants, one can design an arbitrary acoustic index profile while keeping the optical index profile unchanged [Fig.~\ref{fig:AO}].\cite{gray2006high, gray2007502, li2007ge, robin2011acoustically} This allows the acoustic field to be guided away from the optical intensity distribution. 

In 2014, an advanced structural design incorporating three co-dopants was introduced to minimize the acousto-optic overlap in a large-core active fiber with gain confinement.\cite{robin2014modal} This design enabled the development of an 811-W single-frequency laser, a record that remained unbroken until 2025.\cite{robin2014modal, li2025functional}

\textbf{Temperature or strain gradient.}
The Brillouin frequency $\nu_{\rm B}$ of a fiber depends on both temperature and strain, so applying a longitudinal gradient of temperature or strain can cause the local Brillouin frequency $\nu_{\rm B}(z)$ to vary along the fiber.\cite{horiguchi2002development} This broadens the effective gain bandwidth and consequently reduces the peak gain.\cite{imai1993dependence, yoshizawa1993stimulated, boggio2005experimental} This effect is analogous to that observed in a tapered fiber, where $\Omega_{\rm B}$ is similarly shifted by the change in the fiber diameter. 

Heating can originate internally from the quantum defect accompanying stimulated emission, or it can be applied externally using temperature-controlled stages. As the signal is amplified along the fiber, the strongest heating occurs at which the signal power increases most rapidly.\cite{dajani2010stimulated, hansen2011thermo, zhang2013170} The nonuniform heating leads to longitudinal temperature gradients, observed in nearly all fiber amplifiers.\cite{jeong2007power, hildebrandt2008brillouin, dajani2010stimulated, hansen2011thermo, zeringue2011pump, zhang2013170, theeg2015core, dixneuf2020ultra} Active control of the temperature gradients can be achieved by placing different fiber segments on separate temperature-controlled stages.\cite{zeringue2011pump, theeg2012all, wellmann2019high} A temperature change, $\Delta T = T' - T$, induces a linear shift in the Brillouin frequency, from $\nu_{\rm B}(T)$ to $\nu_{\rm B}(T')$. This relationship is given by $\nu_{\rm B}(T') = \nu_{\rm B}(T) [1+C_{\rm temp} \, \Delta T]$, where the scaling factor $C_{\rm temp}$ is typically $10^{-4} \text{ K}^{-1}$.\cite{horiguchi2002development} Given that $\nu_{\rm B} \sim$ 10~GHz, the Brillouin frequency shifts at a rate of $\sim$1~MHz per kelvin. A temperature gradient of 50--100~K can provide $\sim$100~MHz of gain bandwidth broadening, effectively suppressing the SBS.\cite{imai1993dependence, hansryd2001increase, hildebrandt2008brillouin, law2011acoustic, dragic2013brillouin2} When using this method, long-term operation of typical fiber amplifiers should be kept below 80$^\circ$C to prevent accelerated aging of the polymer coating(s).\cite{theeg2015core} 

To apply a longitudinal strain gradient, the fiber is usually held by a series of movable fixtures.\cite{zhang2013170, huang2016414, balliu2022power} Each section of the fiber is subjected to a different tensile strain $\epsilon$, creating a staircase-like gradient along the fiber. 
This strain induces a linear shift in the Brillouin frequency given by $\nu_{\rm B}(\epsilon) = \nu_{\rm B}(0) [1+C_{\rm strain} \epsilon]$, where $C_{\rm strain}$ is around 5 for silica fibers.\cite{horiguchi2002development} Excessive strain can damage the fiber in various ways, so the applied strain is typically limited to a few percent, which corresponds to a shift in the Brillouin frequency of up to $\sim$1~GHz.\cite{balliu2022power} 

\textbf{Broadening laser linewidth.} 
To date, single-frequency fiber amplifiers have barely reached 1~kW, with the path to access the multi-kilowatt regime requiring a significant broadening of the laser linewidth to 1--10~GHz, much wider than the Brillouin gain bandwidth.\cite{montoya2017photonic, creeden2018advanced, nicholson20235, edgecumbe2024fiber} An increase in linewidth ($\Delta \nu_{\rm s}$) reduces the coherence length ($L_{\rm c} \sim c/\Delta \nu_{\rm s}$) to $\sim$1--10~cm. While highly effective for SBS suppression and capable of retaining a useful degree of coherence for certain applications, it clearly goes beyond what is typically defined as a single-frequency laser (linewidth of MHz or below). This linewidth broadening is achieved by intentionally introducing frequency noise by phase-modulating the seed laser. Several modulation schemes are employed, including adding white noise,\cite{williamson2007laser, zeringue2012theoretical, anderson2015comparison} sinusoidal phase modulation,\cite{flores2012experimental} and pseudo-random binary sequence (PRBS).\cite{flores2014pseudo, anderson2015comparison} Among them, PRBS offers the greatest flexibility and control, allowing for optimal tailoring of the laser spectrum to maximize the SBS suppression.

\section{Mitigation of Transverse Mode Instability (TMI)} \label{sec:tmiMain}

\subsection{What is TMI?}
Transverse mode instability (TMI) refers to the dynamic coupling of modes caused by nonlinear thermo-optical scattering.\cite{ward2012origin, hansen2013theoretical, naderi2013investigations, dong2013stimulated, kong2016direct, jauregui2020transverse, jauregui2026recent} Although single-mode fibers are designed to support only the fundamental mode, higher order modes can still exist, albeit with high loss (i.e., the power is rapidly attenuated over short distances). These modes can be excited by fiber imperfections, bending, non-ideal input coupling, or thermally induced refractive-index changes. At high power, even a very small higher-order-mode content is sufficient to trigger instability.  

\begin{figure}
\centerline{\includegraphics[width=\textwidth]{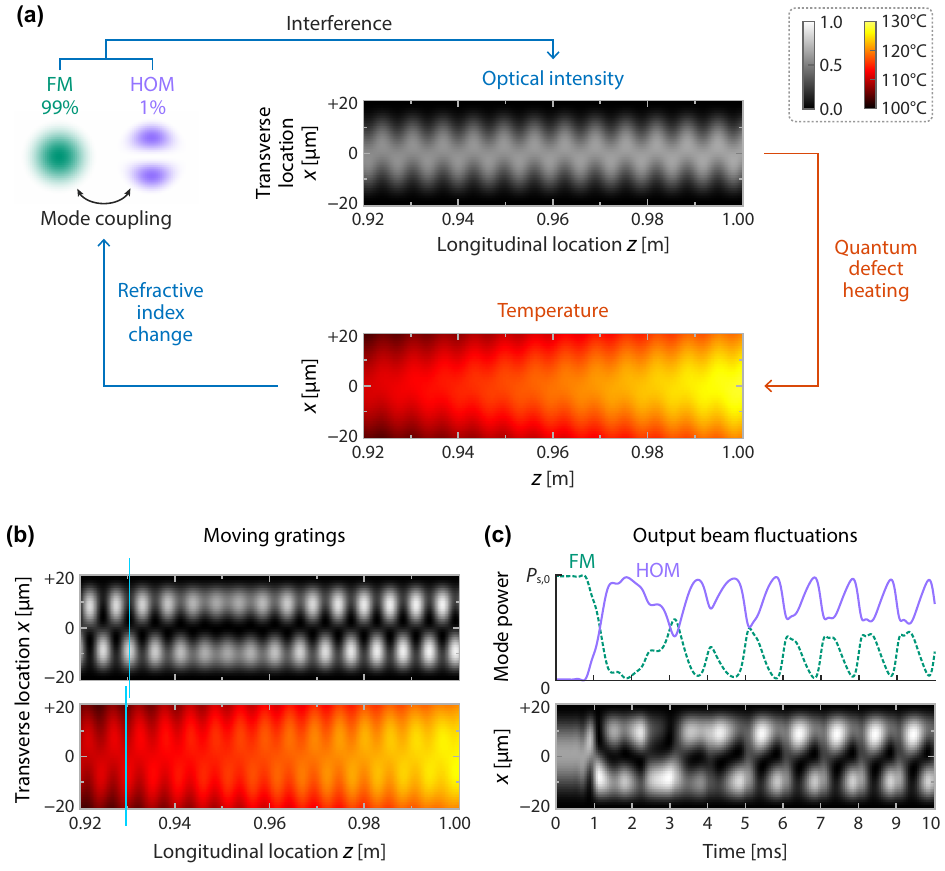}}
\caption{
\textbf{Transverse mode instability (TMI) in fiber amplifier.}\cite{chen2023suppressing} (a) Thermo-optical feedback mechanism underlying TMI. Interference of fundamental mode (FM) and higher-order mode (HOM, 1\% of input power) forms an intensity grating, which is converted into a temperature grating via quantum-defect heating. This, in turn, induces a refractive-index modulation that can transfer power between modes, altering the intensity distribution. This feedback amplifies temporal power noise in HOM and leads to TMI at high power. 
(b) Snapshot of TMI in fiber. Delayed thermal responses at millisecond scale create local offsets between the gratings (indicated by blue lines), causing them to ``chase'' each other. 
(c) Dynamic mode coupling (upper panel) and beam fluctuations (lower panel) at fiber output. All data are simulated using open-source codes developed by Chen \textit{et al.}\cite{chen2024codes, chen2025output} Ambient temperature is set at 20$^{\circ}$C.
}
\label{fig:tmi} 
\end{figure}

Consider the excitation of a single higher-order mode with power that seem negligible compared to the fundamental mode [Fig.~\ref{fig:tmi}(a)]. Their interference creates a shallow intensity grating. Quantum-defect heating converts the intensity grating into a temperature grating, with the brighter regions generating more heat. In silica glass, the refractive index increases with temperature, thus inducing a refractive-index grating. At low power, the heating is weak, so the induced index modulation is insufficient to cause appreciable scattering, keeping the amplifier stable. However, as the power is increased, the temperature contrast in the grating grows, which enhances the index contrast. This intensified index grating scatters light more efficiently, causing power transfer between the modes, which in turn alters the intensity distribution. The redistribution of heat sources, corresponding to areas of high intensity, affects the temperature distribution. While the light responds almost instantaneously to changes in the refractive index, the temperature distribution reacts more slowly due to heat diffusion. This delayed response causes dynamic power transfer between the modes on the millisecond scale [Fig.~\ref{fig:tmi}(c)]. As a result, the intensity and temperature gratings begin to ``chase'' each other along the fiber [Fig.~\ref{fig:tmi}(b)]. The output beam comprises a coherent superposition of the two modes that changes randomly and rapidly [Fig.~\ref{fig:tmi}(c)].

Similar to SBS, this nonlinear thermo-optical scattering acts as a noise amplifier. When no noise is present and the pump power is held constant, the beam fluctuations described above eventually diminish as the amplifier reaches a steady state. In this state, the intensity and temperature gratings are static and longitudinally aligned [Fig.~\ref{fig:tmi}(a)]. As can be inferred from the temperature profile, the induced index modulation is asymmetric in the transverse dimension. Within each grating period, the asymmetric index modulation in the first half scatters light from one mode to the other. In the second half, the index gradient is reversed, scattering back the same amount of light. The net power transfer is zero, and thus the overall mode content remains constant throughout the fiber. Like the input, the output power is predominantly in the fundamental mode, yielding a clean stable beam, even at high power.

However, any small temporal noise in the seed or pump power can be amplified through the thermo-optical feedback [Fig.~\ref{fig:tmi}(a)].\cite{stihler2020intensity} This noise causes signal intensity fluctuations and thus a dynamic heat load. Because the thermal response cannot follow these fluctuations instantaneously, the temperature grating lags behind, shifting it out of alignment with the intensity grating. This misalignment breaks the longitudinal symmetry of the mode-coupling process observed in the static case and yields a non-zero net power transfer. As a result, power is progressively coupled from the fundamental mode into the higher-order mode. This effectively amplifies the noise in the higher-order mode. Power then flows back and forth between the two modes, manifesting as a moving intensity grating that persists with the noise [Fig.~\ref{fig:tmi}(b,c)]. The resulting fluctuations in the output beam profile led to the initial discovery of TMI [Fig.~\ref{fig:tmi}(c)].\cite{eidam2011experimental}

The onset of TMI can be observed using a high-speed camera to monitor the entire beam\cite{eidam2011experimental, christensen2020experimental} or a photodiode to track the power fluctuations in a portion of the beam\cite{otto2012temporal, johansen2013frequency}. The fluctuations are often quantified by their standard deviation. In many cases, a clear and abrupt increase in the standard deviation is observed with increasing pump power, and the corresponding output power is taken as the TMI threshold.\cite{otto2012temporal, johansen2013frequency} This threshold power depends on the noise sources, their spectral characteristics and magnitudes, and the amplifier design. 

Since TMI arises from the presence of higher order modes, the most straightforward and intuitive solution is to minimize their power. This can be achieved through methods that introduce differential loss between modes, such as tight coiling\cite{koplow2000single, tao2016suppressing}, mode filtering\cite{stutzki2014designing}, and confined doping\cite{eidam2011preferential, li2022confined, li2022investigation}. As these methods have already been discussed in Sec.~\ref{sec:largecore}, we will not revisit them in the following subsections.

\subsection{Preventing multimode guiding at high temperature}
In a typical fiber, the refractive index increases with temperature.\cite{brown2002thermal, dragic2018unifieda, dragic2018unifiedb, cavillon2018unified, hawkins2021kilowatt} Since the core heats up more than the cladding, their index difference grows, which makes the NA larger.\cite{dong2023transverse} As a result, higher-order modes become less lossy, gradually turning this single-mode fiber \textit{multimode} as power rises. 

\textbf{Compositional engineering.}
One way to address this is by designing the fiber glass so that it stays single-mode or becomes even more single-mode at high temperature.\cite{dragic2018unifiedb, dragic2024low, hawkins2021kilowatt} Temperature affects the refractive index of glass ($n$) primarily through changes in polarizability and density.\cite{brown2002thermal, cavillon2017additivity} Heating generally increases polarizability, raising the index, while thermal expansion reduces density ($\rho$), lowering the index. In a fiber, the core and cladding are made of different materials, so they expand at different rates when heated. If the core has a higher thermal expansion coefficient ($\alpha$) than the cladding, the cladding restricts the core's expansion, creating compressive stress inside the core.\cite{dragic2018unifiedb} For common materials like silica, which have positive photoelastic coefficients,\cite{bertholds2002determination} this compression increases the core's refractive index. The combined effect determines the thermo-optic coefficient (${\rm d}n/{\rm d}T$). By properly mixing materials with positive ${\rm d}n/{\rm d}T$ (e.g., SiO$_2$, Al$_2$O$_3$, and GeO$_2$) with those having negative ${\rm d}n/{\rm d}T$ (e.g., P$_2$O$_5$ and B$_2$O$_3$), the refractive index can be tailored to remain nearly constant or decrease upon heating [Table~\ref{tab:TMI}].\cite{cavillon2017additivity, dragic2018unifieda, dragic2018unifiedb, cavillon2018unified, dragic2024low, hawkins2021kilowatt}

\textbf{Structural engineering.}
Photonic-crystal fibers offer another solution.\cite{laurila2012distributed, Jansen:13} The fiber structure can be designed so that, at room temperature, the signal lies outside the bandgap and thus cannot be guided. As the fiber heats up, the bandgap blue-shifts, bringing the signal into a regime where only the fundamental mode is supported. This fiber is effectively single-mode within a specific temperature range. However, two potential issues arise. First, if the fiber has not yet reached the required operating temperature, the signal power may be insufficient to suppress ASE, increasing the risk of parasitic lasing. Second, if the temperature becomes too high, excessive bandgap shift can move the signal into a multimode regime.

\begin{table}[ht]
\tbl{Effects of various dopants on the physical properties of silicate glass relevant to TMI.\cite{dragic2018unifiedb, cavillon2018unified, hawkins2021kilowatt} Arrows indicate increases ($\uparrow$), decreases ($\downarrow$), or negligible changes ($\approx$) relative to SiO$_2$ upon doping.}
{\begin{tabular}{@{}lccccc@{}} \toprule
Compound & $n$ & $\rho$ & $\alpha$ & ${\rm d}n/{\rm d}T$* \\ \colrule
SiO$_2$ & 1.45 & 2200~kg/m$^3$ & $0.6$$\times$$10^{-6}$~K$^{-1}$ & $+1.0$$\times$$10^{-5}$~K$^{-1}$ \\
Al$_2$O$_3$ & $\uparrow$ & $\uparrow$ & $\uparrow$ & $\approx$ \\
P$_2$O$_5$ & $\uparrow$ & $\uparrow$ & $\uparrow$ & $\downarrow$ \\
AlPO$_4$ & $\approx$ & $\approx$ & & $\downarrow$ \\
GeO$_2$ & $\uparrow$ & $\uparrow$ & $\uparrow$ & $\uparrow$ \\
B$_2$O$_3$ & $\downarrow$ & $\downarrow$ & $\uparrow$ & $\downarrow$ \\
BaO & $\uparrow$ & $\uparrow$ & $\uparrow$ & $\uparrow$ \\ 
SrO & $\uparrow$ & $\uparrow$ & $\uparrow$ & $\downarrow$ \\
Yb$_2$O$_3$ & $\uparrow$ & $\uparrow$ & & \\
\botrule
\end{tabular}
}
\begin{tabnote}
*In the ${\rm d}n/{\rm d}T$ column, downward arrows incidentally correspond to negative values.
\end{tabnote}
\label{tab:TMI}
\end{table}

\subsection{Gain saturation}   \label{sec:tmiGS}
The most common way to raise the TMI threshold is by saturating the gain with sufficient seed power to minimize local heat load.\cite{smith2013increasing, hansen2014impact, ward2016theory, li2017experimental, dong2022accurate, wisal2024theorytmi} Gain saturation refers to the condition where increasing the signal power yields only a marginal increase in gain, because most excited ions are already contributing to amplification. This effectively caps the amplification, preventing excessive local energy drain and the associated heating. 

The degree of gain saturation can be controlled by the pump power, which sets the level of population inversion. At low pump levels, only a small signal is needed to saturate the gain. This is why counter-pumping is generally preferred over co-pumping.\cite{smith2013increasing, hansen2014impact, li2017experimental} By launching the pump from the end opposite the seed input, both pump and signal ``increase'' in power along the fiber, so the gain remains well saturated throughout. In contrast, co-pumping leaves the input end less saturated and gradually increases saturation as the signal moves down the fiber. For a double-clad fiber with a fixed signal core size, enlarging the inner cladding increases gain saturation and thereby raises the TMI threshold.\cite{smith2013increasing, li2017experimental} 

However, there is a practical limit to increasing gain saturation.\cite{smith2013increasing} Stronger saturation lowers the effective gain per unit length, requiring a longer fiber to reach the target output power. This comes at the cost of a reduced SBS threshold, so properly balancing the saturation and fiber length is critical. 

An effective strategy is dual-wavelength pumping.\cite{shi2022435, jiang2022650, shi2022700} For a Yb-doped fiber amplifier, this requires pairing a strongly absorbed pump near 976~nm with a more weakly absorbed pump near 915~nm or 940~nm [Fig.~\ref{fig:spectra}]. By adjusting their power ratio, the degree of gain saturation can be precisely controlled. Using this approach, an all-fiber single-frequency amplifier achieving 703~W of output power was demonstrated in 2022, even under co-pumping.\cite{shi2022700} To mitigate the intense heating near the input inherent to co-pumping,\cite{hansen2011thermo} a longitudinally segmented doping design was employed.\cite{shi2022435, shi2022700, shi2022high, ren2025425} They spliced two fiber sections together: the first with low Yb doping to limit the peak local heat load, and the second with high Yb doping to achieve strong amplification within a short length.

The preceding sections have covered some of the mainstream strategies used in recent years to mitigate SBS and TMI. Many other innovative approaches have also been developed but are not discussed here. Examples include two-tone seeding for SBS suppression,\cite{dajani2009investigation, dajani2010stimulated, zeringue2011pump} dynamic seed or pump modulation for TMI mitigation,\cite{otto2013controlling, montoya2017photonic, jauregui2018pump} and single-crystalline fibers.\cite{kim2012single, dubinskii2018low, dong2023power}  
Details can be found in the references cited.

\section{New Strategies}
Bringing single-frequency fiber lasers into the multi-kilowatt regime presents a significant technological challenge, but overcoming this barrier will push the frontiers of many laser-based technologies. To make this leap, we will likely require disruptive ideas that can fundamentally change the way we approach laser amplification. Two such promising approaches are described below. 

\subsection{Multimode excitation with wavefront shaping} \label{sec:multimode}
Adaptive multimode fiber amplifiers have recently emerged as a promising route for further power scaling.\cite{rothe2025wavefront, huang2025high} This is largely because they are expected to be inherently more resistant to nonlinear optical effects than single-mode fibers.\cite{wisal2024theorysbs, wisal2024theorytmi} Analogous to chirped pulse amplification (CPA), which stretches laser pulses in time to reduce peak power, the use of multimode fibers spreads the power in space and across many spatial modes. Using a larger core reduces the intensity, which weakens light--matter interactions. Distributing the power over multiple modes further reduces nonlinear effects, as nonlinear \textit{intermodal} scattering is generally weaker than nonlinear \textit{intramodal} scattering within the fundamental mode.  A simplified picture is the small spatial overlap between different modes, although a full analysis must also take into account mediating fields, such as acoustic modes in the case of SBS.\cite{wisal2024theorysbs}

\begin{figure}
\centerline{\includegraphics[width=\textwidth]{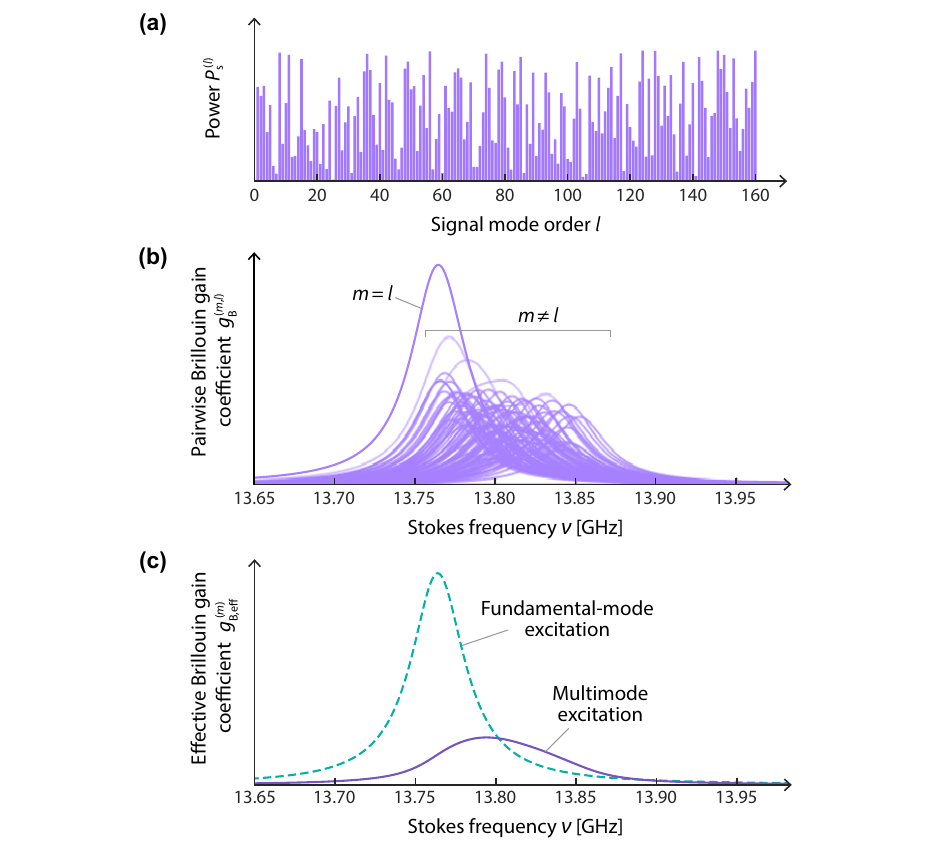}}
\caption{
\textbf{Stimulated Brillouin scattering (SBS) under multimode excitation.}\cite{chen2023mitigating,wisal2024theorysbs,wisal2024optimal,rothe2025wavefront} 
(a) Signal power $P_{\rm s}^{(l)}$ randomly distributed among 160 fiber modes.
(b) Pairwise Brillouin gain spectra $g_{\rm B}^{(m,l)}(\nu)$ experienced by a specific Stokes mode $m$, showing the individual contributions from various signal modes $l$. The intramodal gain ($m=l$) is stronger than the intermodal gain ($m \neq l$).
The Stokes shift is expressed in linear frequency $\nu$ to follow experimental literature conventions.
(c) Effective Brillouin gain spectrum $g_{\rm B,eff}^{(m)}(\nu)$ for fundamental-mode excitation ($l=1$) versus multimode excitation, where $m$ is the mode that maximizes the effective Brillouin gain spectrum. Multimode excitation yields a broad Brillouin gain spectrum with a suppressed peak relative to the fundamental-mode-only excitation (as seen in single-mode amplifiers).
All data were simulated using open-source codes developed by Wisal~\textit{et al.}\cite{wisal2023codes} 
}
\label{fig:SBSmm} 
\end{figure}

\textbf{SBS mitigation.}
It has been shown, both experimentally and theoretically, that multimode excitation leads to a much higher SBS threshold than excitation of the fundamental mode alone [Fig.~\ref{fig:SBSmm}].\cite{chen2023mitigating, wisal2024theorysbs, wisal2024optimal, rothe2025wavefront, smith2025wavefront} 
In a multimode fiber, the signal, Stokes, and acoustic fields may each populate multiple guided modes, depending on the excitation conditions.\cite{wisal2024theorysbs, wisal2024optimal} As a result, SBS can occur both between modes of the same order (intramodal scattering) and between modes of different orders (intermodal scattering). The scattering from a signal mode of order $l$ to a Stokes mode of order $m$ is mediated primarily by one or two specific acoustic modes of $j$. The corresponding Brillouin gain coefficient is determined by the overlap integral of these three fields:\cite{wisal2024theorysbs} 
\begin{equation}
O_{mlj} = \langle (\vec{\psi}_{{\rm s}}^{(l)} \cdot \vec{\psi}_{{\rm St}}^{(m)*}) \xi^{(j)*} \rangle,
\end{equation}
and is maximized when all are in their fundamental modes ($m=l=j=1$), where their spatial profiles closely match. The scattering between the same higher-order optical modes ($m=l>1$) is still primarily mediated by the fundamental acoustic mode ($j=1$). However, the reduced spatial overlap between them leads to a lower intramodal gain than in the fundamental-mode case. For scattering between modes of different orders ($m \neq l$), the overlap is further reduced by mismatches in their spatial amplitude and polarization distributions. Also, the scattering of different signal--Stokes mode pairs $(m,l)$ may be predominantly mediated by different acoustic modes, each characterized by its own phonon lifetime and therefore a distinct Brillouin frequency $\Omega_{\rm B}^{(m,l)}$. As a result, intermodal SBS is not only weaker than intramodal gain but also spectrally separated [Fig.~\ref{fig:SBSmm}(b)]. 

When the signal power is distributed among multiple modes $P_{\rm s,0}~=~\sum_l P_{\rm s}^{(l)}(L)$ [Fig.~\ref{fig:SBSmm}(a)], the effective Brillouin gain spectrum of each Stokes mode $g_{\rm B,eff}^{(m)}(\Omega)$ is a superposition of intramodal and intermodal gain spectra $g_{\rm B}^{(m,l)}(\Omega)$, each weighted by the corresponding signal power fraction:\cite{wisal2024theorysbs, wisal2024optimal}
\begin{equation}
g_{\rm B,eff}^{(m)}(\Omega) = \sum_{l} g_{\rm B}^{(m,l)}(\Omega) \, P_{\rm s}^{(l)}(L) / P_{\rm s,0}.
\end{equation}
In each Stokes mode, the noise-seeded power $P_{\rm St}^{(m)}(\Omega, L)$ propagates backward from the distal end ($z=L$) and is amplified by both linear gain $g_{\rm L}$ via stimulated emission and nonlinear gain $g_{\rm B,eff}^{(m)}(\Omega)$ via SBS, leading to exponential growth toward the proximal end ($z=0$):\cite{wisal2024theorysbs, wisal2024optimal}
\begin{equation}
P_{\rm St}^{(m)}(\Omega, 0) = P_{\rm St}^{(m)}(\Omega, L) \, e^{g_{\rm L} L} \, e^{g_{\rm B,eff}^{(m)}(\Omega) P_{\rm s,0} L_{\rm eff}}. 
\label{eq:SBSgrowth}
\end{equation} 
The Stokes mode with the largest Brillouin gain, $\max_{\Omega,m}g_{\rm B,eff}^{(m)}(\Omega)$, dominates the growth of the total Stokes power.\cite{wisal2024theorysbs, wisal2024optimal}

Under multimode excitation, both the reduced scattering strengths and the spectral separation among pairwise Brillouin gain spectra contribute to a broadened effective spectrum with a low peak gain [Fig.~\ref{fig:SBSmm}(b,c)]. 
This stands in stark contrast to SBS in a single-mode fiber, which exhibits a strong and narrow Brillouin gain spectrum arising exclusively from fundamental-mode scattering [Fig.~\ref{fig:SBSmm}(c)]. Note also that the signal linewidth remains unaffected by multimode excitation, preserving laser coherence.\cite{chen2023mitigating, rothe2025wavefront}

\textbf{TMI mitigation.}
Counterintuitively, multimode excitation does not enhance TMI but instead mitigates it, as shown by numerical and theoretical studies\cite{chen2023suppressing, wisal2024theorytmi, wisal2024optimal} with emerging experimental evidence\cite{zhang2019bending, wen2022experimental, li2023mitigation}. 

Similar to Eq.~\ref{eq:SBSgrowth} for SBS, the output noise power of mode $m$, $P_{\rm N}^{(m)}(\Omega', L)$, can be described as the input noise power $P_{\rm N}^{(m)}(\Omega', 0)$ in that mode amplified by both the linear optical gain $g_{\rm L}$ and the nonlinear thermo-optical gain $g_{\rm T,eff}^{(m)}(\Omega')$:\cite{wisal2024theorytmi, wisal2024optimal} 
\begin{equation}
P_{\rm N}^{(m)}(\Omega', L) = P_{\rm N}^{(m)}(\Omega', 0) \, e^{g_{\rm L} L} \, e^{g_{\rm T,eff}^{(m)}(\Omega') P_{\rm s,0} L_{\rm eff}}. 
\label{eq:TMIgrowth}
\end{equation} 
Here the noise is also at frequencies Stokes-shifted from the signal, $\omega_{\rm N} = \omega_{\rm s} - \Omega'$. 
This nonlinear gain coefficient $g_{\rm T,eff}^{(m)}(\Omega')$ is given by the sum of signal powers across all contributing modes $P_{\rm s}^{(l)}(\Omega', L)$ (with $l \neq m$), weighted by the intermodal thermo-optical coupling coefficient $g_{\rm T}^{(m,l)}(\Omega')$, and normalized by the total signal power $P_{\rm s,0}$:
\begin{equation}
g_{\rm T,eff}^{(m)}(\Omega') = \sum_{l \neq m} g_{\rm T}^{(m,l)}(\Omega') \, P_{\rm s}^{(l)}(L) / P_{\rm s,0}. 
\label{eq:TMIgain}
\end{equation} 

While common strategies for mitigating TMI involve reducing the input noise power in higher-order modes $P_{\rm N}^{(m)}(\Omega', 0)$ by making them extremely lossy (see Sec.~\ref{sec:tmiMain}), the strategy discussed in this section focuses on reducing the nonlinear gain coefficient $g_{\rm T,eff}^{(m)}(\Omega')$. Consider a fiber amplifier supporting only two modes, where the input noise power is distributed according to their signal power ratio. 
Equal excitation of both modes produces intensity and temperature gratings with a much higher contrast than those formed under fundamental-mode excitation (with 1\% power in the higher-order mode). Although the large induced index modulation suggests stronger scattering, TMI can, in fact, be better suppressed.\cite{chen2023suppressing} 

The efficacy of this approach lies in the trade-off between the input noise prefactor $P_{\rm N}^{(2)}(\Omega', 0)$ and the exponential gain term $e^{g_{\rm T,eff}^{(2)}(\Omega') P_{\rm s,0} L_{\rm eff}}$ [Eq.~\ref{eq:TMIgrowth}]. Compared to the fundamental-mode excitation, equal mode excitation increases the input noise power of the higher-order mode (by a factor of 50). However, the nonlinear gain coefficient in the exponent is halved, because the power available for dynamic transfer (i.e., the signal power in the fundamental mode) is halved, $P_{\rm s}^{(1)}(L) = P_{\rm s,0}/2$ [Eq.~\ref{eq:TMIgain}]. For fibers of a few meters in length (large $L_{\rm eff}$), this reduction significantly slows down the exponential growth of the noise power, readily outweighing the impact of the increased input noise. 
The TMI threshold is thereby increased, with this effect becoming even more pronounced when multiple modes are excited (as discussed shortly). Below this threshold, the output beam is stable but distorted by significant higher-order mode content. This distortion can then be corrected using one or two phase masks.\cite{rothe2025output}  

It is important to note, however, that this argument (currently supported only by simulations\cite{chen2023suppressing}) contradicts recent experimental results reported by Jauregui \textit{et al.}\cite{jauregui2026recent} They observed a decrease in the TMI threshold when more power was allocated to the higher-order mode. This discrepancy likely arises from their use of a short fiber with a large core (and thus a long grating period), a regime which can be dominated by the input noise power rather than the nonlinear gain.

\begin{figure}
\centerline{\includegraphics[width=\textwidth]{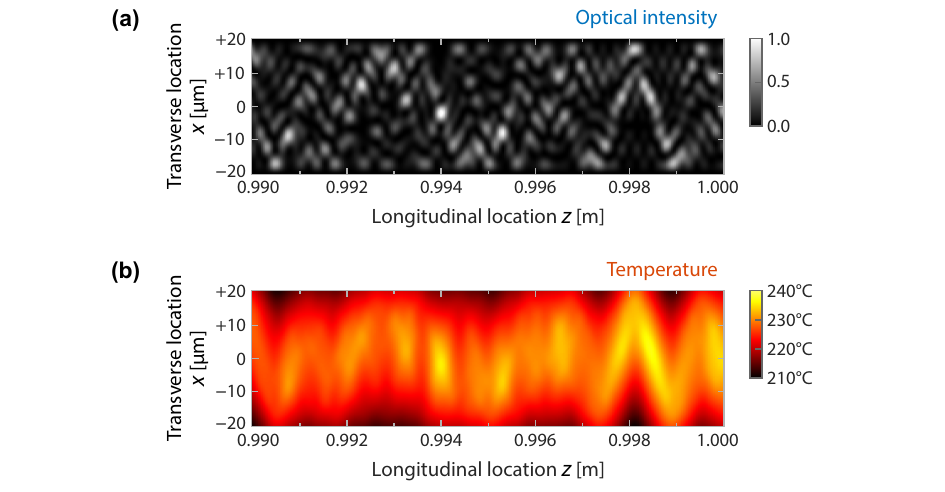}}
\caption{ 
\textbf{Transverse mode instability (TMI) under multimode excitation.}\cite{chen2023suppressing, wisal2024theorytmi} 
(a) Optical intensity and (b) temperature distributions in a fiber amplifier. Temperature speckles are much larger than intensity speckles because transverse heat diffusion smooths out fine spatial modulations. This suppresses coupling between modes with large propagation-constant difference, greatly enhancing the stability. 
Data are simulated using open-source codes developed by Chen \textit{et al.}\cite{chen2024codes, chen2025output} Ambient temperature is set at 20$^{\circ}$C.
}
\label{fig:tmiMM} 
\end{figure}

When multiple modes are excited, the optical intensity distribution in the fiber amplifier becomes speckled [Fig.~\ref{fig:tmiMM}(a)]. These intensity speckles act as localized heat sources, which generate corresponding temperature speckles. However, transverse heat diffusion smooths out the fine temperature variations, leaving only a pseudo-random thermal grating with a relatively large period [Fig.~\ref{fig:tmiMM}(b)]. As a result, the induced refractive-index grating can only efficiently couple modes with a propagation constant difference $\Delta\beta$ small enough that their beating period, $\Lambda = 2\pi/\Delta \beta$, matches the large grating period. This largely restricts the nonlinear thermo-optical scattering between neighboring modes.

In a conventional step-index fiber, each mode thermo-optically interacts with only about six effective nearest neighbors.\cite{chen2023suppressing, wisal2024theorytmi} If the total signal power is distributed equally among $M$ modes ($P_{\rm s}^{(l)}=P_{\rm s,0}/M$), the nonlinear gain in Eq.~\ref{eq:TMIgain} reduces to $g_{\rm T,eff}^{(m)} \approx 6 g_{\rm T}^{(2,1)} P_{\rm s,0}/M$, where $g_{\rm T}^{(2,1)}$ is the coupling coefficient between nearest neighbors. Accordingly, the TMI threshold under multimode excitation is about $M/6$ times that under fundamental-mode excitation (as in single-mode fiber amplifiers). This indicates that the threshold increases linearly with the effective number of excited modes. For instance, exciting a few tens of modes can raise the TMI threshold by roughly an order of magnitude, from hundreds of watts to kilowatts. 
\newpage

\begin{figure}
\centerline{\includegraphics[width=\textwidth]{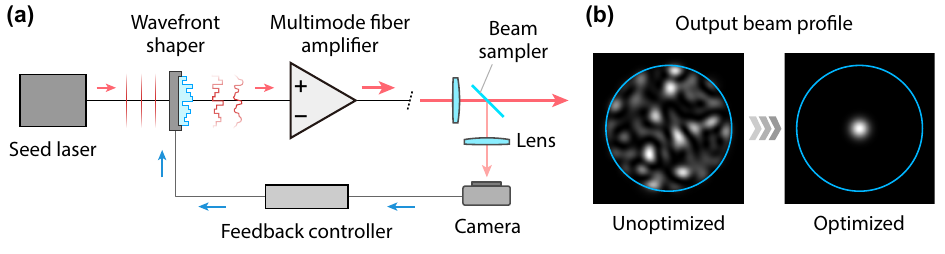}}
\caption{
\textbf{Adaptive multimode fiber amplifier.}\cite{rothe2025output, rothe2025wavefront} (a) Input wavefront is optimized for a target output beam profile via a camera-based feedback loop. (b) Simulated output intensity profile before and after optimization of the input wavefront for a diffraction-limited output beam. Blue circles mark the core area. Mode field profiles used in the simulation were generated with open-source codes developed by Michael Hughes.\cite{huges2020codes}
}
\label{fig:MMFA} 
\end{figure}

\textbf{Wavefront shaping.}
Contrary to the common belief that multimode fiber amplifiers necessarily produce random speckled output, a diffraction-limited beam can be produced through coherent control of multimode amplification [Fig.~\ref{fig:MMFA}]. The key condition is that the signal linewidth must be narrower than the spectral correlation width of the output field, which is typically on the order of gigahertz for a few meters of step-index fiber.\cite{redding2013all} By shaping the input seed wavefront (e.g., using a spatial light modulator), a set of modes with tailored amplitudes and phases can be selectively excited. These mode fields stay mutually coherent throughout amplification and interfere to form a desired output beam. The output beam can be diffraction-limited or, more generally, an arbitrary complex shape, provided that it can be synthesized through coherent superposition of the supported modes. The number of modes excited depends on the target beam profile and position, the fiber structure (core size, NA, and shape), and the strength of mode coupling in the fiber. Although the signal power is distributed across multiple modes within the fiber, the output field can, in principle, couple entirely into the fundamental Gaussian mode in free space and thus propagate exactly as a standard diffraction-limited beam.

Following several prior studies\cite{florentin2017shaping, florentin2019shaping, chen2023suppressing, chen2023mitigating, wisal2024theorysbs, wisal2024theorytmi, chen2025output, rothe2025output}, the first experimental demonstrations of single-frequency multimode fiber amplifier were reported last year (2025), achieving output powers up to 503~W.\cite{rothe2025wavefront, huang2025high} By optimizing the seed wavefront to maximize the output power within a target focal area, the amplifier produces an output beam with $M^2$ of 1.35 that remained stable for at least several hours. Beam quality approaching the diffraction limit ($M^2 \approx 1.0$) is anticipated as the experimental apparatus and optimization algorithms are further refined.

\subsection{Anti-Stokes fluorescence cooling}
\label{sec:ASF}
One way to suppress transverse mode instability (TMI) is to reduce the quantum defect, which is the energy difference between the pump and signal photons.\cite{zhou2017high, ma2018high, jauregui2020transverse, yu2022optically, ballato2024prospects} The resulting heat load can be written as $Q = g_{\rm L}(\lambda_{\rm s} / \lambda_{\rm p}-1)I_{\rm s}$, where $g_{\rm L}$ is the gain coefficient, $\lambda_{\rm s}$ and $\lambda_{\rm p}$ are the signal and pump wavelengths, and $I_{\rm s}$ is the local signal intensity. This heating decreases as the pump wavelength moves closer to the signal wavelength. While high-power diode lasers at wavelengths longer than the conventional pump wavelength (e.g., 976~nm for Yb, where absorption peaks) are technically feasible, they are rarely available commercially because there is no significant market for them. Consequently, this approach is usually implemented using tandem pumping, where one or several fiber lasers are used to pump a fiber amplifier.\cite{richardson2010high, zhou2017high, ma2018high, yu2022optically, baer2024ultra} 

\begin{figure}
\centerline{\includegraphics[width=\textwidth]{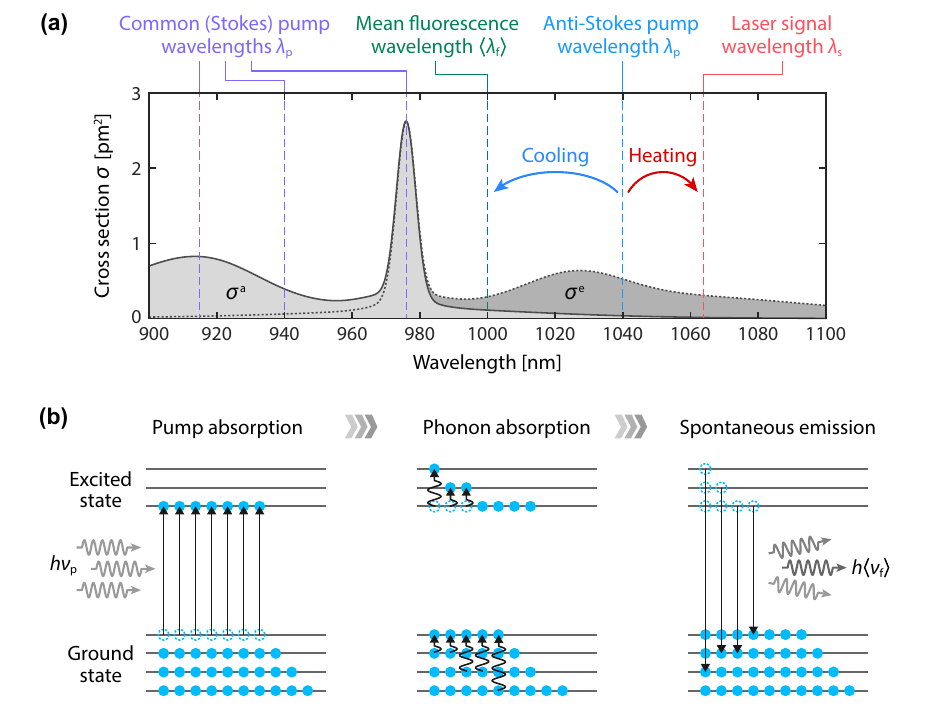}}
\caption{
\textbf{Anti-Stokes fluorescence cooling.} (a) Absorption ($\sigma^{\rm a}$) and emission ($\sigma^{\rm e}$) spectra of Yb$^{3+}$-doped silica fiber. Dashed lines are representative wavelengths for conventional and radiation-balanced laser amplification. 
These include common pump wavelengths $\lambda_{\rm p}$ at 915, 940, and 976~nm, mean fluorescence wavelength $\langle \lambda_{\rm f} \rangle$ at $\approx$1000~nm, and signal wavelength $\lambda_{\rm s}$ at 1064~nm. Stokes pumping ($\lambda_{\rm p} < \langle \lambda_{\rm f} \rangle$) offers high absorption but causes heating. By contrast, anti-Stokes pumping at $\lambda_{\rm p} =$ 1040~nm $> \langle \lambda_{\rm f} \rangle$ yields a \textit{negative} quantum defect for spontaneous emission, which can induce cooling to counteract the heating caused by stimulated emission. Spectra were generated using open-source codes developed by Luke Rumbaugh.\cite{rumbaugh2013codes} 
(b) Energy diagrams illustrating anti-Stokes fluorescence process. Low-energy pump photons excite electrons into lower sublevels of the excited state. Electrons then thermalize to higher sublevels by absorbing phonons until the Boltzmann distribution is reached. Spontaneous emission carries away more energy than supplied by the pump, leading to cooling. 
}
\label{fig:asf} 
\end{figure}

Pushing this idea one step further, it is natural to ask whether a laser could operate without heating at all. This led to the concept of radiation-balanced lasers, proposed in 1999 and demonstrated in a Yb:YAG rod in 2010 by Bowman \textit{et al.}\cite{bowman2002lasers, bowman2010minimizing} Extending this idea to fiber lasers proved much more challenging and was not achieved until 2021, when Knall \textit{et al.} demonstrated it in a Yb-doped silicate fiber, following major advances in fiber materials and fabrication.\cite{knall2021radiationLaser, knall2021radiationAmp} 

Consider a Yb-doped silicate fiber amplifier operating at a signal wavelength of $\lambda_{\rm s}=1064$~nm, with a mean fluorescence wavelength $\langle \lambda_{\rm f} \rangle$ typically between 990 and 1010~nm.\cite{chen2025emerging} Although it is commonly pumped at $\lambda_{\rm p}=$ 915, 940, or 976~nm to ensure efficient absorption, these wavelengths are shorter than $\lambda_{\rm s}$ and $\langle \lambda_{\rm f} \rangle$ (i.e., Stokes pumping) [Fig.~\ref{fig:asf}(a)].\cite{richardson2010high, jauregui2013high, zervas2014high2} The quantum defect is \textit{positive} for both spontaneous and stimulated emission, and this excess photon energy is dissipated as heat.\cite{knall2021design} 

In a radiation-balanced amplifier, the pump wavelength is red-shifted beyond the mean fluorescence wavelength (anti-Stokes) but remains shorter than the signal wavelength, $\langle \lambda_{\rm f} \rangle < \lambda_{\rm p} < \lambda_{\rm s}$, as depicted in Fig.~\ref{fig:asf}(a).\cite{knall2021design} The anti-Stokes pump wavelength is usually selected in the range of 1030--1050~nm for optimal performance.\cite{knall2018model, knall2020laser} These low-energy pump photons can only excite electrons from the top of the Yb$^{3+}$ ground-state manifold to the bottom of the excited-state manifold [Fig.~\ref{fig:asf}(b)]. The electrons then acquire energy from the phonon bath, allowing them to populate higher sublevels within the manifold and reach the Boltzmann distribution. Subsequent radiative relaxation to the ground state via spontaneous emission carries this energy out of the fiber. Since the average energy of the emitted photons ($h\langle \nu_{\rm f} \rangle$) exceeds that of the absorbed pump photons ($h\nu_{\rm p}$), the fiber is cooled. This process is commonly referred to as anti-Stokes fluorescence cooling.
The maximum heat extraction rate per unit length is given by:
\begin{equation}
\left( \frac{{\rm d}Q}{{\rm d}t} \right)_{\rm max} = \left( \frac{\tau_{\rm rad}}{\tau(N_0)} h \nu_{\rm p} - h \langle \nu_{\rm f} \rangle \right) \frac{\sigma_{\rm p}^{\rm a}}{\sigma_{\rm p}^{\rm a} + \sigma_{\rm p}^{\rm e}} \frac{A_0 N_0}{\tau_{\rm rad}},
\label{eq:asf}
\end{equation}
where $N_0$ is the Yb$^{3+}$ concentration, $A_0$ is the doped area, $\tau_{\rm rad}$ is the radiative lifetime, $\tau(N_0)$ is the concentration-dependent excited-state lifetime (as shortened by quenching), $\sigma_{\rm p}^{\rm a}$ and $\sigma_{\rm p}^{\rm e}$ are the absorption and emission cross-sections at the pump wavelength.  
When this cooling balances the heat generated during signal amplification, the amplifier can operate at room temperature with zero net heating.

Net cooling by anti-Stokes pumping does not occur in all fiber amplifiers. In most commercial fibers, parasitic heating from impurity absorption, concentration quenching, and photodarkening can readily outpace heat extraction by anti-Stokes fluorescence, causing the fiber to heat up even at low pump powers.\cite{knall2020experimental, chen2025observation, meehan2024impact} With advances in materials engineering,\cite{chen2024optical, chen2025emerging} radiation-balanced amplification of single-frequency lasers has been demonstrated at output powers of up to 600~mW.\cite{knall2021radiationLaser, knall2021radiationAmp, balliu2024high, balliu2024single, chen2025advancing} Even when scaling to the 1--10~W level, the temperature rise of these fiber amplifiers remains within a few kelvins when suspended in air, as parasitic effects are effectively suppressed.\cite{chen2026600mw} 

Scaling to hundred-watt output comparable to state-of-the-art amplifiers is a non-trivial task. One major challenge is the need for much stronger cooling. Another is ensuring sufficient gain, which likely requires a longer fiber (potentially tens of meters) because the absorption cross-section $\sigma_{\rm p}^{\rm a}$ at anti-Stokes pump wavelengths is much lower than that at conventional (Stokes) pump wavelengths.\cite{knall2021design} However, increasing the fiber length lowers the SBS threshold.

Equation~\ref{eq:asf} suggests two pathways for enhancing cooling. The first is to further increase the rare-earth doping concentration $N_0$ beyond the typical range of $10^{25}$--$10^{27}$~Yb/m$^3$, which increases both cooling and gain per unit length. This, however, often comes with stronger quenching and higher density of impurities. Effective suppression of parasitic effects at very high doping levels is essential and requires optimization of both glass composition and fiber fabrication.\cite{ballato2024prospects, chen2025emerging} The second is to increase the doped area $A_0$, which enhances not only the achievable cooling but also the pump saturation power, allowing higher pump injection. This inevitably results in multimode operation. As discussed in Sec.~\ref{sec:multimode}, high beam quality can be maintained through seed-wavefront shaping, and the SBS threshold is effectively increased due to reduced intensity and multimode excitation.\cite{rothe2025wavefront} 

Along similar lines, excitation-balanced amplification has been proposed to achieve athermal operation.\cite{yu2021reduced, yu2023fdtd} The active fiber is pumped at two wavelengths: one shorter (Stokes) and one longer (anti-Stokes) than the signal wavelength. The quantum-defect heating from the former is balanced by the cooling from the latter. While this concept has been validated numerically and supported by preliminary experimental checks, it still awaits further experimental demonstration to confirm its practical efficacy.  

\section{Beyond Current Limits}
While the mitigation of nonlinear acousto- and thermo-optical scattering remains the primary focus for high-power single-frequency fiber amplifiers, continued power scaling will inevitably bring other nonlinear physical processes into play. Most of these nonlinear optical effects are essentially a result of optical noise amplification driven by nonlinear physical processes.\cite{hildebrandt2008brillouin, hansen2013theoretical, stihler2020intensity, wisal2024theorysbs, wisal2024theorytmi} Even if the nonlinear effects are not severe enough to prevent increased output power from the amplifier, their presence can still lead to noisy output, which significantly degrades system performance.\cite{hochheim2021single, tao2024experimental, rothe2025wavefront} Therefore, bold ideas for mitigating nonlinear effects are always welcomed. 

\newpage

The goal of designing high-power single-frequency fiber amplifiers shares a similar spirit with the peace sought in this classic bedtime ritual, where every unwanted disturbance must be hushed: \\
\newline
\textit{Goodnight stars,}\\
\textit{Goodnight air,}\\
\textit{Goodnight noises everywhere.}\\
\newline
--------- \textit{Goodnight Moon} by Margaret Wise Brown\\

\section{Acknowledgments}
  I would like to thank Prof.\ Hui Cao at Yale University, Dr.\ Peyman Ahmadi at Tescan, and Dr.\ Kabish Wisal at ASML for their feedback during the writing of this chapter and Dr.\ Ting-Mao Feng at National Sun Yat-sen University for his assistance in preparing the figures. 
  
  Beyond these specific contributions, I am sincerely appreciative of the support and guidance provided by Prof.\ Hui Cao, Prof.\ Michel J.\ F.\ Digonnet at Stanford University, Prof.\ Iam Choon Khoo at Penn State University, and Prof.\ Tsung-Hsien Lin at National Sun Yat-sen University over the years. I am also grateful to all the brilliant minds I've had the privilege to work with and learn from. 
  
  This book chapter is dedicated to my partner Shuo, my sisters Yu-yu and Ning-ning, and my children Bailey, Mochi, Rary (Laurie), and Robin.

\bibliographystyle{ws-rv-van}
\bibliography{fiber_lasers}

@incollection{agrawal2000nonlinear,
  title={Nonlinear fiber optics},
  author={Agrawal, Govind P},
  booktitle={Nonlinear Science at the Dawn of the 21st Century},
  pages={195--211},
  year={2000},
  publisher={Springer}
}

@article{kobyakov2009stimulated,
  title={{Stimulated Brillouin scattering in optical fibers}},
  author={Kobyakov, Andrey and Sauer, Michael and Chowdhury, Dipak},
  journal={Advances in optics and photonics},
  volume={2},
  number={1},
  pages={1--59},
  year={2009},
  publisher={Optical Society of America}
}

@article{eggleton2013inducing,
  title={{Inducing and harnessing stimulated Brillouin scattering in photonic integrated circuits}},
  author={Eggleton, Benjamin J and Poulton, Christopher G and Pant, Ravi},
  journal={Advances in Optics and Photonics},
  volume={5},
  number={4},
  pages={536--587},
  year={2013},
  publisher={Optical Society of America}
}

@article{willke2008stabilized,
  title={{Stabilized lasers for advanced gravitational wave detectors}},
  author={Willke, Benno and Danzmann, Karsten and Frede, Maik and King, Peter and Kracht, D and Kwee, Patrick and Puncken, O and Savage, RL and Schulz, Bastian and Seifert, Frank and others},
  journal={Classical and Quantum Gravity},
  volume={25},
  number={11},
  pages={114040},
  year={2008},
  publisher={IOP Publishing}
}

@article{willke2010stabilized,
  title={{Stabilized lasers for advanced gravitational wave detectors}},
  author={Willke, Benno},
  journal={Laser \& Photonics Reviews},
  volume={4},
  number={6},
  pages={780--794},
  year={2010},
  publisher={Wiley Online Library}
}

@article{lyons2000shot,
  title={{Shot noise in gravitational-wave detectors with Fabry--Perot arms}},
  author={Lyons, Torrey T and Regehr, Martin W and Raab, Frederick J},
  journal={Applied Optics},
  volume={39},
  number={36},
  pages={6761--6770},
  year={2000},
  publisher={Optical Society of America}
}

@article{kwee2012stabilized,
  title={{Stabilized high-power laser system for the gravitational wave detector advanced LIGO}},
  author={Kwee, Patrick and Bogan, C and Danzmann, K and Frede, M and Kim, H and King, P and P{\"o}ld, J and Puncken, O and Savage, Rick L and Seifert, F and others},
  journal={Optics Express},
  volume={20},
  number={10},
  pages={10617--10634},
  year={2012},
  publisher={Optical Society of America}
}

@inproceedings{hall2016laser,
  title={{Laser Frequency and Intensity Stabilization for Advanced LIGO}},
  author={Hall, Evan},
  booktitle={18th Coherent Laser Radar Conference and the Lidar Working Group on Space Based Winds, CLRC 2016},
  year={2016}
}

@article{jia2021point,
  title={{Point absorber limits to future gravitational-wave detectors}},
  author={Jia, Wenxuan and Yamamoto, Hiroaki and Kuns, Kevin and Effler, Anamaria and Evans, Matthew and Fritschel, Peter and Abbott, R and Adams, C and Adhikari, Rana X and Ananyeva, A and others},
  journal={Physical Review Letters},
  volume={127},
  number={24},
  pages={241102},
  year={2021},
  publisher={APS}
}

@article{cahillane2021laser,
  title={{Laser frequency noise in next generation gravitational-wave detectors}},
  author={Cahillane, Craig and Mansell, Georgia L and Sigg, Daniel},
  journal={Optics Express},
  volume={29},
  number={25},
  pages={42144--42161},
  year={2021},
  publisher={Optical Society of America}
}

@article{de2017single,
  title={{Single-frequency fiber amplifier at 1.5 \textmu m with 100 W in the linearly-polarized TEM$_{00}$ mode for next-generation gravitational wave detectors}},
  author={De Varona, Omar and Fittkau, Willy and Booker, Phillip and Theeg, Thomas and Steinke, Michael and Kracht, Dietmar and Neumann, J{\"o}rg and Wessels, Peter},
  journal={Optics Express},
  volume={25},
  number={21},
  pages={24880--24892},
  year={2017},
  publisher={Optical Society of America}
}

@article{steinke2017single,
  title={{Single-frequency fiber amplifiers for next-generation gravitational wave detectors}},
  author={Steinke, Michael and Tuennermann, Henrik and Kuhn, Vincent and Theeg, Thomas and Karow, Malte and de Varona, Omar and Jahn, Philipp and Booker, Phillip and Neumann, Joerg and Wessels, Peter and others},
  journal={IEEE Journal of Selected Topics in Quantum Electronics},
  volume={24},
  number={3},
  pages={1--13},
  year={2017},
  publisher={IEEE}
}

@article{buikema2019narrow,
  title={{Narrow-linewidth fiber amplifier for gravitational-wave detectors}},
  author={Buikema, Aaron and Jose, Franklin and Augst, Steven J and Fritschel, Peter and Mavalvala, Nergis},
  journal={Optics Letters},
  volume={44},
  number={15},
  pages={3833--3836},
  year={2019},
  publisher={Optical Society of America}
}

@article{kapasi2020tunable,
  title={{Tunable narrow-linewidth laser at 2 $\mu$m wavelength for gravitational wave detector research}},
  author={Kapasi, DP and Eichholz, Johannes and McRae, Terry and Ward, RL and Slagmolen, BJJ and Legge, Samuel and Hardman, KS and Altin, PA and McClelland, DE},
  journal={Optics Express},
  volume={28},
  number={3},
  pages={3280--3288},
  year={2020},
  publisher={Optical Society of America}
}

@article{wellmann2021low,
  title={{Low noise 400 W coherently combined single frequency laser beam for next generation gravitational wave detectors}},
  author={Wellmann, Felix and Bode, Nina and Wessels, Peter and Overmeyer, Ludger and Neumann, J{\"o}rg and Willke, Benno and Kracht, Dietmar},
  journal={Optics Express},
  volume={29},
  number={7},
  pages={10140--10149},
  year={2021},
  publisher={Optical Society of America}
}

@article{meylahn2022stabilized,
  title={{Stabilized laser system at 1550 nm wavelength for future gravitational-wave detectors}},
  author={Meylahn, Fabian and Knust, Nicole and Willke, Benno},
  journal={Physical Review D},
  volume={105},
  number={12},
  pages={122004},
  year={2022},
  publisher={APS}
}

@article{richardson2010high,
  title={{High power fiber lasers: current status and future perspectives}},
  author={Richardson, Davis J and Nilsson, John and Clarkson, William A},
  journal={Journal of the Optical Society of America B},
  volume={27},
  number={11},
  pages={B63--B92},
  year={2010},
  publisher={OSA}
}

@article{jauregui2013high,
  title={{High-power fibre lasers}},
  author={Jauregui, Cesar and Limpert, Jens and T{\"u}nnermann, Andreas},
  journal={Nature Photonics},
  volume={7},
  number={11},
  pages={861--867},
  year={2013},
  publisher={Nature Publishing Group UK London}
}

@article{zervas2014high2,
  title={{High power ytterbium-doped fiber lasers---fundamentals and applications}},
  author={Zervas, Michalis N},
  journal={International Journal of Modern Physics B},
  volume={28},
  number={12},
  pages={1442009},
  year={2014},
  publisher={World Scientific}
}

@article{zervas2014high,
  title={{High power fiber lasers: a review}},
  author={Zervas, Michalis N and Codemard, Christophe A},
  journal={IEEE Journal of selected topics in Quantum Electronics},
  volume={20},
  number={5},
  pages={219--241},
  year={2014},
  publisher={IEEE}
}

@article{dong2025past,
  title={{Past, present, and future of fiber lasers and amplifiers}},
  author={Dong, Liang and Zervas, Michalis N},
  journal={Optics Communications},
  volume={577},
  pages={131419},
  year={2025},
  publisher={Elsevier}
}

@book{digonnet2001rare,
  title={{Rare-earth-doped fiber lasers and amplifiers, revised and expanded}},
  author={Digonnet, Michel JF},
  year={2001},
  publisher={CRC press}
}

@article{zhu2010high,
  title={{High-power ZBLAN glass fiber lasers: Review and prospect}},
  author={Zhu, Xiushan and Peyghambarian, N},
  journal={Advances in OptoElectronics},
  volume={2010},
  number={1},
  pages={501956},
  year={2010},
  publisher={Wiley Online Library}
}

@article{jackson2012towards,
  title={{Towards high-power mid-infrared emission from a fibre laser}},
  author={Jackson, Stuart D},
  journal={Nature Photonics},
  volume={6},
  number={7},
  pages={423--431},
  year={2012},
  publisher={Nature Publishing Group UK London}
}

@inproceedings{bernier2025get,
  title={{Get the most out of a fiber laser: from high-power CW to ultrafast operation, from visible to mid-infrared emission}},
  author={Bernier, Martin and Aydin, Yigit Ozan and Paradis, Pascal and Fortin, Vincent and Bisson, William and Michaud, Alexandre and Michaud, Louis-Charles and Boilard, Tommy and Lemieux-Tanguay, Maxime and Talbot, Lauris and others},
  booktitle={Fiber Lasers XXII: Technology and Systems},
  volume={13342},
  pages={10--12},
  year={2025},
  organization={SPIE}
}

@article{tanabe2002rare,
  title={{Rare-earth-doped glasses for fiber amplifiers in broadband telecommunication}},
  author={Tanabe, Setsuhisa},
  journal={Comptes Rendus Chimie},
  volume={5},
  number={12},
  pages={815--824},
  year={2002},
  publisher={Elsevier}
}

@article{paschotta2002ytterbium,
  title={{Ytterbium-doped fiber amplifiers}},
  author={Paschotta, Rudiger and Nilsson, Johan and Tropper, Anne C and Hanna, David C},
  journal={IEEE Journal of Quantum Electronics},
  volume={33},
  number={7},
  pages={1049--1056},
  year={2002},
  publisher={IEEE}
}

@book{becker1999erbium,
  title={{Erbium-doped fiber amplifiers: fundamentals and technology}},
  author={Becker, Philippe M and Olsson, Anders A and Simpson, Jay R},
  year={1999},
  publisher={Elsevier}
}

@article{sincore2017high,
  title={{High average power thulium-doped silica fiber lasers: review of systems and concepts}},
  author={Sincore, Alex and Bradford, Joshua D and Cook, Justin and Shah, Lawrence and Richardson, Martin C},
  journal={IEEE Journal of selected topics in quantum electronics},
  volume={24},
  number={3},
  pages={1--8},
  year={2017},
  publisher={IEEE}
}

@article{gaida2018ultrafast,
  title={{Ultrafast thulium fiber laser system emitting more than 1 kW of average power}},
  author={Gaida, Christian and Gebhardt, M and Heuermann, T and Stutzki, F and Jauregui, C and Limpert, J},
  journal={Optics Letters},
  volume={43},
  number={23},
  pages={5853--5856},
  year={2018},
  publisher={Optical Society of America}
}

@article{ballato2013rethinking,
  title={{Rethinking optical fiber: new demands, old glasses}},
  author={Ballato, John and Dragic, Peter},
  journal={Journal of the American Ceramic Society},
  volume={96},
  number={9},
  pages={2675--2692},
  year={2013},
  publisher={Wiley Online Library}
}

@article{dragic2018materials,
  title={{Materials for optical fiber lasers: A review}},
  author={Dragic, Peter D and Cavillon, M and Ballato, JJAPR},
  journal={Applied Physics Reviews},
  volume={5},
  number={4},
  year={2018},
  publisher={AIP Publishing}
}

@article{lucas1999infrared,
  title={{Infrared glasses}},
  author={Lucas, Jacques},
  journal={Current Opinion in Solid State and Materials Science},
  volume={4},
  number={2},
  pages={181--187},
  year={1999},
  publisher={Elsevier}
}

@article{schuster2014material,
  title={{Material and technology trends in fiber optics}},
  author={Schuster, Kay and Unger, Sonja and Aichele, Claudia and Lindner, Florian and Grimm, Stephan and Litzkendorf, Doris and Kobelke, Jens and Bierlich, J{\"o}rg and Wondraczek, Katrin and Bartelt, Hartmut},
  journal={Advanced Optical Technologies},
  volume={3},
  number={4},
  pages={447--468},
  year={2014},
  publisher={De Gruyter}
}

@article{meehan2025insights,
  title={{Insights into draw-induced refractive index changes in intrinsically low nonlinearity MCVD preforms and optical fibers}},
  author={Meehan, Bailey and Pietros, Alexander R and Topper, Brian and Karki, Peshal and Rao, Apparao M and Liu, Jiahui and Urban, Marek and Jercinovic, Michael J and Youngman, Randall and Hawkins, Thomas W and others},
  journal={Optical Materials Express},
  volume={15},
  number={4},
  pages={661--673},
  year={2025},
  publisher={Optica Publishing Group}
}

@article{koester1964amplification,
  title={{Amplification in a fiber laser}},
  author={Koester, Charles J and Snitzer, Elias},
  journal={Applied Optics},
  volume={3},
  number={10},
  pages={1182--1186},
  year={1964},
  publisher={OSA}
}

@inproceedings{snitzer1988double,
  title={{Double clad, offset core Nd fiber laser}},
  author={Snitzer, E and Po, Hong and Hakimi, Fateme and Tumminelli, Richard and McCollum, BC},
  booktitle={Optical fiber sensors},
  pages={PD5},
  year={1988},
  organization={Optica Publishing Group}
}

@inproceedings{po1989double,
  title={{Double clad high brightness Nd fiber laser pumped by GaAlAs phased array}},
  author={Po, H and Snitzer, E and Tumminelli, R and Zenteno, L and Hakimi, F and Cho, N M and Haw, T},
  booktitle={Optical Fiber Communication Conference},
  pages={PD7},
  year={1989},
  organization={Optica Publishing Group}
}

@article{zenteno2002high,
  title={{High-power double-clad fiber lasers}},
  author={Zenteno, Luis},
  journal={Journal of Lightwave Technology},
  volume={11},
  number={9},
  pages={1435--1446},
  year={2002},
  publisher={IEEE}
}

@article{mortensen2007air,
  title={{Air-clad fibers: pump absorption assisted by chaotic wave dynamics?}},
  author={Mortensen, Niels Asger},
  journal={Optics Express},
  volume={15},
  number={14},
  pages={8988--8996},
  year={2007},
  publisher={Optical Society of America}
}

@article{doya2001optimized,
  title={{Optimized absorption in a chaotic double-clad fiber amplifier}},
  author={Doya, Val{\'e}rie and Legrand, Olivier and Mortessagne, Fabrice},
  journal={Optics Letters},
  volume={26},
  number={12},
  pages={872--874},
  year={2001},
  publisher={OSA}
}

@article{leproux2001modeling,
  title={Modeling and optimization of double-clad fiber amplifiers using chaotic propagation of the pump},
  author={Leproux, Philippe and F{\'e}vrier, S{\'e}bastien and Doya, Val{\'e}rie and Roy, Philippe and Pagnoux, Dominique},
  journal={Optical Fiber Technology},
  volume={7},
  number={4},
  pages={324--339},
  year={2001},
  publisher={Elsevier}
}

@article{li2004high,
  title={{High absorption and low splice loss properties of hexagonal double-clad fiber}},
  author={Li, Yahua and Jackson, Stuart D and Fleming, Simon},
  journal={IEEE Photonics Technology Letters},
  volume={16},
  number={11},
  pages={2502--2504},
  year={2004},
  publisher={IEEE}
}

@article{kouznetsov2001efficiency1,
  title={{Efficiency of pump absorption in double-clad fiber amplifiers. I. Fiber with circular symmetry}},
  author={Kouznetsov, Dmitrii and Moloney, Jerome V and Wright, Ewan M},
  journal={Journal of the Optical Society of America B},
  volume={18},
  number={6},
  pages={743--749},
  year={2001},
  publisher={Optical Society of America}
}

@article{kouznetsov2002efficiency2,
  title={{Efficiency of pump absorption in double-clad fiber amplifiers. II. Broken circular symmetry}},
  author={Kouznetsov, Dmitrii and Moloney, Jerome V},
  journal={Journal of the Optical Society of America B},
  volume={19},
  number={6},
  pages={1259--1263},
  year={2002},
  publisher={Optical Society of America}
}

@article{kouznetsov2002efficiency3,
  title={{Efficiency of pump absorption in double-clad fiber amplifiers. III. Calculation of modes}},
  author={Kouznetsov, Dmitrii and Moloney, Jerome V},
  journal={Journal of the Optical Society of America B},
  volume={19},
  number={6},
  pages={1304--1309},
  year={2002},
  publisher={Optical Society of America}
}

@article{kovska2016enhanced,
  title={{Enhanced pump absorption efficiency in coiled and twisted double-clad thulium-doped fibers}},
  author={Ko{\v{s}}ka, Pavel and Peterka, Pavel and Aubrecht, Jan and Podrazk{\`y}, Ond{\v{r}}ej and Todorov, Filip and Becker, Martin and Baravets, Yauhen and Honz{\'a}tko, Pavel and Ka{\v{s}}{\'\i}k, Ivan},
  journal={Optics Express},
  volume={24},
  number={1},
  pages={102--107},
  year={2016},
  publisher={Optical Society of America}
}

@article{tunnermann2010fiber,
  title={{Fiber lasers and amplifiers: an ultrafast performance evolution}},
  author={T{\"u}nnermann, Andreas and Schreiber, Thomas and Limpert, Jens},
  journal={Applied Optics},
  volume={49},
  number={25},
  pages={F71--F78},
  year={2010},
  publisher={OSA}
}

@inproceedings{stachowiak2018high,
  title={{High-power passive fiber components for all-fiber lasers and amplifiers application—design and fabrication}},
  author={Stachowiak, Dorota},
  booktitle={Photonics},
  volume={5},
  number={4},
  pages={38},
  year={2018},
  organization={MDPI}
}

@article{liu2021high,
  title={{High power pump and signal combiner for backward pumping structure with two different fused fiber bundle designs by means of pretapered pump fibers}},
  author={Liu, Yili and Liu, Kai and Yang, Yifeng and Liu, Meizhong and He, Bing and Zhou, Jun},
  journal={Optics Express},
  volume={29},
  number={9},
  pages={13344--13358},
  year={2021},
  publisher={Optical Society of America}
}

@article{majumder2022design,
  title={{Design and fabrication of a tapered fiber bundle for a pump combiner with a uniform splicing region}},
  author={Majumder, Debparna and Das Chowdhury, Sourav and Pal, Atasi},
  journal={Journal of the Optical Society of America B},
  volume={39},
  number={7},
  pages={1871--1878},
  year={2022},
  publisher={Optica Publishing Group}
}

@article{jauregui2010side,
  title={{Side-pump combiner for all-fiber monolithic fiber lasers and amplifiers}},
  author={Jauregui, Cesar and B{\"o}hme, Steffen and Wenetiadis, Georgios and Limpert, Jens and T{\"u}nnermann, Andreas},
  journal={Journal of the Optical Society of America B},
  volume={27},
  number={5},
  pages={1011--1015},
  year={2010},
  publisher={Optical Society of America}
}

@article{magnan2020fuseless,
  title={{Fuseless side-pump combiner for efficient fluoride-based double-clad fiber pumping}},
  author={Magnan-Saucier, S{\'e}bastien and Duval, Simon and Matte-Breton, Charles and Aydin, Yi{\u{g}}it Ozan and Fortin, Vincent and LaRochelle, Sophie and Bernier, Martin and Vall{\'e}e, R{\'e}al},
  journal={Optics Letters},
  volume={45},
  number={20},
  pages={5828--5831},
  year={2020},
  publisher={Optical Society of America}
}

@inproceedings{brockmuller2025side,
  title={{Side-fused signal-pump combiner for triple clad fibers}},
  author={Brockm{\"u}ller, Eike and Kranert, Fabian and Lachmayer, Roland and Neumann, J{\"o}rg and Kracht, Dietmar},
  booktitle={Fiber Lasers XXII: Technology and Systems},
  volume={13342},
  pages={132--138},
  year={2025},
  organization={SPIE}
}

@article{beier2017single,
  title={{Single mode 4.3 kW output power from a diode-pumped Yb-doped fiber amplifier}},
  author={Beier, Franz and Hupel, Christian and Kuhn, Stefan and Hein, Sigrun and Nold, Johannes and Proske, Fritz and Sattler, Bettina and Liem, Andreas and Jauregui, C{\'e}sar and Limpert, Jens and others},
  journal={Optics Express},
  volume={25},
  number={13},
  pages={14892--14899},
  year={2017},
  publisher={Optical Society of America}
}

@inproceedings{smith2025wavefront,
  title={{Wavefront Shaping for Near-Diffraction Limited Multimode Output in a Record Peak Power, Single-Frequency, 1.5 $\mu$m Fiber Amplifier}},
  author={Smith, Darcy L and Henderson-Sapir, Ori and Singh, Jassimar and Wei, Shuen and Nguyen, Linh V and Ebendorff-Heidepriem, Heike and Warren-Smith, Stephen C and Ottaway, David J},
  booktitle={2025 Conference on Lasers and Electro-Optics (CLEO)},
  pages={1--2},
  year={2025},
  organization={IEEE}
}

@article{hildebrandt2008brillouin,
  title={{Brillouin scattering spectra in high-power single-frequency ytterbium doped fiber amplifiers}},
  author={Hildebrandt, Matthias and B{\"u}sche, Sebastian and We{\ss}els, Peter and Frede, Maik and Kracht, Dietmar},
  journal={Optics Express},
  volume={16},
  number={20},
  pages={15970--15979},
  year={2008},
  publisher={Optical Society of America}
}

@article{kanskar200573,
  title={{73\% CW power conversion efficiency at 50 W from 970 nm diode laser bars}},
  author={Kanskar, M and Earles, T and Goodnough, TJ and Stiers, E and Botez, D and Mawst, LJ},
  journal={Electronics Letters},
  volume={41},
  number={5},
  pages={245--247},
  year={2005},
  publisher={IET}
}

@inproceedings{berishev2005algainas,
  title={{AlGaInAs/GaAs record high-power conversion efficiency and record high-brightness coolerless 915-nm multimode pumps}},
  author={Berishev, Igor and Komissarov, Alexey and Moshegov, Nikolay and Trubenko, Pavel and Wright, Lisa and Berezin, Andrei and Todorov, Svetlan and Ovtchinnikov, Alexander},
  booktitle={Novel In-Plane Semiconductor Lasers IV},
  volume={5738},
  pages={25--32},
  year={2005},
  organization={SPIE}
}

@inproceedings{samson2007diode,
  title={{Diode pump requirements for high power fiber lasers}},
  author={Samson, Bryce and Frith, Gavin},
  booktitle={International Congress on Applications of Lasers \& Electro-Optics},
  volume={2007},
  number={1},
  pages={501},
  year={2007},
  organization={Laser Institute of America}
}

@inproceedings{zucker2014advancements,
  title={{Advancements in laser diode chip and packaging technologies for application in kW-class fiber laser pumping}},
  author={Zucker, Erik and Zou, Daniel and Zavala, Laura and Yu, Hongbo and Yalamanchili, Prasad and Xu, Lei and Xu, Hui and Venables, David and Skidmore, Jay and Rossin, Victor and others},
  booktitle={High-Power Diode Laser Technology and Applications XII},
  volume={8965},
  pages={38--51},
  year={2014},
  organization={SPIE}
}

@inproceedings{gapontsev2017highly,
  title={{Highly-efficient high-power pumps for fiber lasers}},
  author={Gapontsev, V and Moshegov, N and Berezin, I and Komissarov, A and Trubenko, P and Miftakhutdinov, D and Berishev, I and Chuyanov, V and Raisky, O and Ovtchinnikov, A},
  booktitle={High-Power Diode Laser Technology XV},
  volume={10086},
  pages={16--25},
  year={2017},
  organization={SPIE}
}

@article{volodin2004wavelength,
  title={{Wavelength stabilization and spectrum narrowing of high-power multimode laser diodes and arrays by use of volume Bragg gratings}},
  author={Volodin, BL and Dolgy, SV and Melnik, ED and Downs, E and Shaw, J and Ban, VS},
  journal={Optics Letters},
  volume={29},
  number={16},
  pages={1891--1893},
  year={2004},
  publisher={OSA}
}

@article{talbot2022wavelength,
  title={{Wavelength stabilization of high-power laser diodes using Bragg gratings inscribed in their highly multimode fiber pigtails}},
  author={Talbot, Lauris and Pelletier-Ouellet, Samantha and Tr{\'e}panier, Fran{\c{c}}ois and Bernier, Martin},
  journal={Optics Letters},
  volume={47},
  number={3},
  pages={633--636},
  year={2022},
  publisher={Optical Society of America}
}

@article{zhou2017high,
  title={{High-power fiber lasers based on tandem pumping}},
  author={Zhou, Pu and Xiao, Hu and Leng, Jinyong and Xu, Jiangmin and Chen, Zilun and Zhang, Hanwei and Liu, Zejin},
  journal={Journal of the Optical Society of America B},
  volume={34},
  number={3},
  pages={A29--A36},
  year={2017},
  publisher={OSA}
}

@article{baer2024ultra,
  title={{Ultra-low-noise, single-frequency, all-PM Thulium-and Holmium-doped Fiber Amplifiers at 1950 nm and 2090 nm for third-generation Gravitational Wave Detectors}},
  author={Baer, Patrick and Cebeci, Pelin and Reiter, Melina and Bontke, Florian and Giesberts, Martin and Hoffmann, Hans-Dieter},
  journal={IEEE Photonics Journal},
  volume={16},
  number={1},
  pages={1--9},
  year={2024},
  publisher={IEEE}
}

@article{chen2024915,
  title={{915 nm pumping kilowatt fiber oscillator with high optical-to-optical efficiency}},
  author={Chen, Xin and Yang, Yide and Gong, Mali and Su, Ping and Ma, Jianshe},
  journal={Scientific Reports},
  volume={14},
  number={1},
  pages={26331},
  year={2024},
  publisher={Nature Publishing Group UK London}
}

@article{tao2023over,
  title={{Over 250 W low noise core-pumped single-frequency all-fiber amplifier}},
  author={Tao, Yue and Jiang, Man and Liu, Liu and Li, Can and Zhou, Pu and Jiang, Zongfu},
  journal={Optics Express},
  volume={31},
  number={6},
  pages={10586--10595},
  year={2023},
  publisher={Optica Publishing Group}
}

@article{kotov2022high,
  title={{High-energy single-frequency core-pumped Er-doped fiber amplifiers}},
  author={Kotov, Leonid V and Temyanko, Valery and Bubnov, Mikhail M and Lipatov, Denis S and Lobanov, Alexey S and Abramov, Alexey and Aleshkina, Svetlana S and Guryanov, Alexey N and Likhachev, Mikhail E},
  journal={Journal of Lightwave Technology},
  volume={41},
  number={5},
  pages={1526--1532},
  year={2022},
  publisher={IEEE}
}

@article{li2007ge,
  title={{Al/Ge co-doped large mode area fiber with high SBS threshold}},
  author={Li, Ming-Jun and Chen, Xin and Wang, Ji and Gray, Stuart and Liu, Anping and Demeritt, Jeffrey A and Ruffin, A Boh and Crowley, Alana M and Walton, Donnell T and Zenteno, Luis A},
  journal={Optics Express},
  volume={15},
  number={13},
  pages={8290--8299},
  year={2007},
  publisher={Optical Society of America}
}

@inproceedings{gray2006high,
  title={{High power, narrow linewidth fiber amplifiers}},
  author={Gray, Stuart and Walton, Donnell and Wang, Ji and Li, Ming-Jun and Chen, Xin and Ruffin, A Boh and Demeritt, Jeffrey and Zenteno, Luis},
  booktitle={Optical Amplifiers and their Applications},
  pages={OSuB1},
  year={2006},
  organization={Optica Publishing Group}
}

@article{nilsson2003high,
  title={{High-power and tunable operation of erbium-ytterbium co-doped cladding-pumped fiber lasers}},
  author={Nilsson, Johan and Alam, S-U and Alvarez-Chavez, Jose A and Turner, Paul W and Clarkson, W Andrew and Grudinin, Anatoly B},
  journal={IEEE Journal of Quantum Electronics},
  volume={39},
  number={8},
  pages={987--994},
  year={2003},
  publisher={IEEE}
}

@article{fermann1988efficient,
  title={{Efficient operation of an Yb-sensitised Er fibre laser at 1.56 $\mu$m}},
  author={Fermann, ME and Hanna, DC and Shepherd, DP and Suni, PJ and Townsend, JE},
  journal={Electronics Letters},
  volume={24},
  number={18},
  pages={1135--1136},
  year={1988},
  publisher={IET}
}

@incollection{pollnau2008advances,
  title={{Advances in mid-infrared fiber lasers}},
  author={Pollnau, Markus and Jackson, Stuart D},
  booktitle={Mid-infrared coherent sources and applications},
  pages={315--346},
  year={2008},
  publisher={Springer}
}

@article{xiang2025fully,
  title={{Fully recoverable fiber lasers under radiation enabled by in-situ blue light photobleaching}},
  author={Xiang, Guangbiao and Zhang, Hanwei and Wang, Xiaolin and Zhang, Jiangbin and Chen, Hongwei and Lin, Chao and Ye, Yun and Hua, Weihong and Chen, Jinbao},
  journal={Photonics Research},
  volume={13},
  number={8},
  pages={2362--2370},
  year={2025},
  publisher={Chinese Laser Press and Optica Publishing Group}
}

@article{engholm2009improved,
  title={{Improved photodarkening resistivity in ytterbium-doped fiber lasers by cerium codoping}},
  author={Engholm, Magnus and Jelger, P{\"a}r and Laurell, Fredrik and Norin, L},
  journal={Optics Letters},
  volume={34},
  number={8},
  pages={1285--1287},
  year={2009},
  publisher={Optical Society of America}
}

@article{zhao2017mitigation,
  title={{Mitigation of photodarkening effect in Yb-doped fiber through Na$^+$ ions doping}},
  author={Zhao, Nan and Liu, Yehui and Li, Miao and Li, Jiaming and Peng, Jinggang and Yang, Luyun and Dai, Nengli and Li, Haiqing and Li, Jinyan},
  journal={Optics Express},
  volume={25},
  number={15},
  pages={18191--18196},
  year={2017},
  publisher={Optical Society of America}
}

@article{ballato2018unified,
  title={{A unified materials approach to mitigating optical nonlinearities in optical fiber. I. Thermodynamics of optical scattering}},
  author={Ballato, John and Cavillon, Maxime and Dragic, Peter},
  journal={International Journal of Applied Glass Science},
  volume={9},
  number={2},
  pages={263--277},
  year={2018},
  publisher={Wiley Online Library}
}

@article{dragic2018unifieda,
  title={{A unified materials approach to mitigating optical nonlinearities in optical fiber. II. A. Material additivity models and basic glass properties}},
  author={Dragic, Peter D and Cavillon, Maxime and Ballato, Arthur and Ballato, John},
  journal={International Journal of Applied Glass Science},
  volume={9},
  number={2},
  pages={278--287},
  year={2018},
  publisher={Wiley Online Library}
}

@article{dragic2018unifiedb,
  title={{A unified materials approach to mitigating optical nonlinearities in optical fiber. II. B. The optical fiber, material additivity and the nonlinear coefficients}},
  author={Dragic, Peter D and Cavillon, Maxime and Ballato, Arthur and Ballato, John},
  journal={International Journal of Applied Glass Science},
  volume={9},
  number={3},
  pages={307--318},
  year={2018},
  publisher={Wiley Online Library}
}

@article{cavillon2018unified,
  title={{A unified materials approach to mitigating optical nonlinearities in optical fiber. III. Canonical examples and materials road map}},
  author={Cavillon, Maxime and Kucera, Courtney and Hawkins, Thomas and Dawson, Jay and Dragic, Peter D and Ballato, John},
  journal={International Journal of Applied Glass Science},
  volume={9},
  number={4},
  pages={447--470},
  year={2018},
  publisher={Wiley Online Library}
}

@article{cavillon2017additivity,
  title={{Additivity of the coefficient of thermal expansion in silicate optical fibers}},
  author={Cavillon, M and Dragic, PD and Ballato, J},
  journal={Optics Letters},
  volume={42},
  number={18},
  pages={3650--3653},
  year={2017},
  publisher={Optical Society of America}
}

@incollection{dragic2024low,
  title={{Low-nonlinearity optical fibers and their applications}},
  author={Dragic, Peter D and Ballato, John and Hawkins, Thomas W},
  booktitle={Specialty Optical Fibers},
  pages={303--344},
  year={2024},
  publisher={Elsevier}
}

@article{hawkins2021kilowatt,
  title={{Kilowatt power scaling of an intrinsically low Brillouin and thermo-optic Yb-doped silica fiber}},
  author={Hawkins, TW and Dragic, PD and Yu, N and Flores, A and Ballato, J},
  journal={Journal of the Optical Society of America B},
  volume={38},
  number={12},
  pages={F38--F49},
  year={2021},
  publisher={OSA}
}

@article{bertholds2002determination,
  title={{Determination of the individual strain-optic coefficients in single-mode optical fibres}},
  author={Bertholds, Axel and Dandliker, Rene},
  journal={Journal of Lightwave Technology},
  volume={6},
  number={1},
  pages={17--20},
  year={2002},
  publisher={IEEE}
}

@article{melloni1998direct,
  title={{Direct measurement of electrostriction in optical fibers}},
  author={Melloni, A and Frasca, M and Garavaglia, A and Tonini, A and Martinelli, Mario},
  journal={Optics Letters},
  volume={23},
  number={9},
  pages={691--693},
  year={1998},
  publisher={Optical Society of America}
}

@article{law2011acoustic,
  title={{Acoustic coefficients of P$_2$O$_5$-doped silica fiber: acoustic velocity, acoustic attenuation, and thermo-acoustic coefficient}},
  author={Law, Pi-Cheng and Liu, Yuh-Shiuan and Croteau, Andr{\'e} and Dragic, Peter D},
  journal={Optical Materials Express},
  volume={1},
  number={4},
  pages={686--699},
  year={2011},
  publisher={Optical Society of America}
}

@article{dragic2011brillouin,
  title={{Brillouin Gain Reduction Via B$_2$O$_3$ Doping}},
  author={Dragic, Peter D},
  journal={Journal of Lightwave Technology},
  volume={29},
  number={7},
  pages={967--973},
  year={2011},
  publisher={IEEE}
}

@article{dragic2013pockels,
  title={{Pockels’ coefficients of alumina in aluminosilicate optical fiber}},
  author={Dragic, Peter D and Ballato, John and Morris, Stephanie and Hawkins, Thomas},
  journal={Journal of the Optical Society of America B},
  volume={30},
  number={2},
  pages={244--250},
  year={2013},
  publisher={Optical Society of America}
}

@article{dragic2013brillouin,
  title={{Brillouin spectroscopy of a novel baria-doped silica glass optical fiber}},
  author={Dragic, P and Kucera, C and Furtick, J and Guerrier, J and Hawkins, T and Ballato, J},
  journal={Optics Express},
  volume={21},
  number={9},
  pages={10924--10941},
  year={2013},
  publisher={Optical Society of America}
}

@article{cavillon2016brillouin,
  title={{Brillouin properties of a novel strontium aluminosilicate glass optical fiber}},
  author={Cavillon, Maxime and Furtick, Joshua and Kucera, Courtney J and Ryan, Colin and Tuggle, Matthew and Jones, Maxwell and Hawkins, Thomas Wade and Dragic, Peter and Ballato, John},
  journal={Journal of Lightwave Technology},
  volume={34},
  number={6},
  pages={1435--1441},
  year={2016},
  publisher={OSA}
}

@article{dragic2012sapphire,
  title={{Sapphire-derived all-glass optical fibres}},
  author={Dragic, P and Hawkins, T and Foy, P and Morris, S and Ballato, J},
  journal={Nature Photonics},
  volume={6},
  number={9},
  pages={627--633},
  year={2012},
  publisher={Nature Publishing Group UK London}
}

@article{dragic2010brillouin,
  title={{Brillouin spectroscopy of YAG-derived optical fibers}},
  author={Dragic, P and Law, P-C and Ballato, J and Hawkins, T and Foy, P},
  journal={Optics Express},
  volume={18},
  number={10},
  pages={10055--10067},
  year={2010},
  publisher={Optical Society of America}
}

@article{dragic2013brillouin2,
  title={{The Brillouin gain coefficient of Yb-doped aluminosilicate glass optical fibers}},
  author={Dragic, Peter D and Ballato, John and Morris, Stephanie and Hawkins, Thomas},
  journal={Optical Materials},
  volume={35},
  number={9},
  pages={1627--1632},
  year={2013},
  publisher={Elsevier}
}

@article{yu2019alpo,
  title={{AlPO$_4$ in silica glass optical fibers: deduction of additional material properties}},
  author={Yu, Nanjie and Hawkins, Thomas W and Bui, Thao-Vien and Cavillon, Maxime and Ballato, John and Dragic, Peter D},
  journal={IEEE Photonics Journal},
  volume={11},
  number={5},
  pages={1--13},
  year={2019},
  publisher={IEEE}
}

@inproceedings{shcherbakov2013industrial,
  title={{Industrial grade 100 kW power CW fiber laser}},
  author={Shcherbakov, EA and Fomin, VV and Abramov, AA and Ferin, AA and Mochalov, DV and Gapontsev, Valentin P},
  booktitle={Advanced Solid State Lasers},
  pages={ATh4A--2},
  year={2013},
  organization={Optica Publishing Group}
}

@article{kawahito2018ultra,
  title={{Ultra high power (100 kW) fiber laser welding of steel}},
  author={Kawahito, Yousuke and Wang, Hongze and Katayama, Seiji and Sumimori, Daichi},
  journal={Optics Letters},
  volume={43},
  number={19},
  pages={4667--4670},
  year={2018},
  publisher={Optical Society of America}
}

@article{sun2022100,
  title={{100 kW ultra high power fiber laser}},
  author={Sun, Jiapo and Liu, Lie and Han, Lianghua and Zhu, Qixin and Shen, Xiang and Yang, Ke},
  journal={Optics Continuum},
  volume={1},
  number={9},
  pages={1932--1938},
  year={2022},
  publisher={Optica Publishing Group}
}

@article{bakhtari2024review,
  title={{A Review on Laser Beam Shaping Application in Laser-Powder Bed Fusion}},
  author={Bakhtari, Ahmad Reshad and Sezer, Huseyin Kursad and Canyurt, Olcay Ersel and Eren, O{\u{g}}ulcan and Shah, Mussadiq and Marimuthu, Sundar},
  journal={Advanced Engineering Materials},
  volume={26},
  number={14},
  pages={2302013},
  year={2024},
  publisher={Wiley Online Library}
}

@article{rochat2001fiber,
  title={{Fiber amplifiers for coherent space communication}},
  author={Rochat, Etienne and Dandliker, Rene and Haroud, Karim and Czichy, Reinhard H and Roth, Ulrich and Costantini, D and Holzner, Reto},
  journal={IEEE Journal of Selected Topics in Quantum Electronics},
  volume={7},
  number={1},
  pages={64--80},
  year={2001},
  publisher={IEEE}
}

@inproceedings{anthony2019fiber,
  title={{Fiber lasers and amplifiers for space Lidar applications}},
  author={Anthony, W Yu},
  booktitle={Sixth International Workshop on Specialty Optical Fibers and Their Applications (WSOF 2019)},
  volume={11206},
  pages={1120606--1},
  year={2019}
}

@article{cariou2006laser,
  title={{Laser source requirements for coherent lidars based on fiber technology}},
  author={Cariou, Jean-Pierre and Augere, B{\'e}atrice and Valla, Matthieu},
  journal={Comptes Rendus Physique},
  volume={7},
  number={2},
  pages={213--223},
  year={2006},
  publisher={Elsevier}
}

@inproceedings{canat2016high,
  title={{High peak power single-frequency MOPFA for lidar applications}},
  author={Canat, G and Aug{\`e}re, B and Besson, C and Dolfi-Bouteyre, A and Durecu, A and Goular, D and Le Gou{\"e}t, J and Lombard, L and Planchat, C and Valla, M},
  booktitle={CLEO: Applications and Technology},
  pages={AM3K--4},
  year={2016},
  organization={Optica Publishing Group}
}

@article{wang20201645,
  title={{1645 nm coherent Doppler wind lidar with a single-frequency Er: YAG laser}},
  author={Wang, KaiXin and Gao, ChunQing and Lin, ZhiFeng and Wang, Qing and Gao, MingWei and Huang, Shuai and Chen, ChaoYong},
  journal={Optics Express},
  volume={28},
  number={10},
  pages={14694--14704},
  year={2020},
  publisher={Optical Society of America}
}

@article{bayer2021single,
  title={{Single-shot ranging and velocimetry with a CW lidar far beyond the coherence length of the CW laser}},
  author={Bayer, Mustafa Mert and Li, Xun and Guentchev, George Nikolaev and Torun, Rasul and Velazco, Jose E and Boyraz, Ozdal},
  journal={Optics Express},
  volume={29},
  number={26},
  pages={42343--42354},
  year={2021},
  publisher={Optical Society of America}
}

@article{atalar20243d,
  title={{3D coherent single shot lidar imaging beyond coherence length}},
  author={Atalar, Ataberk and Margison, Christian Joseph and Bayer, Mustafa Mert and Li, Xun and Boyraz, Ozan Berk and Boyraz, Ozdal},
  journal={Optics Express},
  volume={32},
  number={23},
  pages={40783--40793},
  year={2024},
  publisher={Optica Publishing Group}
}

@article{fu2017review,
  title={{Review of recent progress on single-frequency fiber lasers}},
  author={Fu, Shijie and Shi, Wei and Feng, Yan and Zhang, Lei and Yang, Zhongmin and Xu, Shanhui and Zhu, Xiushan and Norwood, Robert A and Peyghambarian, N},
  journal={Journal of the Optical Society of America B},
  volume={34},
  number={3},
  pages={A49--A57},
  year={2017},
  publisher={OSA}
}

@article{li2023high,
  title={{High-power single-frequency fiber amplifiers: progress and challenge}},
  author={Li, Can and Tao, Yue and Jiang, Man and Ma, Pengfei and Liu, Wei and Su, Rongtao and Xu, Jiangming and Leng, Jinyong and Zhou, Pu},
  journal={Chinese Optics Letters},
  volume={21},
  number={9},
  pages={090002},
  year={2023}
}

@article{xie2020single,
  title={{A single-frequency 1064-nm Yb3+-doped fiber laser tandem-pumped at 1018 nm}},
  author={Xie, Zhaoxin and Shi, Chaodu and Sheng, Quan and Fu, Shijie and Shi, Wei and Yao, Jianquan},
  journal={Optics Communications},
  volume={461},
  pages={125262},
  year={2020},
  publisher={Elsevier}
}

@article{di2010simple,
  title={{Simple approach to the relation between laser frequency noise and laser line shape}},
  author={Di Domenico, Gianni and Schilt, St{\'e}phane and Thomann, Pierre},
  journal={Applied Optics},
  volume={49},
  number={25},
  pages={4801--4807},
  year={2010},
  publisher={Optical Society of America}
}

@article{karr2024new,
  title={{The new laser weapons}},
  author={Karr, Thomas and Trebes, James},
  journal={Physics Today},
  volume={77},
  number={1},
  pages={32--38},
  year={2024},
  publisher={AIP Publishing}
}

@inproceedings{fathi2021towards,
  title={{Towards ultimate high-power scaling: Coherent beam combining of fiber lasers}},
  author={Fathi, Hossein and N{\"a}rhi, Mikko and Gumenyuk, Regina},
  booktitle={Photonics},
  volume={8},
  number={12},
  pages={566},
  year={2021},
  organization={MDPI}
}

@article{linslal2022challenges,
  title={{Challenges in coherent beam combining of high power fiber amplifiers: a review}},
  author={Linslal, Charles Lailabai and Ayyaswamy, Padmanabhan and Maji, Satyajit and Sooraj, Mundakkolly Sureshbabu and Dixit, Awakash and Venkitesh, Deepa and Srinivasan, Balaji},
  journal={ISSS Journal of Micro and Smart Systems},
  volume={11},
  number={1},
  pages={277--293},
  year={2022},
  publisher={Springer}
}

@inproceedings{creeden2018advanced,
  title={{Advanced packaging and power scaling of narrow linewidth fiber amplifiers}},
  author={Creeden, Daniel and Underwood, Mitchell and D'Alberto, Tiffanie G and Tero, Tony and Hosmer, David and Basque, Ronald and Galipeau, Joshua and Sears, Jill and Paquette, David and Ebert, Chris},
  booktitle={Laser Technology for Defense and Security XIV},
  volume={10637},
  pages={8--14},
  year={2018},
  organization={SPIE}
}

@inproceedings{nicholson20235,
  title={{5 kW single-mode output power from Yb-doped fibers with increased higher-order mode loss}},
  author={Nicholson, Jeffrey W and Pincha, Jose and Kansal, Ishu and Windeler, Robert S and Monberg, Eric and Lukonin, Vasiliy and Hariharan, Anand and Williams, Gregory and Rosales-Garcia, Andrea and Bansal, Lalitkumar and others},
  booktitle={Fiber Lasers XX: Technology and Systems},
  volume={12400},
  pages={1240002},
  year={2023},
  organization={SPIE}
}

@inproceedings{edgecumbe2024fiber,
  title={{Fiber lasers for directed energy}},
  author={Edgecumbe, JP and Martz, DH},
  booktitle={High-Power Laser Ablation VIII},
  volume={12939},
  pages={75--80},
  year={2024},
  organization={SPIE}
}

@inproceedings{holten2024beam,
  title={{Beam-combinable high-power fiber laser sources for directed energy applications}},
  author={Holten, Roger and Flores, Angel and Ehrenreich, Thomas and Anderson, Brian and Dajani, Iyad},
  booktitle={High-Power Laser Ablation VIII},
  volume={12939},
  pages={81--88},
  year={2024},
  organization={SPIE}
}

@article{grandidier2021laser,
  title={{Laser power beaming for lunar night and permanently shadowed regions}},
  author={Grandidier, Jonathan and Jaffe, Paul and Roberts, William T and Wright, Malcolm W and Fraeman, Abigail A and Raymond, Carol A and Austin, Alex and Lubin, Philip and Sunada, Eric T and Jones, John-Paul and others},
  journal={Jet Propulsion},
  volume={818},
  pages={354--1566},
  year={2021}
}

@article{kikuchi2015fundamentals,
  title={{Fundamentals of coherent optical fiber communications}},
  author={Kikuchi, Kazuro},
  journal={Journal of lightwave technology},
  volume={34},
  number={1},
  pages={157--179},
  year={2015},
  publisher={IEEE}
}

@article{drever1983laser,
  title={{Laser phase and frequency stabilization using an optical resonator}},
  author={Drever, Ronald WP and Hall, John L and Kowalski, Frank V and Hough, James and Ford, GM and Munley, AJ and Ward, Hywel},
  journal={Applied Physics B},
  volume={31},
  number={2},
  pages={97--105},
  year={1983},
  publisher={Springer}
}

@article{salomon1988laser,
  title={{Laser stabilization at the millihertz level}},
  author={Salomon, Ch and Hils, D and Hall, JL},
  journal={Journal of the Optical Society of America B},
  volume={5},
  number={8},
  pages={1576--1587},
  year={1988},
  publisher={Optical Society of America}
}

@article{lee2013spiral,
  title={{Spiral resonators for on-chip laser frequency stabilization}},
  author={Lee, Hansuek and Suh, Myoung-Gyun and Chen, Tong and Li, Jiang and Diddams, Scott A and Vahala, Kerry J},
  journal={Nature Communications},
  volume={4},
  number={1},
  pages={2468},
  year={2013},
  publisher={Nature Publishing Group UK London}
}

@article{hirata2014sub,
  title={{Sub-hertz-linewidth diode laser stabilized to an ultralow-drift high-finesse optical cavity}},
  author={Hirata, Shu and Akatsuka, Tomoya and Ohtake, Yurie and Morinaga, Atsuo},
  journal={Applied Physics Express},
  volume={7},
  number={2},
  pages={022705},
  year={2014},
  publisher={IOP Publishing}
}

@article{liang2015ultralow,
  title={{Ultralow noise miniature external cavity semiconductor laser}},
  author={Liang, Wei and Ilchenko, VS and Eliyahu, D and Savchenkov, AA and Matsko, AB and Seidel, D and Maleki, L},
  journal={Nature Communications},
  volume={6},
  number={1},
  pages={7371},
  year={2015},
  publisher={Nature Publishing Group UK London}
}

@article{morton2018high,
  title={{High-power, ultra-low noise hybrid lasers for microwave photonics and optical sensing}},
  author={Morton, Paul A and Morton, Michael J},
  journal={Journal of Lightwave Technology},
  volume={36},
  number={21},
  pages={5048--5057},
  year={2018},
  publisher={IEEE}
}

@article{jin2021hertz,
  title={{Hertz-linewidth semiconductor lasers using CMOS-ready ultra-high-Q microresonators}},
  author={Jin, Warren and Yang, Qi-Fan and Chang, Lin and Shen, Boqiang and Wang, Heming and Leal, Mark A and Wu, Lue and Gao, Maodong and Feshali, Avi and Paniccia, Mario and others},
  journal={Nature Photonics},
  volume={15},
  number={5},
  pages={346--353},
  year={2021},
  publisher={Nature Publishing Group UK London}
}

@article{zhang2025narrow,
  title={{Narrow-linewidth self-injection locking 1 $\mu$m laser using a fiber Fabry--Perot resonator}},
  author={Zhang, Xiaofan and Jia, Kunpeng and Cheng, Shanshan and Shi, Yongwei and Zhu, Shi-Ning and Xie, Zhenda},
  journal={Journal of the Optical Society of America B},
  volume={42},
  number={5},
  pages={1160--1163},
  year={2025},
  publisher={Optica Publishing Group}
}

@article{kane1987frequency,
  title={{Frequency stability and offset locking of a laser-diode-pumped Nd: YAG monolithic nonplanar ring oscillator}},
  author={Kane, Thomas J and Nilsson, Alan C and Byer, Robert L},
  journal={Optics Letters},
  volume={12},
  number={3},
  pages={175--177},
  year={1987},
  publisher={Optical Society of America}
}

@article{loh19951,
  title={{1.55 \textmu m phase-shifted distributed feedback fibre laser}},
  author={Loh, WH and Laming, RI},
  journal={Electronics Letters},
  volume={31},
  number={17},
  pages={1440--1442},
  year={1995},
  publisher={IET}
}

@article{agger2004single,
  title={{Single-frequency thulium-doped distributed-feedback fiber laser}},
  author={Agger, S{\o}ren and Povlsen, J{\o}rn Hedegaard and Varming, Poul},
  journal={Optics Letters},
  volume={29},
  number={13},
  pages={1503--1505},
  year={2004},
  publisher={OSA}
}

@article{bernier2014all,
  title={{All-fiber DFB laser operating at 2.8 $\mu$m}},
  author={Bernier, Martin and Michaud-Belleau, Vincent and Levasseur, Simon and Fortin, Vincent and Genest, J{\'e}r{\^o}me and Vall{\'e}e, R{\'e}al},
  journal={Optics Letters},
  volume={40},
  number={1},
  pages={81--84},
  year={2015},
  publisher={Optical Society of America}
}

@article{fu2021diode,
  title={{Diode-pumped 1.15 W linearly polarized single-frequency Yb3+-doped phosphate fiber laser}},
  author={Fu, Shijie and Zhu, Xiushan and Zong, Jie and Norwood, Robert A and Peyghambarian, Nasser},
  journal={Optics Express},
  volume={29},
  number={19},
  pages={30637--30643},
  year={2021},
  publisher={Optical Society of America}
}

@article{zhang20222,
  title={{2.56 W Single-Frequency All-Fiber Oscillator at 1720 nm}},
  author={Zhang, Junxiang and Sheng, Quan and Zhang, Lu and Shi, Chaodu and Sun, Shuai and Shi, Wei and Yao, Jianquan},
  journal={Advanced Photonics Research},
  volume={3},
  number={2},
  pages={2100256},
  year={2022},
  publisher={Wiley Online Library}
}

@article{tao2022high,
  title={{High power and high efficiency single-frequency 1030 nm DFB fiber laser}},
  author={Tao, Yue and Zhang, Song and Jiang, Man and Li, Can and Zhou, Pu and Jiang, Zongfu},
  journal={Optics \& Laser Technology},
  volume={145},
  pages={107519},
  year={2022},
  publisher={Elsevier}
}

@article{li2023high1,
  title={{High-power, high-efficiency single-frequency DBR fiber laser at 1064 nm based on Yb3+-doped silica fiber}},
  author={Li, Yanyan and Deng, Xun and Fu, Shijie and Sheng, Quan and Shi, Chaodu and Zhang, Junxiang and Zhang, Lu and Shi, Wei and Yao, Jianquan},
  journal={Optics Letters},
  volume={48},
  number={3},
  pages={598--601},
  year={2023},
  publisher={Optica Publishing Group}
}

@article{panbhiharwala2018investigation,
  title={{Investigation of temporal dynamics due to stimulated Brillouin scattering using statistical correlation in a narrow-linewidth cw high power fiber amplifier}},
  author={Panbhiharwala, Yusuf and Harish, Achar V and Venkitesh, Deepa and Nilsson, Johan and Srinivasan, Balaji},
  journal={Optics Express},
  volume={26},
  number={25},
  pages={33409--33417},
  year={2018},
  publisher={Optical Society of America}
}

@article{wisal2024theorysbs,
  title={{Theory of stimulated Brillouin scattering in fibers for highly multimode excitations}},
  author={Wisal, Kabish and Warren-Smith, Stephen C and Chen, Chun-Wei and Cao, Hui and Stone, A Douglas},
  journal={Physical Review X},
  volume={14},
  number={3},
  pages={031053},
  year={2024},
  publisher={APS}
}

@article{boyd1990noise,
  title={Noise initiation of stimulated Brillouin scattering},
  author={Boyd, Robert W and Rzaewski, Kazimierz and Narum, Paul},
  journal={Physical review A},
  volume={42},
  number={9},
  pages={5514},
  year={1990},
  publisher={APS}
}

@article{gaeta1991stochastic,
  title={{Stochastic dynamics of stimulated Brillouin scattering in an optical fiber}},
  author={Gaeta, Alexander L and Boyd, Robert W},
  journal={Physical Review A},
  volume={44},
  number={5},
  pages={3205},
  year={1991},
  publisher={APS}
}

@article{strzelecki2002investigation,
  title={{Investigation of tunable single frequency diode lasers for sensor applications}},
  author={Strzelecki, EM and Cohen, DA and Coldren, LA},
  journal={Journal of Lightwave Technology},
  volume={6},
  number={10},
  pages={1610--1618},
  year={2002},
  publisher={IEEE}
}

@article{salhi2006single,
  title={{Single-frequency Sb-based distributed-feedback lasers emitting at 2.3 $\mu$m above room temperature for application in tunable diode laser absorption spectroscopy}},
  author={Salhi, Abdelmagid and Barat, David and Romanini, Daniele and Rouillard, Yves and Ouvrard, Aimeric and Werner, Ralph and Seufert, Jochen and Koeth, Johannes and Vicet, Aurore and Garnache, Arnaud},
  journal={Applied Optics},
  volume={45},
  number={20},
  pages={4957--4965},
  year={2006},
  publisher={Optical Society of America}
}

@article{jimenez2017narrow,
  title={{Narrow-line external cavity diode laser micro-packaging in the NIR and MIR spectral range}},
  author={Jim{\'e}nez, A and Milde, T and Staacke, N and Assmann, C and Carpintero, G and Sacher, J},
  journal={Applied Physics B},
  volume={123},
  number={7},
  pages={207},
  year={2017},
  publisher={Springer}
}

@article{davis1998thermal,
  title={{Thermal effects in doped fibers}},
  author={Davis, MK and Digonnet, MJF and Pantell, Richard H and Fellow, Life},
  journal={Journal of Lightwave Technology},
  volume={16},
  number={6},
  pages={1013},
  year={1998},
  publisher={OSA}
}

@article{brown2001thermal,
  title={{Thermal, stress, and thermo-optic effects in high average power double-clad silica fiber lasers}},
  author={Brown, David C and Hoffman, Hanna J},
  journal={IEEE Journal of Quantum Electronics},
  volume={37},
  number={2},
  pages={207--217},
  year={2001},
  publisher={IEEE}
}

@article{wang2004thermal,
  title={{Thermal effects in kilowatt fiber lasers}},
  author={Wang, Yong and Xu, Chang-Qing and Po, Hong},
  journal={IEEE Photonics Technology Letters},
  volume={16},
  number={1},
  pages={63--65},
  year={2004},
  publisher={IEEE}
}

@article{galvanauskas2004high,
  title={{High power fiber lasers}},
  author={Galvanauskas, Almantas},
  journal={Optics and photonics news},
  volume={15},
  number={7},
  pages={42--47},
  year={2004},
  publisher={OSA}
}

@article{li20053,
  title={{3-Dimensional thermal analysis and active cooling of short-length high-power fiber lasers}},
  author={Li, L and Li, H and Qiu, T and Temyanko, VL and Morrell, MM and Sch{\"u}lzgen, A and Mafi, A and Moloney, JV and Peyghambarian, N},
  journal={Optics Express},
  volume={13},
  number={9},
  pages={3420--3428},
  year={2005},
  publisher={Optical Society of America}
}

@inproceedings{lapointe2009thermal,
  title={{Thermal effects in high-power CW fiber lasers}},
  author={Lapointe, Marc-Andr{\'e} and Chatigny, Stephane and Pich{\'e}, Michel and Cain-Skaff, Michael and Maran, Jean-No{\"e}l},
  booktitle={Fiber lasers VI: technology, systems, and applications},
  volume={7195},
  pages={430--440},
  year={2009},
  organization={SPIE}
}

@article{stolov2009thermal,
  title={{Thermal stability of specialty optical fibers}},
  author={Stolov, Andrei A and Simoff, Debra A and Li, Jie},
  journal={Journal of Lightwave Technology},
  volume={26},
  number={20},
  pages={3443--3451},
  year={2009},
  publisher={IEEE}
}

@article{fan2011thermal,
  title={{Thermal effects in kilowatt all-fiber MOPA}},
  author={Fan, Yuanyuan and He, Bing and Zhou, Jun and Zheng, Jituo and Liu, Houkang and Wei, Yunrong and Dong, Jingxing and Lou, Qihong},
  journal={Optics Express},
  volume={19},
  number={16},
  pages={15162--15172},
  year={2011},
  publisher={Optical Society of America}
}

@article{dong2016thermal,
  title={{Thermal lensing in optical fibers}},
  author={Dong, Liang},
  journal={Optics Express},
  volume={24},
  number={17},
  pages={19841--19852},
  year={2016},
  publisher={Optical Society of America}
}

@article{daniel2016metal,
  title={{Metal clad active fibres for power scaling and thermal management at kW power levels}},
  author={Daniel, Jae MO and Simakov, Nikita and Hemming, Alexander and Clarkson, W Andrew and Haub, John},
  journal={Optics Express},
  volume={24},
  number={16},
  pages={18592--18606},
  year={2016},
  publisher={Optical Society of America}
}

@inproceedings{charles2016diode,
  title={{Diode-pumped narrow linewidth multi-kW metalized Yb fiber amplifier}},
  author={Charles, X Yu and Shatrovoy, Oleg and Fan, TY and Taunay, Thierry},
  booktitle={Advanced Solid State Lasers},
  pages={ATu6A--1},
  year={2016},
  organization={Optica Publishing Group}
}

@article{ballato2024prospects,
  title={{Prospects and challenges for all-optical thermal management of fiber lasers}},
  author={Ballato, John and Dragic, Peter D and Digonnet, Michel JF},
  journal={Journal of Physics D: Applied Physics},
  volume={57},
  number={16},
  pages={162001},
  year={2024},
  publisher={IOP Publishing}
}

@article{sinha2007investigation,
  title={{Investigation of the suitability of silicate bonding for facet termination in active fiber devices}},
  author={Sinha, Supriyo and Urbanek, Karel E and Krzywicki, Alan and Byer, Robert L},
  journal={Optics Express},
  volume={15},
  number={20},
  pages={13003--13022},
  year={2007},
  publisher={Optical Society of America}
}

@article{aydin2019endcapping,
  title={{Endcapping of high-power 3 $\mu$m fiber lasers}},
  author={Aydin, Yigit Ozan and Maes, Fr{\'e}d{\'e}ric and Fortin, Vincent and Bah, Souleymane T and Vall{\'e}e, R{\'e}al and Bernier, Martin},
  journal={Optics Express},
  volume={27},
  number={15},
  pages={20659--20669},
  year={2019},
  publisher={Optical Society of America}
}

@article{nicholson2015axicons,
  title={{Axicons for mode conversion in high peak power, higher-order mode, fiber amplifiers}},
  author={Nicholson, JW and DeSantolo, A and Westbrook, PS and Windeler, RS and Kremp, T and Headley, C and DiGiovanni, DJ},
  journal={Optics Express},
  volume={23},
  number={26},
  pages={33849--33860},
  year={2015},
  publisher={Optical Society of America}
}

@article{han2019fiber,
  title={{Fiber combiner terminated with quartz block head for direct beam combining of fiber coupled laser diodes}},
  author={Han, Lixiang and Hao, Mingming and Tao, Lili and Li, Jingbo},
  journal={Optik},
  volume={184},
  pages={35--39},
  year={2019},
  publisher={Elsevier}
}

@article{braglia2014architectures,
  title={{Architectures and components for high power CW fiber lasers}},
  author={Braglia, Andrea and Califano, Alessio and Liu, Yu and Perrone, Guido},
  journal={International Journal of Modern Physics B},
  volume={28},
  number={12},
  pages={1442001},
  year={2014},
  publisher={World Scientific}
}

@article{chen2023functional,
  title={{Functional fibers and functional fiber-based components for high-power lasers}},
  author={Chen, Xiao and Yao, Tianfu and Huang, Liangjin and An, Yi and Wu, Hanshuo and Pan, Zhiyong and Zhou, Pu},
  journal={Advanced Fiber Materials},
  volume={5},
  number={1},
  pages={59--106},
  year={2023},
  publisher={Springer}
}

@article{boyd2016co2,
  title={{CO$_2$ laser-fabricated cladding light strippers for high-power fiber lasers and amplifiers}},
  author={Boyd, Keiron and Simakov, Nikita and Hemming, Alexander and Daniel, Jae and Swain, Robert and Mies, Eric and Rees, Simon and Andrew Clarkson, W and Haub, John},
  journal={Applied Optics},
  volume={55},
  number={11},
  pages={2915--2920},
  year={2016},
  publisher={Optical Society of America}
}

@article{yan2017kilowatt,
  title={{Kilowatt-level cladding light stripper for high-power fiber laser}},
  author={Yan, Ping and Sun, Junyi and Huang, Yusheng and Li, Dan and Wang, Xuejiao and Xiao, Qirong and Gong, Mali},
  journal={Applied Optics},
  volume={56},
  number={7},
  pages={1935--1939},
  year={2017},
  publisher={Optical Society of America}
}

@article{liu20202,
  title={{> 2 kW high stability robust fiber cladding mode stripper with moderate package temperature rising}},
  author={Liu, Yu and Huang, Shan and Wu, Wenjie and Zhao, Pengfei and Tang, Xuan and Feng, Xi and Li, Min and Shen, Benjian and Song, Huaqing and Tao, Rumao and others},
  journal={IEEE Photonics Technology Letters},
  volume={32},
  number={18},
  pages={1151--1154},
  year={2020},
  publisher={IEEE}
}

@inproceedings{morasse2024efficient,
  title={{Efficient 1018nm high power fiber laser using intracavity tilted FBG ASE filters}},
  author={Morasse, Bertrand and Perron, Alexandre and Faucher, Dominic and Belzile, Pierre-Michel and Desch{\^e}nes, Mathieu Vollant and Faucher, Fr{\'e}d{\'e}ric and Brochu, Guillaume and Tr{\'e}panier, Fran{\c{c}}ois and Deladurantaye, Pascal},
  booktitle={Fiber Lasers XXI: Technology and Systems},
  volume={12865},
  pages={40--49},
  year={2024},
  organization={SPIE}
}

@article{leleux2025high,
  title={{High power neodymium-doped single-mode all-fiber laser at 922 nm using chirped tilted fiber Bragg gratings}},
  author={Leleux, Alban and Aydin, Yigit Ozan and Bisson, William and Michaud, Alexandre and Karim, Andrew and Mehaboob, Anees and Carvalho Pinto, Iago and Boessinger, Nicolas and Esclingand, Mathieu and Moriamez, Jean and others},
  journal={Optics Express},
  volume={33},
  number={22},
  pages={46485--46493},
  year={2025},
  publisher={Optica Publishing Group}
}

@article{zeng2023simultaneous,
  title={{Simultaneous achievement of power boost and low-frequency intensity noise suppression in a bidirectional pumping fiber amplifier based on saturated even-distribution gain}},
  author={Zeng, Chun and Peng, Wenkun and Zhao, Qilai and Lin, Wei and Yang, Changsheng and Sun, Yuxin and Wang, Changhe and Feng, Zhouming and Yang, Zhongmin and Xu, Shanhui},
  journal={Optics Express},
  volume={31},
  number={3},
  pages={5122--5130},
  year={2023},
  publisher={Optica Publishing Group}
}

@article{zeringue2012theoretical,
  title={{A theoretical study of transient stimulated Brillouin scattering in optical fibers seeded with phase-modulated light}},
  author={Zeringue, Clint and Dajani, Iyad and Naderi, Shadi and Moore, Gerald T and Robin, Craig},
  journal={Optics Express},
  volume={20},
  number={19},
  pages={21196--21213},
  year={2012},
  publisher={Optical Society of America}
}

@misc{williamson2007laser,
  title={{Laser coherence control using homogeneous linewidth broadening}},
  author={Williamson III, Robert S},
  year={2007},
  publisher={Google Patents},
  note={US Patent 7,280,568}
}

@inproceedings{flores2012experimental,
  title={{Experimental and theoretical studies of phase modulation in Yb-doped fiber amplifiers}},
  author={Flores, Angel and Lu, Chunte and Robin, Craig and Naderi, Shadi and Vergien, Christopher and Dajani, Iyad},
  booktitle={Laser Technology for Defense and Security VIII},
  volume={8381},
  pages={271--278},
  year={2012},
  organization={SPIE}
}

@article{anderson2015comparison,
  title={{Comparison of phase modulation schemes for coherently combined fiber amplifiers}},
  author={Anderson, Brian and Flores, Angel and Holten, Roger and Dajani, Iyad},
  journal={Optics Express},
  volume={23},
  number={21},
  pages={27046--27060},
  year={2015},
  publisher={Optical Society of America}
}

@article{flores2014pseudo,
  title={{Pseudo-random binary sequence phase modulation for narrow linewidth, kilowatt, monolithic fiber amplifiers}},
  author={Flores, Angel and Robin, Craig and Lanari, Ann and Dajani, Iyad},
  journal={Optics Express},
  volume={22},
  number={15},
  pages={17735--17744},
  year={2014},
  publisher={Optical Society of America}
}

@article{montoya2017photonic,
  title={{Photonic lantern kW-class fiber amplifier}},
  author={Montoya, Juan and Hwang, Christopher and Martz, Dale and Aleshire, Christopher and Fan, TY and Ripin, Daniel J},
  journal={Optics Express},
  volume={25},
  number={22},
  pages={27543--27550},
  year={2017},
  publisher={Optical Society of America}
}

@article{liem2003100,
  title={{100-W single-frequency master-oscillator fiber power amplifier}},
  author={Liem, A and Limpert, J and Zellmer, H and T{\"u}nnermann, A},
  journal={Optics Letters},
  volume={28},
  number={17},
  pages={1537--1539},
  year={2003},
  publisher={Optical Society of America}
}

@article{jeong2005single,
  title={{Single-frequency, single-mode, plane-polarized ytterbium-doped fiber master oscillator power amplifier source with 264 W of output power}},
  author={Jeong, Y and Nilsson, J and Sahu, JK and Soh, DBS and Alegria, C and Dupriez, P and Codemard, CA and Payne, DN and Horley, R and Hickey, LMB and others},
  journal={Optics Letters},
  volume={30},
  number={5},
  pages={459--461},
  year={2005},
  publisher={Optical Society of America}
}

@inproceedings{mermelstein2007all,
  title={{All-fiber 194 W single-frequency single-mode Yb-doped master-oscillator power-amplifier}},
  author={Mermelstein, MD and Brar, K and Andrejco, MJ and Yablon, AD and Fishteyn, M and Headley, C and DiGiovanni, DJ},
  booktitle={LEOS 2007-IEEE Lasers and Electro-Optics Society Annual Meeting Conference Proceedings},
  pages={382--383},
  year={2007},
  organization={IEEE}
}

@article{wang2012310,
  title={{310 W single-frequency all-fiber laser in master oscillator power amplification configuration}},
  author={Wang, XL and Zhou, P and Xiao, H and Ma, YX and Xu, XJ and Liu, ZJ},
  journal={Laser Physics Letters},
  volume={9},
  number={8},
  pages={591},
  year={2012},
  publisher={IOP Publishing}
}

@article{ma2013single,
  title={{Single-frequency 332 W, linearly polarized Yb-doped all-fiber amplifier with near diffraction-limited beam quality}},
  author={Ma, Pengfei and Zhou, Pu and Ma, Yanxing and Su, Rongtao and Xu, Xiaojun and Liu, Zejin},
  journal={Applied Optics},
  volume={52},
  number={20},
  pages={4854--4857},
  year={2013},
  publisher={Optical Society of America}
}

@article{dixneuf2020ultra,
  title={{Ultra-low intensity noise, all fiber 365 W linearly polarized single frequency laser at 1064 nm}},
  author={Dixneuf, Cl{\'e}ment and Guiraud, Germain and Bardin, Yves-Vincent and Rosa, Quentin and Goeppner, Mathieu and Hilico, Ad{\`e}le and Pierre, Christophe and Boullet, Johan and Traynor, Nicholas and Santarelli, Giorgio},
  journal={Optics Express},
  volume={28},
  number={8},
  pages={10960--10969},
  year={2020},
  publisher={Optical Society of America}
}

@inproceedings{zhu2011single,
  title={{Single-frequency and single-transverse mode Yb-doped CCC fiber MOPA with robust polarization SBS-free 511W output}},
  author={Zhu, Cheng and Hu, I-Ning and Ma, Xiuquan and Galvanauskas, A},
  booktitle={Advanced Solid-State Photonics},
  pages={AMC5},
  year={2011},
  organization={Optica Publishing Group}
}

@article{hochheim2020single,
  title={{Single-frequency chirally coupled-core all-fiber amplifier with 100 W in a linearly polarized TEM00 mode}},
  author={Hochheim, Sven and Steinke, Michael and Wessels, Peter and De Varona, Omar and Koponen, Joona and Lowder, Tyson and Novotny, Steffen and Neumann, J{\"o}rg and Kracht, Dietmar},
  journal={Optics Letters},
  volume={45},
  number={4},
  pages={939--942},
  year={2020},
  publisher={Optical Society of America}
}

@article{hochheim2021single,
  title={{Single-frequency 336 W spliceless all-fiber amplifier based on a chirally-coupled-core fiber for the next generation of gravitational wave detectors}},
  author={Hochheim, Sven and Brockm{\"u}ller, Eike and Wessels, Peter and Koponen, Joona and Lowder, Tyson and Novotny, Steffen and Willke, Benno and Neumann, J{\"o}rg and Kracht, Dietmar},
  journal={Journal of Lightwave Technology},
  volume={40},
  number={7},
  pages={2136--2143},
  year={2021},
  publisher={IEEE}
}

@article{li2025functional,
  title={{Functional Yb-doped fiber with a bat-type refractive index distribution for beyond kilowatt all-fiber single-frequency laser amplification}},
  author={Li, Wei and Liu, Wei and Deng, Yu and Chen, Yisha and Yang, Huan and Chen, Qi and Zheng, Junjie and Xiao, Hu and Chen, Zilun and Pan, Zhiyong and Ma, Pengfei and Wang, Zefeng and Si, Lei and Xu, Shanhui and Chen, Jinbao},
  journal={Light: Science \& Applications},
  volume={14},
  number={1},
  pages={271},
  year={2025},
  publisher={Nature Publishing Group UK London}
}

@article{robin2011acoustically,
  title={{Acoustically segmented photonic crystal fiber for single-frequency high-power laser applications}},
  author={Robin, Craig and Dajani, Iyad},
  journal={Optics Letters},
  volume={36},
  number={14},
  pages={2641--2643},
  year={2011},
  publisher={Optical Society of America}
}

@article{robin2014modal,
  title={{Modal instability-suppressing, single-frequency photonic crystal fiber amplifier with 811 W output power}},
  author={Robin, Craig and Dajani, Iyad and Pulford, Benjamin},
  journal={Optics Letters},
  volume={39},
  number={3},
  pages={666--669},
  year={2014},
  publisher={Optical Society of America}
}

@article{pulford2015400,
  title={{400-W near diffraction-limited single-frequency all-solid photonic bandgap fiber amplifier}},
  author={Pulford, Benjamin and Ehrenreich, Thomas and Holten, Roger and Kong, Fanting and Hawkins, Thomas W and Dong, Liang and Dajani, Iyad},
  journal={Optics Letters},
  volume={40},
  number={10},
  pages={2297--2300},
  year={2015},
  publisher={Optical Society of America}
}

@article{matniyaz2022high,
  title={{High-power single-frequency single-mode all-solid photonic bandgap fiber laser with kHz linewidth}},
  author={Matniyaz, Turghun and Bingham, Samuel P and Kalichevsky-Dong, Monica T and Hawkins, Thomas W and Pulford, Benjamin and Dong, Liang},
  journal={Optics Letters},
  volume={47},
  number={2},
  pages={377--380},
  year={2022},
  publisher={Optica Publishing Group}
}

@article{stutzki2014designing,
  title={{Designing advanced very-large-mode-area fibers for power scaling of fiber-laser systems}},
  author={Stutzki, Fabian and Jansen, Florian and Otto, Hans-J{\"u}rgen and Jauregui, Cesar and Limpert, Jens and T{\"u}nnermann, Andreas},
  journal={Optica},
  volume={1},
  number={4},
  pages={233--242},
  year={2014},
  publisher={Optical Society of America}
}

@article{sousa1999multimode,
  title={{Multimode Er-doped fiber for single-transverse-mode amplification}},
  author={Sousa, JM and Okhotnikov, OG},
  journal={Applied Physics Letters},
  volume={74},
  number={11},
  pages={1528--1530},
  year={1999},
  publisher={American Institute of Physics}
}

@inproceedings{cooper2022confined,
  title={{Confined doping LMA fibers for high power single frequency lasers}},
  author={Cooper, Matthew A and Gausmann, Stefan and Antonio-Lopez, Jose E and Sch{\"u}lzgen, Axel and Correa, Rodrigo Amezcua},
  booktitle={Fiber Lasers XIX: Technology and Systems},
  volume={11981},
  pages={23--29},
  year={2022},
  organization={SPIE}
}

@article{li2022confined,
  title={{Confined-doped active fiber enabled all-fiber high-power single-frequency laser}},
  author={Li, Wei and Yan, Zhiping and Ren, Shuai and Deng, Yu and Chen, Yisha and Ma, Pengfei and Liu, Wei and Huang, Liangjin and Pan, Zhiyong and Zhou, Pu and others},
  journal={Optics Letters},
  volume={47},
  number={19},
  pages={5024--5027},
  year={2022},
  publisher={Optica Publishing Group}
}

@inproceedings{kruska2024high,
  title={{High-power single-frequency depressed-cladding, confined-doping Yb3+ fiber amplifier}},
  author={Kruska, Kristopher and Booker, Phillip and We{\ss}els, Peter and Neumann, J{\"o}rg and Kracht, Dietmar},
  booktitle={Fiber Lasers XXI: Technology and Systems},
  volume={12865},
  pages={185--190},
  year={2024},
  organization={SPIE}
}

@article{eidam2011preferential,
  title={{Preferential gain photonic-crystal fiber for mode stabilization at high average powers}},
  author={Eidam, Tino and H{\"a}drich, Steffen and Jansen, Florian and Stutzki, Fabian and Rothhardt, Jan and Carstens, Henning and Jauregui, Cesar and Limpert, Jens and T{\"u}nnermann, Andreas},
  journal={Optics Express},
  volume={19},
  number={9},
  pages={8656--8661},
  year={2011},
  publisher={Optical Society of America}
}

@article{li2022investigation,
  title={{Investigation of the confined-doped fiber on single-mode operating and power scaling in all-fiber single-frequency amplifiers}},
  author={Li, Wei and Ren, Shuai and Deng, Yu and Chen, Yisha and Lu, Yao and Liu, Wei and Ma, Pengfei and Pan, Zhiyong and Chen, Zilun and Si, Lei and others},
  journal={Frontiers in Physics},
  volume={10},
  pages={1016047},
  year={2022},
  publisher={Frontiers Media SA}
}

@article{horiguchi2002development,
  title={{Development of a distributed sensing technique using Brillouin scattering}},
  author={Horiguchi, Tsuneo and Shimizu, Kaoru and Kurashima, Toshio and Tateda, Mitsuhiro and Koyamada, Yahei},
  journal={Journal of Lightwave Technology},
  volume={13},
  number={7},
  pages={1296--1302},
  year={2002},
  publisher={IEEE}
}

@article{yoshizawa1993stimulated,
  title={{Stimulated Brillouin scattering suppression by means of applying strain distribution to fiber with cabling}},
  author={Yoshizawa, Nobuyuki and Imai, Takeshi},
  journal={Journal of Lightwave Technology},
  volume={11},
  number={10},
  pages={1518--1522},
  year={1993},
  publisher={IEEE}
}

@article{boggio2005experimental,
  title={{Experimental and numerical investigation of the SBS-threshold increase in an optical fiber by applying strain distributions}},
  author={Boggio, JM Chavez and Marconi, JD and Fragnito, HL},
  journal={Journal of Lightwave Technology},
  volume={23},
  number={11},
  pages={3808--3814},
  year={2005},
  publisher={IEEE}
}

@article{huang2016414,
  title={{414 W near-diffraction-limited all-fiberized single-frequency polarization-maintained fiber amplifier}},
  author={Huang, Long and Wu, Hanshuo and Li, Ruixian and Li, Lei and Ma, Pengfei and Wang, Xiaolin and Leng, Jinyong and Zhou, Pu},
  journal={Optics Letters},
  volume={42},
  number={1},
  pages={1--4},
  year={2016},
  publisher={Optical Society of America}
}

@article{zhang2013170,
  title={{170 W, single-frequency, single-mode, linearly-polarized, Yb-doped all-fiber amplifier}},
  author={Zhang, Lei and Cui, Shuzhen and Liu, Chi and Zhou, Jun and Feng, Yan},
  journal={Optics Express},
  volume={21},
  number={5},
  pages={5456--5462},
  year={2013},
  publisher={Optical Society of America}
}

@phdthesis{balliu2022power,
  title={{Power Scaling of Highly Compact Single-Frequency Yb-Doped Fiber Amplifiers}},
  author={Balliu, Enkeleda},
  year={2022},
  school={Mid Sweden University}
}

@article{hansen2011thermo,
  title={{Thermo-optical effects in high-power ytterbium-doped fiber amplifiers}},
  author={Hansen, Kristian Rymann and Alkeskjold, Thomas Tanggaard and Broeng, Jes and L{\ae}gsgaard, Jesper},
  journal={Optics Express},
  volume={19},
  number={24},
  pages={23965--23980},
  year={2011},
  publisher={Optical Society of America}
}

@article{imai1993dependence,
  title={{Dependence of stimulated Brillouin scattering on temperature distribution in polarization-maintaining fibers}},
  author={Imai, Yoh and Shimada, Noriaki},
  journal={IEEE photonics technology letters},
  volume={5},
  number={11},
  pages={1335--1337},
  year={1993},
  publisher={IEEE}
}

@article{hansryd2001increase,
  title={{Increase of the SBS threshold in a short highly nonlinear fiber by applying a temperature distribution}},
  author={Hansryd, J and Dross, F and Westlund, M and Andrekson, PA and Knudsen, SN},
  journal={Journal of Lightwave Technology},
  volume={19},
  number={11},
  pages={1691},
  year={2001},
  publisher={OSA}
}

@article{jeong2007power,
  title={{Power scaling of single-frequency ytterbium-doped fiber master-oscillator power-amplifier sources up to 500 W}},
  author={Jeong, Yoonchan and Nilsson, Johan and Sahu, Jayanta K and Payne, David N and Horley, R and Hickey, LMB and Turner, PW},
  journal={IEEE Journal of Selected Topics in Quantum Electronics},
  volume={13},
  number={3},
  pages={546--551},
  year={2007},
  publisher={IEEE}
}

@article{theeg2012all,
  title={{All-fiber counter-propagation pumped single frequency amplifier stage with 300-W output power}},
  author={Theeg, Thomas and Sayinc, Hakan and Neumann, J{\"o}rg and Kracht, Dietmar},
  journal={IEEE Photonics Technology Letters},
  volume={24},
  number={20},
  pages={1864--1867},
  year={2012},
  publisher={IEEE}
}

@article{zeringue2011pump,
  title={{Pump-limited, 203 W, single-frequency monolithic fiber amplifier based on laser gain competition}},
  author={Zeringue, Clint and Vergien, Christopher and Dajani, Iyad},
  journal={Optics Letters},
  volume={36},
  number={5},
  pages={618--620},
  year={2011},
  publisher={Optical Society of America}
}

@article{dajani2010stimulated,
  title={{Stimulated Brillouin scattering suppression through laser gain competition: scalability to high power}},
  author={Dajani, Iyad and Zeringue, Clint and Lu, Chunte and Vergien, Christopher and Henry, Leanne and Robin, Craig},
  journal={Optics Letters},
  volume={35},
  number={18},
  pages={3114--3116},
  year={2010},
  publisher={Optical Society of America}
}

@article{dajani2009investigation,
  title={{Investigation of nonlinear effects in multitone-driven narrow-linewidth high-power amplifiers}},
  author={Dajani, Iyad and Zeringue, Clint and Shay, Thomas M},
  journal={IEEE Journal of Selected Topics in Quantum Electronics},
  volume={15},
  number={2},
  pages={406--414},
  year={2009},
  publisher={IEEE}
}

@article{dajani2009experimental,
  title={{Experimental and theoretical investigations of photonic crystal fiber amplifier with 260 W output}},
  author={Dajani, Iyad and Vergien, Christopher and Robin, Craig and Zeringue, Clint},
  journal={Optics Express},
  volume={17},
  number={26},
  pages={24317--24333},
  year={2009},
  publisher={OSA}
}

@article{theeg2015core,
  title={{Core-pumped single-frequency fiber amplifier with an output power of 158 W}},
  author={Theeg, Thomas and Ottenhues, Christoph and Sayinc, Hakan and Neumann, J{\"o}rg and Kracht, Dietmar},
  journal={Optics Letters},
  volume={41},
  number={1},
  pages={9--12},
  year={2015},
  publisher={Optical Society of America}
}

@article{wellmann2019high,
  title={{High power, single-frequency, monolithic fiber amplifier for the next generation of gravitational wave detectors}},
  author={Wellmann, Felix and Steinke, Michael and Meylahn, Fabian and Bode, Nina and Willke, Benno and Overmeyer, Ludger and Neumann, J{\"o}rg and Kracht, Dietmar},
  journal={Optics Express},
  volume={27},
  number={20},
  pages={28523--28533},
  year={2019},
  publisher={Optical Society of America}
}

@article{gray2007502,
  title={{502 Watt, single transverse mode, narrow linewidth, bidirectionally pumped Yb-doped fiber amplifier}},
  author={Gray, Stuart and Liu, Anping and Walton, Donnell T and Wang, Ji and Li, Ming-Jun and Chen, Xin and Ruffin, A Boh and DeMeritt, Jeffrey A and Zenteno, Luis A},
  journal={Optics Express},
  volume={15},
  number={25},
  pages={17044--17050},
  year={2007},
  publisher={Optical Society of America}
}

@article{shi2022435,
  title={{435 W single-frequency all-fiber amplifier at 1064 nm based on cascaded hybrid active fibers}},
  author={Shi, Chaodu and Fu, Shijie and Deng, Xun and Sheng, Quan and Xu, Yang and Fang, Qiang and Sun, Shuai and Zhang, Junxiang and Shi, Wei and Yao, Jianquan},
  journal={Optics Communications},
  volume={502},
  pages={127428},
  year={2022},
  publisher={Elsevier}
}

@article{shi2022700,
  title={{700 W single-frequency all-fiber amplifier at 1064 nm with kHz-level spectral linewidth}},
  author={Shi, Chaodu and Deng, Xun and Fu, Shijie and Sheng, Quan and Jiang, Peiheng and Shi, Zheng and Li, Yanyan and Shi, Wei and Yao, Jianquan},
  journal={Frontiers in Physics},
  volume={10},
  pages={982900},
  year={2022},
  publisher={Frontiers Media SA}
}

@article{shi2022high,
  title={{High-energy single-frequency pulsed fiber MOPA at 1064 nm based on a hybrid active-fiber}},
  author={Shi, Chaodu and Tian, Hao and Fu, Shijie and Sheng, Quan and Shi, Zheng and Jiang, Peiheng and Shi, Wei and Yao, Jianquan},
  journal={Optics Express},
  volume={30},
  number={9},
  pages={15575--15582},
  year={2022},
  publisher={Optica Publishing Group}
}

@article{ren2025425,
  title={{425-W kilohertz-linewidth single-frequency Tm-fiber MOPA enabled by distributed cladding-pumping}},
  author={Ren, Chuanyong and Hu, Yongxiang and Song, Yi and Wang, Haotian and Guo, Jun and Wang, Fei and Shen, Deyuan},
  journal={Optics Express},
  volume={33},
  number={17},
  pages={36358--36367},
  year={2025},
  publisher={Optica Publishing Group}
}

@inproceedings{kruska2025power,
  title={{Power scaling of single-frequency Yb$^{3+}$ fiber amplifiers with highly absorbing standard LMA fibers}},
  author={Kruska, Kristopher and We{\ss}els, Peter and Neumann, J{\"o}rg and Kracht, Dietmar},
  booktitle={Fiber Lasers XXII: Technology and Systems},
  volume={13342},
  pages={1334214},
  year={2025},
  organization={SPIE}
}

@article{shiraki1995suppression,
  title={{Suppression of stimulated Brillouin scattering in a fibre by changing the core radius}},
  author={Shiraki, Kazuyuki and Ohashi, M and Tateda, M},
  journal={Electronics letters},
  volume={31},
  number={8},
  pages={668--669},
  year={1995},
  publisher={IET}
}

@article{trikshev2013160,
  title={{A 160 W single-frequency laser based on an active tapered double-clad fiber amplifier}},
  author={Trikshev, AI and Kurkov, AS and Tsvetkov, VB and Filatova, SA and Kertulla, J and Filippov, Valery and Chamorovskiy, Yu K and Okhotnikov, Oleg G},
  journal={Laser Physics Letters},
  volume={10},
  number={6},
  pages={065101},
  year={2013},
  publisher={IOP Publishing}
}

@inproceedings{pierre2018200,
  title={{200-W single frequency laser based on short active double clad tapered fiber}},
  author={Pierre, Christophe and Guiraud, Germain and Yehouessi, Jean-Paul and Santarelli, Giorgio and Boullet, Johan and Traynor, Nicholas and VINCONT, Cyril},
  booktitle={Fiber Lasers XV: Technology and Systems},
  volume={10512},
  pages={403--403},
  year={2018},
  organization={SPIE}
}

@article{lai2020550,
  title={{550 W single frequency fiber amplifiers emitting at 1030 nm based on a tapered Yb-doped fiber}},
  author={Lai, Wenchang and Ma, Pengfei and Liu, Wei and Huang, Long and Li, Can and Ma, Yanxing and Zhou, Pu},
  journal={Optics Express},
  volume={28},
  number={14},
  pages={20908--20919},
  year={2020},
  publisher={Optical Society of America}
}

@inproceedings{jiang2022650,
  title={{650 W all-fiber single-frequency polarization-maintaining fiber amplifier based on hybrid wavelength pumping and tapered Yb-doped fibers}},
  author={Jiang, Wanpeng and Yang, Changsheng and Zhao, Qilai and Gu, Quan and Huang, Jiamin and Jiang, Kui and Zhou, Kaijun and Feng, Zhouming and Yang, Zhongmin and Xu, Shanhui},
  booktitle={Photonics},
  volume={9},
  number={8},
  pages={518},
  year={2022},
  organization={MDPI}
}

@article{jauregui2026recent,
  title={{Recent developments in the understanding and passive mitigation of transverse mode instability}},
  author={Jauregui, C{\'e}sar and Tu, Yiming and Kholaif, Sobhy and M{\"o}ller, Friedrich and Palma-Vega, Gonzalo and Haarlammert, Nicoletta and Walbaum, Till and Schreiber, Thomas and Limpert, Jens},
  journal={Optical Fiber Technology},
  volume={96},
  pages={104496},
  year={2026},
  publisher={Elsevier}
}

@article{jauregui2020transverse,
  title={{Transverse mode instability}},
  author={Jauregui, Cesar and Stihler, Christoph and Limpert, Jens},
  journal={Advances in Optics and Photonics},
  volume={12},
  number={2},
  pages={429--484},
  year={2020},
  publisher={Optical Society of America}
}

@article{hansen2013theoretical,
  title={{Theoretical analysis of mode instability in high-power fiber amplifiers}},
  author={Hansen, Kristian Rymann and Alkeskjold, Thomas Tanggaard and Broeng, Jes and L{\ae}gsgaard, Jesper},
  journal={Optics Express},
  volume={21},
  number={2},
  pages={1944--1971},
  year={2013},
  publisher={Optical Society of America}
}

@article{stihler2020intensity,
  title={{Intensity noise as a driver for transverse mode instability in fiber amplifiers}},
  author={Stihler, Christoph and Jauregui, Cesar and Kholaif, Sobhy E and Limpert, Jens},
  journal={PhotoniX},
  volume={1},
  number={1},
  pages={8},
  year={2020},
  publisher={Springer}
}

@article{otto2012temporal,
  title={{Temporal dynamics of mode instabilities in high-power fiber lasers and amplifiers}},
  author={Otto, Hans-J{\"u}rgen and Stutzki, Fabian and Jansen, Florian and Eidam, Tino and Jauregui, Cesar and Limpert, Jens and T{\"u}nnermann, Andreas},
  journal={Optics Express},
  volume={20},
  number={14},
  pages={15710--15722},
  year={2012},
  publisher={Optical Society of America}
}

@article{dong2023transverse,
  title={{Transverse mode instability considering bend loss and heat load}},
  author={Dong, Liang},
  journal={Optics Express},
  volume={31},
  number={12},
  pages={20480--20488},
  year={2023},
  publisher={Optica Publishing Group}
}

@article{ward2012origin,
  title={{Origin of thermal modal instabilities in large mode area fiber amplifiers}},
  author={Ward, B and Robin, C and Dajani, I},
  journal={Optics Express},
  volume={20},
  number={10},
  pages={11407--11422},
  year={2012},
  publisher={Optical Society of America}
}

@article{dong2013stimulated,
  title={Stimulated thermal Rayleigh scattering in optical fibers},
  author={Dong, Liang},
  journal={Optics Express},
  volume={21},
  number={3},
  pages={2642--2656},
  year={2013},
  publisher={Optical Society of America}
}

@article{kong2016direct,
  title={{Direct experimental observation of stimulated thermal Rayleigh scattering with polarization modes in a fiber amplifier}},
  author={Kong, Fanting and Xue, Junwen and Stolen, Roger H and Dong, Liang},
  journal={Optica},
  volume={3},
  number={9},
  pages={975--978},
  year={2016},
  publisher={Optical Society of America}
}

@article{kholaif2025influence,
  title={{Influence of core size on the transverse mode instability threshold of fiber amplifiers}},
  author={Kholaif, Sobhy and Jauregui, Cesar and Nold, Johannes and Haarlammert, Nicoletta and Kuhn, Stefan and Schreiber, Thomas and Limpert, Jens},
  journal={Optics Express},
  volume={33},
  number={26},
  pages={54245--54256},
  year={2025},
  publisher={Optica Publishing Group}
}

@article{eidam2011experimental,
  title={{Experimental observations of the threshold-like onset of mode instabilities in high power fiber amplifiers}},
  author={Eidam, Tino and Wirth, Christian and Jauregui, Cesar and Stutzki, Fabian and Jansen, Florian and Otto, Hans-J{\"u}rgen and Schmidt, Oliver and Schreiber, Thomas and Limpert, Jens and T{\"u}nnermann, Andreas},
  journal={Optics Express},
  volume={19},
  number={14},
  pages={13218--13224},
  year={2011},
  publisher={Optical Society of America}
}

@article{johansen2013frequency,
  title={{Frequency resolved transverse mode instability in rod fiber amplifiers}},
  author={Johansen, Mette Marie and Laurila, Marko and Maack, Martin D and Noordegraaf, Danny and Jakobsen, Christian and Alkeskjold, Thomas Tanggaard and L{\ae}gsgaard, Jesper},
  journal={Optics Express},
  volume={21},
  number={19},
  pages={21847--21856},
  year={2013},
  publisher={Optical Society of America}
}

@article{christensen2020experimental,
  title={{Experimental investigations of seeding mechanisms of TMI in rod fiber amplifier using spatially and temporally resolved imaging}},
  author={Christensen, Simon L and Johansen, Mette M and Michieletto, Mattia and Triches, Marco and Maack, Martin D and L{\ae}gsgaard, Jesper},
  journal={Optics Express},
  volume={28},
  number={18},
  pages={26690--26705},
  year={2020},
  publisher={Optical Society of America}
}

@inproceedings{tao2024experimental,
  title={{Experimental Study on Transverse Mode Instability of All-Fiber Single-Frequency Amplifier Based on Tapered Yb-Doped Fiber}},
  author={Tao, Yue and Mo, Zhengfei and Kang, Pengrui and Jiang, Man and Li, Can and Leng, Jinyong and Zhou, Pu and Jiang, Zongfu},
  booktitle={Photonics},
  volume={11},
  number={8},
  pages={696},
  year={2024},
  organization={MDPI}
}

@article{dong2007leakage,
  title={{Leakage channel optical fibers with large effective area}},
  author={Dong, Liang and Peng, Xiang and Li, Jun},
  journal={Journal of the Optical Society of America B},
  volume={24},
  number={8},
  pages={1689--1697},
  year={2007},
  publisher={Optical Society of America}
}

@article{limpert2012yb,
  title={{Yb-doped large-pitch fibres: effective single-mode operation based on higher-order mode delocalisation}},
  author={Limpert, Jens and Stutzki, Fabian and Jansen, Florian and Otto, Hans-J{\"u}rgen and Eidam, Tino and Jauregui, Cesar and T{\"u}nnermann, Andreas},
  journal={Light: Science \& Applications},
  volume={1},
  number={4},
  pages={e8--e8},
  year={2012},
  publisher={Nature Publishing Group}
}

@article{cheng2024design,
  title={{Design and fabrication of all-solid anti-resonant silicate fibers for Yb ASE suppression in Er/Yb fiber amplifier}},
  author={Cheng, Yue and Yang, Qiubai and Zhu, Yiming and Wu, Dakun and Yu, Chunlei and Sun, Yan and Chen, Yichong and Zhou, Qinling and Wang, Xin and Yu, Fei and others},
  journal={Optics Express},
  volume={32},
  number={19},
  pages={33962--33973},
  year={2024},
  publisher={Optica Publishing Group}
}

@article{brown2002thermal,
  title={{Thermal, stress, and thermo-optic effects in high average power double-clad silica fiber lasers}},
  author={Brown, David C and Hoffman, Hanna J},
  journal={IEEE Journal of Quantum Electronics},
  volume={37},
  number={2},
  pages={207--217},
  year={2002},
  publisher={IEEE}
}

@article{laurila2012distributed,
  title={{Distributed mode filtering rod fiber amplifier delivering 292W with improved mode stability}},
  author={Laurila, Marko and J{\o}rgensen, Mette M and Hansen, Kristian R and Alkeskjold, Thomas T and Broeng, Jes and L{\ae}gsgaard, Jesper},
  journal={Optics Express},
  volume={20},
  number={5},
  pages={5742--5753},
  year={2012},
  publisher={Optical Society of America}
}

@article{Jansen:13,
author = {Florian Jansen and Fabian Stutzki and Hans-J\"{u}rgen Otto and Cesar Jauregui and Jens Limpert and Andreas T\"{u}nnermann},
journal = {Opt. Lett.},
number = {4},
pages = {510--512},
publisher = {Optica Publishing Group},
title = {{High-power thermally guiding index-antiguiding-core fibers}},
volume = {38},
month = {Feb},
year = {2013},
}

@article{tao2016suppressing,
  title={{Suppressing mode instabilities by optimizing the fiber coiling methods}},
  author={Tao, Rumao and Su, Rongtao and Ma, Pengfei and Wang, Xiaolin and Zhou, Pu},
  journal={Laser Physics Letters},
  volume={14},
  number={2},
  pages={025101},
  year={2016},
  publisher={IOP Publishing}
}

@article{koplow2000single,
  title={{Single-mode operation of a coiled multimode fiber amplifier}},
  author={Koplow, Jeffrey P and Kliner, Dahv AV and Goldberg, Lew},
  journal={Optics Letters},
  volume={25},
  number={7},
  pages={442--444},
  year={2000},
  publisher={Optical Society of America}
}

@article{otto2013controlling,
  title={{Controlling mode instabilities by dynamic mode excitation with an acousto-optic deflector}},
  author={Otto, Hans-J{\"u}rgen and Jauregui, Cesar and Stutzki, Fabian and Jansen, Florian and Limpert, Jens and T{\"u}nnermann, Andreas},
  journal={Optics Express},
  volume={21},
  number={14},
  pages={17285--17298},
  year={2013},
  publisher={Optical Society of America}
}

@article{jauregui2018pump,
  title={{Pump-modulation-induced beam stabilization in high-power fiber laser systems above the mode instability threshold}},
  author={Jauregui, Cesar and Stihler, Christoph and T{\"u}nnermann, Andreas and Limpert, Jens},
  journal={Optics Express},
  volume={26},
  number={8},
  pages={10691--10704},
  year={2018},
  publisher={Optical Society of America}
}

@inproceedings{kim2012single,
  title={{Single crystal fibers for high power lasers}},
  author={Kim, W and Florea, C and Baker, C and Gibson, D and Shaw, LB and Bowman, S and O'Connor, S and Villalobos, G and Bayya, S and Aggarwal, ID and others},
  booktitle={High-Power Lasers 2012: Technology and Systems},
  volume={8547},
  pages={123--128},
  year={2012},
  organization={SPIE}
}

@article{dubinskii2018low,
  title={{Low-loss ‘crystalline-core/crystalline-clad’(C4) fibers for highly power scalable high efficiency fiber lasers}},
  author={Dubinskii, Mark and Zhang, Jun and Fromzel, Viktor and Chen, Youming and Yin, Stuart and Luo, Claire},
  journal={Optics Express},
  volume={26},
  number={4},
  pages={5092--5101},
  year={2018},
  publisher={OSA}
}

@article{dong2023power,
  title={{Power scaling limits of diffraction-limited fiber amplifiers considering transverse mode instability}},
  author={Dong, Liang and Ballato, John and Kolis, Joseph},
  journal={Optics Express},
  volume={31},
  number={4},
  pages={6690--6703},
  year={2023},
  publisher={Optica Publishing Group}
}

@article{zervas2019transverse,
  title={{Transverse mode instability, thermal lensing and power scaling in Yb$^{3+}$-doped high-power fiber amplifiers}},
  author={Zervas, Michalis N},
  journal={Optics Express},
  volume={27},
  number={13},
  pages={19019--19041},
  year={2019},
  publisher={Optical Society of America}
}

@article{young2022tradeoff,
  title={{Tradeoff between the Brillouin and transverse mode instabilities in Yb-doped fiber amplifiers}},
  author={Young, J\_T and Goers, A\_J and Brown, D\_M and Dennis, M\_L and Lehr, K and Wei, C and Menyuk, C\_R and Hu, J},
  journal={Optics Express},
  volume={30},
  number={22},
  pages={40691--40703},
  year={2022},
  publisher={Optica Publishing Group}
}

@article{lee2015transverse,
  title={{Transverse mode instability induced by stimulated Brillouin scattering in a pulsed single-frequency large-core fiber amplifier}},
  author={Lee, Kyung-Hyun and Lee, Kangin and Kim, Yonghee and Cha, Yong-Ho and Lim, Gwon and Park, Hyunmin and Cho, Hyuck and Jeong, Do-Young},
  journal={Applied Optics},
  volume={54},
  number={2},
  pages={189--194},
  year={2015},
  publisher={Optical Society of America}
}

@article{smith2013increasing,
  title={{Increasing mode instability thresholds of fiber amplifiers by gain saturation}},
  author={Smith, Arlee V and Smith, Jesse J},
  journal={Optics Express},
  volume={21},
  number={13},
  pages={15168--15182},
  year={2013},
  publisher={Optical Society of America}
}

@article{hansen2014impact,
  title={{Impact of gain saturation on the mode instability threshold in high-power fiber amplifiers}},
  author={Hansen, Kristian Rymann and L{\ae}gsgaard, Jesper},
  journal={Optics Express},
  volume={22},
  number={9},
  pages={11267--11278},
  year={2014},
  publisher={Optical Society of America}
}

@article{ward2016theory,
  title={{Theory and modeling of photodarkening-induced quasi static degradation in fiber amplifiers}},
  author={Ward, Benjamin},
  journal={Optics Express},
  volume={24},
  number={4},
  pages={3488--3501},
  year={2016},
  publisher={Optical Society of America}
}

@article{naderi2013investigations,
  title={{Investigations of modal instabilities in fiber amplifiers through detailed numerical simulations}},
  author={Naderi, Shadi and Dajani, Iyad and Madden, Timothy and Robin, Craig},
  journal={Optics Express},
  volume={21},
  number={13},
  pages={16111--16129},
  year={2013},
  publisher={Optical Society of America}
}

@article{li2017experimental,
  title={{Experimental demonstration of transverse mode instability enhancement by a counter-pumped scheme in a 2 kW all-fiberized laser}},
  author={Li, Zebiao and Huang, Zhihua and Xiang, Xiaoyu and Liang, Xiaobao and Lin, Honghuan and Xu, Shanhui and Yang, Zhongmin and Wang, Jianjun and Jing, Feng},
  journal={Photonics Research},
  volume={5},
  number={2},
  pages={77--81},
  year={2017},
  publisher={Chinese Laser Press and Optical Society of America}
}

@article{dong2022accurate,
  title={{Accurate modeling of transverse mode instability in fiber amplifiers}},
  author={Dong, Liang},
  journal={Journal of Lightwave Technology},
  volume={40},
  number={14},
  pages={4795--4803},
  year={2022}
}

@article{ma2018high,
  title={{High power all-fiberized and narrow-bandwidth MOPA system by tandem pumping strategy for thermally induced mode instability suppression}},
  author={Ma, Pengfei and Xiao, Hu and Meng, Daren and Liu, Wei and Tao, Rumao and Leng, Jinyong and Ma, Yanxing and Su, Rongtao and Zhou, Pu and Liu, Zejin},
  journal={High Power Laser Science and Engineering},
  volume={6},
  pages={e57},
  year={2018},
  publisher={Cambridge University Press}
}

@article{chen2025output,
  title={{Output control of dissipative nonlinear multimode amplifiers using spacetime symmetry mapping}},
  author={Chen, Chun-Wei and Wisal, Kabish and Fink, Mathias and Stone, A Douglas and Cao, Hui},
  journal={Nature Physics},
  pages={1--7},
  year={2025},
  publisher={Nature Publishing Group UK London}
}

@article{rothe2025output,
  title={{Output beam shaping of a multimode fiber amplifier}},
  author={Rothe, Stefan and Wisal, Kabish and Chen, Chun-Wei and Ercan, Mert and Jesacher, Alexander and Stone, A Douglas and Cao, Hui},
  journal={Optics Communications},
  volume={577},
  pages={131405},
  year={2025},
  publisher={Elsevier}
}

@article{wisal2024theorytmi,
  title={{Theory of transverse mode instability in fiber amplifiers with multimode excitations}},
  author={Wisal, Kabish and Chen, Chun-Wei and Cao, Hui and Stone, A Douglas},
  journal={APL Photonics},
  volume={9},
  number={6},
  year={2024},
  publisher={AIP Publishing}
}


\end{document}